\documentclass[aip,
amsmath,amssymb,
preprint,
floatfix
]{revtex4-1}

\usepackage{graphicx}
\usepackage{dcolumn}
\usepackage{bm}

\usepackage[utf8]{inputenc}
\usepackage[T1]{fontenc}
\usepackage{mathptmx}
\usepackage{etoolbox}

\usepackage{float}
\usepackage{subfig}
\usepackage{multirow}%
\usepackage[mathscr]{euscript}
\usepackage{placeins}

\usepackage{xcolor}

\usepackage[normalem]{ulem}

\makeatletter
\def\@email#1#2{%
 \endgroup
 \patchcmd{\titleblock@produce}
  {\frontmatter@RRAPformat}
  {\frontmatter@RRAPformat{\produce@RRAP{*#1\href{mailto:#2}
  {#2}}}\frontmatter@RRAPformat}
  {}{}
}%
\makeatother
\begin{document}

\preprint{AIP/123-QED}

\title[Assessment of Jet Inflow Conditions on the Development of Supersonic Jet
Flows]{Assessment of Jet Inflow Conditions on\\the Development of Supersonic Jet
Flows}
\author{Diego F. Abreu}
 \altaffiliation[PhD. Candidate at ]{Graduate Program in Space Sciences and
 Technologies.}
 \affiliation{Instituto Tecnológico de Aeronáutica, DCTA/ITA \\
 Praça Marechal Eduardo Gomes, 50, 12228--900, São José dos Campos -- SP, Brazil}
 
\author{João Luiz F. Azevedo}%
 \altaffiliation[Senior Research Engineer, ]{Aerodynamics Division}
 \affiliation{Instituto de Aeronáutica e Espaço, DCTA/IAE/ALA \\
 Praça Marechal Eduardo Gomes, 50, 12228--904, São José dos Campos -- SP, Brazil}

\author{Carlos Junqueira-Junior}
 \altaffiliation[Research Engineer]{}
 \affiliation{DynFluid Laboratory, École Nationale Supérieure d'Arts et Métiers, CNAM \\ 151 Boulevard de l'Hôpital, 75013, Paris, Île-de-France, France}
 \email{junior.junqueira@ensam.eu}

\date{\today}

\begin{abstract}

In the present work, large-eddy simulations of free supersonic jet flows are performed to investigate the influence of inflow conditions on the jet flow field and its turbulent properties. A high-order nodal discontinuous Galerkin method is employed to solve the governing equations on the generated mesh. Three different inflow profiles are evaluated to represent the nozzle-exit conditions, namely, inviscid, steady viscous, and unsteady viscous profiles. Velocity and shear stress tensor component profiles obtained from the simulations are compared with experimental data. Among the investigated profiles, the steady viscous inflow shows the most significant deviation from the inviscid case, particularly in the near-field region of the jet inlet. The steady viscous profile also leads to reduced peak velocity fluctuations, showing better agreement with experimental results. Further downstream, the influence of the inflow condition diminishes, with all three profiles converging toward the experimental reference. In addition, power spectral density analyses of streamwise velocity fluctuations reveal that the inflow conditions have little effect on spectral distributions, with numerical results showing consistent agreement with experimental data within the accessible Strouhal range. Beyond these findings, the study provides a highly detailed, high-fidelity database of supersonic jet flow simulations, encompassing six large-eddy computations with different meshes, polynomial refinements, and inflow conditions. The database includes high-frequency data in relevant regions of the jet flow field and is openly available in the Zenodo repository, ensuring accessibility and reusability for the scientific community.

\end{abstract}

\maketitle

\section*{Nomenclature}
\vspace*{-0.25\baselineskip}

\noindent
\begin{minipage}[t]{0.48\textwidth}
\textbf{Latin Symbols}

\vspace{2mm}
\begin{tabular}{l l}
$c_p$ & specific heat at constant pressure \\
$C_s$ & Smagorinsky constant \\
$D_j$ & jet inlet section diameter \\
$\check{E}$ & filtered total energy per unit of mass \\
$\mathbf{F}$ & flux vector \\
$\mathcal{F}$ & contravariant flux vector \\
$J$ & Jacobian of the coordinate transformation \\
$k$ & thermal conductivity coefficient \\
$\ell$ & Lagrange polynomial \\
$\vec{N}$ & unit normal vector \\
$\bar{p}$ & filtered pressure \\
$Pr$ & Prandtl number \\
$\bar{q}$ & filtered heat flux vector \\
$r$ & radial position \\
$R$ & gas constant \\
$\tilde{S}$ & filtered strain-rate tensor \\
$St$ & Strouhal number \\
$St_i$ & station $i$ \\
$\tilde{T}$ & filtered temperature \\
$\tilde{\boldsymbol{u}}$ & filtered velocity vector (Cartesian) \\
$\tilde{\boldsymbol{u}}_c$ & filtered velocity vector (cylindrical) \\
$u_r$ & radial velocity component fluctuation \\
$u_x$ & axial velocity component fluctuation \\
$U_j$ & jet exit velocity \\
$\langle U_x \rangle$ & mean axial velocity component \\
$\mathbf{U}$ & vector of filtered conserved variables \\
\end{tabular}
\end{minipage}
\hfill
\begin{minipage}[t]{0.48\textwidth}
\textbf{Greek Symbols}

\vspace{2mm}
\begin{tabular}{l l}
$\alpha$ & tripping intensity constant \\
$\gamma$ & ratio of specific heats \\
$\delta$ & Kronecker delta \\
$\delta_{BL}$ & boundary-layer thickness \\
$\Delta$ & filter size \\
$\epsilon$ & random number \\
$\mu$ & dynamic viscosity \\
$\boldsymbol{\xi}$ & reference-space coordinate \\
$\bar{\rho}$ & filtered density \\
$\bar{\tau}$ & filtered stress tensor \\
$\phi$ & interpolating polynomial \\
$\psi$ & test function \\
$\Omega$ & reference element
\end{tabular}

\vspace{4mm}
\textbf{Subscripts}

\vspace{2mm}
\begin{tabular}{l l}
$ff$ & far field \\
$h$ & discrete \\
$jet$ & jet \\
$RANS$ & Reynolds-averaged Navier--Stokes \\
$RMS$ & root mean square \\
$SGS$ & subgrid-scale
\end{tabular}

\vspace{4mm}
\textbf{Superscripts}

\vspace{2mm}
\begin{tabular}{l l}
$*$ & element interface \\
$int$ & internal contribution
\end{tabular}
\end{minipage}

\newpage
\section{Introduction}
\label{sec:intro}

Numerous industrial applications, such as heat transfer, thermal machines, and aerospace engineering, use jet flows. For instance, in aeronautical engineering, the thrust produced by small jet planes and massive launch vehicles is due to jet flows. In such applications, the high-velocity jet flow interacting with the surrounding air creates a turbulent flow that results in significant pressure fluctuations. These fluctuations cause noise and structural stresses. High-fidelity data from jet flows must be available early in the design process to meet performance standards and enable proper structural design around the flow to create the next generation of aircraft, launching vehicles, and engines. The high-fidelity characteristics of jet flows can be studied from physical test benches \cite{Woodmansee2004, BridgesWernet2008, MorrisZaman2010} or by performing numerical simulations \cite{BogeyBailly2010, Debonis2017, Bres2017, Junior2018, Abreuetal2024}.

The present study contributes to ongoing efforts on developing high-fidelity numerical techniques for modeling supersonic jet flows. Large-eddy simulation (LES), which relies on the scale separation of turbulent motions, is employed as it can provide high-fidelity flow field data at a significantly lower computational cost compared to direct numerical simulation (DNS) \cite{Pope2000}, and even relative to the cost and complexity of physical experiments. 
Previous work by \citet{BodonyLele2005}, \citet{BogeyBailly2004}, \citet{DebonisScott2002}, \citet{Mendezetal2012}, \citet{Junior2018}, \citet{Shen_et_al_24}, and \citet{Golliard_Mihaescu_25} demonstrated the capability of LES for jet flow simulations.
It is clear that LES methodologies have advanced substantially and are now widely used in various applications \cite{Chapelier_et_al_24, Yang2020, PradhanGhosh2023, Deng2022, Noah2021}. Despite these developments, the specification of accurate jet inflow conditions remains an open challenge in LES of jet flows \cite{Shen_et_al_24, Lian_et_al_23, Dhamankaretal2017, Bresetal2018, Shen_et_al_24}.

A common feature of numerical studies on supersonic jet flows is the inclusion of the nozzle geometry in the computational domain \cite{BodonyLele2008, BogeyMarsden2016, Bresetal2018, ChauhanMassa2022}. While this approach can improve physical realism, it is computationally expensive for large-eddy simulations, as resolving the nozzle walls under high Reynolds number conditions demands fine mesh resolution near the boundaries to accurately capture high-frequency turbulent structures. Excluding the nozzle from the computational domain can significantly reduce the computational cost. 
According to the LES guidelines of \citet{Larssonetal2016}, the simulation of a nozzle flow producing a perfectly expanded supersonic jet would require approximately 160 million grid points to be considered a wall-resolved large-eddy simulation. This mesh size constraint can also affect the numerical stability of the simulation, limiting the maximum CFL (Courant–Friedrichs–Lewy) number of the time marching scheme. However, removing the nozzle geometry from the computational domain requires prescribing a physically realistic inflow condition at the inlet in order to ensure accurate flow development.
In addition to including the nozzle geometry in the computational domain, generating realistic nozzle-exit conditions requires careful modeling of the flow development inside the nozzle. For subsonic jet flows, the influence of different boundary layer profiles originating from the nozzle has been investigated by \citet{BogeyBailly2010}. Further studies have examined the impact of using tripped boundary layers to initiate a more disturbed inflow, also in the subsonic regime \cite{BogeyMarsdenBailly2011}. To match nozzle-exit conditions observed in experiments, the parameters of the tripping technique were tuned to create highly disturbed boundary layers \cite{BogeyMarsden2016}. In these studies \cite{BogeyBailly2010, BogeyMarsdenBailly2011, BogeyMarsden2016}, the nozzle was modeled as a pipe with constant radius, a simplification also adopted in other efforts in the literature \cite{Lorteauetal2018, Lindbladetal2022}. For supersonic jet simulations, \citet{Bresetal2018} employed synthetic turbulence at the inflow, combined with a tripping mechanism inside the nozzle, to replicate realistic jet inflow conditions.

In the subsonic regime, several strategies have successfully generated realistic inflow conditions for jet flow simulations. However, in the supersonic regime, generating high-resolution data can be challenging. A number of studies have incorporated convergent-divergent nozzles into the computational domain \cite{Mendezetal2012, Bres2017, Langenaisetal2019, ShenMiller2019}. In many of these simulations, insufficient mesh resolution within the nozzle interior prevented the flow from developing into a fully turbulent state, as is commonly observed in physical experiments. In some cases, a laminar jet was observed at the nozzle exit \cite{Mendezetal2012}. 
More recent articles \cite{Junior2018, Junior2020, Abreuetal2024} have reported supersonic jet simulations without including the 
nozzle geometry. In these studies, inviscid inflow profiles were imposed at the inlet, and the resulting velocity distributions near the inlet showed similar trends to those obtained from simulations that included the nozzle geometry.

The present effort gives continuity to the evolution of high-fidelity large-eddy 
simulations by investigating the influence of different inflow conditions on the 
development of the supersonic free jet flows. The main idea is to
progressively incorporate more realistic features into the inflow boundary conditions,
starting from a steady viscous turbulent boundary layer and subsequently introducing a
time-dependent turbulent inlet to assess its effects on the development of a perfectly
expanded supersonic jet flow. Quantifying the effects of each inlet condition 
considered here helps to elucidate the direct impact of each evaluated feature on the 
resulting supersonic jet flow. 
The numerical data obtained with an inviscid profile are compared with data obtained with viscous profiles. The viscous profiles are generated from an independent Reynolds-averaged Navier-Stokes (RANS) simulation of a supersonic nozzle flow. Steady and unsteady versions of the viscous profile are investigated. The steady viscous profile is the one obtained from the nozzle RANS simulation, and the unsteady viscous profile is obtained by superimposing a tripping method \cite{BogeyMarsdenBailly2011} to the boundary layer of the viscous profile.

A resolution assessment for such free supersonic jet flows was performed in previous work \cite{Abreuetal2024}, with an inviscid inflow profile imposed in the jet inlet section. The highly resolved simulations agree well with the experimental data in flow regions away from the very near the inlet section. Hence, the current study applies these previous resolution guidelines for jet flow calculations discussed in the present paper.
The numerical simulations are performed with the discontinuous Galerkin spectral element method (DGSEM) \cite{Gao_mossier_munz_24, Kopriva2010, Hindenlang2012}, which is implemented in the FLEXI numerical tool \cite{Kraisetal2021,Blind_et_al_24}. 
The jet flow of interest is a perfectly expanded supersonic condition with a Mach number $1.4$ and a Reynolds number based on the jet inlet diameter of $1.58 \times 10^6$. 

It is important to emphasize that the present research effort adheres to the open-data philosophy. The numerical results presented here, together with those from the resolution study \cite{Abreuetal2024}, are openly available to the scientific community through the Zenodo repository \cite{database12024, database22024, database32024, database42024, database52024, database62024}. The database provides six high-frequency data sets collected at multiple locations in the flow field, focusing on key regions relevant to jet analysis. Four of them are used on the validation of the FLEXI solver and mesh/polynomial resolution study \cite{database12024, database22024, database32024, database42024}, and the other two \cite{database52024, database62024} present data for calculations using steady and unsteady viscous input boundary conditions. The databases are delivered in an open-source format to support diverse applications, including turbulence modeling and the training of artificial intelligence models.

The present work is structured with the governing equations and the discontinuous Galerkin method outlined in Sect.\ \ref{sec:numfor}. The jet flow model is explained in Sect.\ \ref{sec:supjet}, with information on the physical domain, numerical mesh, simulation settings, and data sampling. The available numerical database is presented and detailed in the Sect.\ \ref{sec:database} with all relevant information for its usage. The numerical results of the simulations are presented in Sect.\ \ref{sec:numres}, and the concluding remarks are discussed in Sect.\ \ref{sec:conclusion}.

\section{Numerical Formulation}
\label{sec:numfor}
\subsection{Filtered Navier-Stokes Equations}
\label{sec:goveq}

The Navier-Stokes equations provide a complete representation of the fluid motions, with the solution of each turbulent structure present in the flow field. For many applications, it is not viable to solve the Navier-Stokes equations due to the restrictive spatial and temporal resolution. The large-eddy simulation (LES) formulation employs a spatial filtering process \cite{Garnieretal2009} to separate the length scales of the turbulent eddies, becoming an alternative capable of solving the highly energetic scales with reduced computational costs. The spatial filtering process separates the variables into a resolved part and a subgrid-scale (SGS) part. The filter does not have the resolution to resolve the SGS part and, therefore, the subgrid-scale terms must be modeled. In the present work, the subgrid-scale models \cite{Junior2018, Aminianetal2019} perform the closure of the filtered Navier-Stokes equations.

The filtered Navier-Stokes equations can be written as 
\begin{equation}
\partial \mathbf{U} / \partial t + \nabla \cdot \mathbf{F} =0,
\label{eq:filteredNS}
\end{equation}
where $\mathbf{U}=\left[ \bar{\rho}, \bar{\rho}\tilde{u}_1, \bar{\rho}\tilde{u}_2, \bar{\rho} \tilde{u}_3, \bar{\rho}\check{E} \right]^T$ is the vector of filtered conserved variables, and $\mathbf{F}$ is the flux vector,
\begin{equation}
\mathbf{F}_i= \left[ \begin{array}{c} 
                    \bar{\rho} \tilde{u}_i 
                    \\ \bar{\rho} \tilde{u}_1 \tilde{u}_i + \delta_{1i} \bar{p} - \bar{\tau}_{1i}
                    \\ \bar{\rho} \tilde{u}_2 \tilde{u}_i + \delta_{2i}\bar{p} - \bar{\tau}_{2i}
                    \\ \bar{\rho} \tilde{u}_3 \tilde{u}_i + \delta_{3i}\bar{p} - \bar{\tau}_{3i}
                    \\ (\bar{\rho} \check{E} + \bar{p}) \tilde{u}_i - \bar{\tau}_{ij} \tilde{u}_j + \bar{q}_i
                   \end{array} \right], \hspace*{20pt} \mbox{for } i = 1,2,3. 
\label{eq:evfluxes}
\end{equation}
In the filtered Navier-Stokes equations, $\bar{\rho}$ represents the filtered density, $\tilde{\boldsymbol{u}}=(\tilde{u}_1, \tilde{u}_2, \tilde{u}_3)$ is the Favre filtered velocity vector, $\bar{p}$ is the filtered pressure, 
$\check{E}$ is the filtered total energy per unit of mass, $\bar{\tau}_{ij}$ is the filtered stress tensor, and $\bar{q}$ is the filtered heat flux vector. The Kronecker delta is represented by $\delta_{ij}$. The total energy per unit of volume is defined based on the proposition of Vreman in the so-called ``system~I'' approach \cite{Vremanetal1995}, and it is calculated by
\begin{equation}
\bar{\rho} \check{E} = \frac{\bar{p}}{\gamma - 1} + \frac{1}{2}\bar{\rho}
\tilde{u}_i\tilde{u}_i.
\label{eq:rhoe}
\end{equation}
The filtered pressure, Favre filtered temperature, and filtered density are correlated using the ideal gas equation of state $\bar{p}= \bar{\rho} R \tilde{T}$, where $R$ is the gas constant.

The SGS contribution is introduced using the Boussinesq hypothesis, which modifies the stress tensor and the heat flux vector. The modification of the terms is performed via the addition of an SGS viscosity coefficient, $\mu_{SGS}$, and an SGS thermal conductivity coefficient, $k_{SGS}$. The filtered stress tensor and the filtered heat flux vector are given by
\begin{equation}
\bar{\tau}_{ij}=(\mu + \mu_{SGS}) \left(\frac{\partial \tilde{u}_i}
{\partial x_j} + \frac{\partial \tilde{u}_j}{\partial x_i} \right) 
 - \frac{2}{3} (\mu + \mu_{SGS}) \left(\frac{\partial \tilde{u}_k}
 {\partial x_k} \right) \delta_{ij},
\label{eq:bartau}
\end{equation}
\begin{equation}
\bar{q}_i=-(k+k_{SGS})\frac{\partial \tilde{T}}{\partial x_i}.
\label{eq:barq}
\end{equation}
The molecular viscosity coefficient, $\mu$, is calculated using Sutherland's law \cite{White2006}. The thermal conductivity coefficient of the fluid, $k$, is calculated by
\begin{equation}
k=\frac{\mu c_p}{Pr}.
\label{eq:kq}
\end{equation}
The specific heat at constant pressure is obtained by $c_p=R\gamma/(\gamma-1)$, with the ratio of specific heats, $\gamma$. In Eq.\ (\ref{eq:kq}), $Pr$ is the Prandtl number. As in many aerodynamic calculations, $\gamma$ and $Pr$ are assumed constant, with respective values of $1.4$ and $0.72$\@.

The selection of an appropriate subgrid-scale (SGS) closure for a given flow configuration remains a nontrivial task, given the wide range of models available in the literature. In the present work, the classical \citet{Smagorinsky1963} model is adopted due to its robustness, simplicity, and low additional computational cost. This choice is further supported by the recent study of \citet{Nabae25}, which demonstrated that the Smagorinsky closure is capable of accurately reproducing mean velocity profiles obtained from direct numerical simulation (DNS) of a high-Reynolds-number channel flow with streamwise traveling wave-like wall deformation. In addition, \citet{uddin15} reported that variations in the Smagorinsky constant have a negligible impact on the Reynolds-stress tensor components in regions sufficiently far from wall boundaries when comparing LES and DNS results. Consistent with these findings, previous investigations performed by the research group \cite{Junior2018} showed that, on sufficiently well-resolved grids, the static Smagorinsky model \citep{Smagorinsky1963}, the dynamic Smagorinsky formulation \citep{Germano_etal_91}, and the Vreman model \citep{Vreman04} yield comparable flow statistics, supporting the suitability of the chosen SGS model for the present simulations.

The SGS viscosity coefficient is calculated using the Smagorinsky model
\cite{Smagorinsky1963}:
\begin{equation}
\mu_{SGS}=\bar{\rho}C_s^2\Delta^2 | \tilde{S} |,
\label{eq:musgs}
\end{equation}
where the filter size, $\Delta$, is obtained by the cubic root of the element volume, and the magnitude of the filtered strain rate, $| \tilde{S} |$, is calculated by 
\begin{equation}
| \tilde{S} | = (2 \tilde{S}_{ij} \tilde{S}_{ij})^{0.5}.
\end{equation}
The filtered strain rate tensor is computed by
\begin{equation}
\tilde{S}_{ij}=\frac{1}{2}\left( \frac{\partial \tilde{u}_i}
{\partial x_j} 
+ \frac{\partial \tilde{u}_j}{\partial x_i} \right).
\label{eq:tildeSij}
\end{equation}
The Smagorinsky model constant, $C_s$, has a value of $0.148$ according to \citet{Garnieretal2009}. The SGS thermal conductivity coefficient is calculated using Eq.\ (\ref{eq:kq}) and substituting the viscosity coefficient and the Prandtl number by the SGS viscosity coefficient, $\mu_{SGS}$, and the SGS Prandtl number, $Pr_{SGS}$. The SGS Prandtl number is also assumed constant with a value of $0.9$\@.

\subsection{Nodal Discontinuous Galerkin Scheme}
\label{sec:nummet}

The nodal discontinuous Galerkin (DG) scheme solves the filtered Navier-Stokes equations. The nodal DG used in the present work is referred to as the discontinuous Galerkin spectral element method (DGSEM) \cite{Kopriva2010, Hindenlang2012}, which is implemented in the FLEXI numerical framework \cite{Kraisetal2021}. The computational framework has been validated using different flow configurations \cite{GassnerBeck2013, Gassner2014, Becketal2014, FladGassner2017, DauricioAzevedo2023}. One characteristic of the method that distinguishes it from other nodal DG schemes is that it solves the equations exclusively in hexahedral elements. The restriction to the use of hexahedral elements increases the computational efficiency and simplifies the implementation.

Hence, the physical domain is divided into multiple non-overlapping hexahedral elements. Each hexahedral element is mapped onto a reference unit cube element, $\Omega=[-1,1]^3$. The filtered Navier-Stokes equations, Eq.\ (\ref{eq:filteredNS}), are mapped to the reference domain by
\begin{equation}
J \partial \mathbf{U} / \partial t + \nabla_{\xi} \cdot \mathcal{F} = 0.
\label{eq:refNS}
\end{equation}
The reference element coordinates are given by $\boldsymbol{\xi}=(\xi,\eta,\zeta)^T$. The divergence operator for the reference element coordinates is represented by $\nabla_{\xi}$. The Jacobian of the coordinate transformation is given by $J= \arrowvert \partial \boldsymbol{x} / \partial \boldsymbol{\xi} \arrowvert$, and the last term is the contravariant flux vector, $\mathcal{F}$.

The solution is approximated by a nodal polynomial interpolation
\begin{equation}
\mathbf{U}(\boldsymbol{\xi}) \approx \sum_{p,q,r=0}^N \mathbf{U}_h
(\xi_p,\eta_q,\zeta_r,t)\phi_{pqr}(\boldsymbol{\xi}),
\label{eq:num_sol}
\end{equation}
where the vector of conserved variables at each solution node is represented by $\mathbf{U}_h (\xi_p,\eta_q,\zeta_r,t)$ and $\phi_{pqr} (\boldsymbol{\xi})$ is the interpolating polynomial. The interpolating polynomial is a tensor product of one-dimensional Lagrange polynomials, $\ell$, in each space direction in the formulation of the present implementation. The Lagrange polynomials are calculated by
\begin{equation}
\phi_{pqr}(\boldsymbol{\xi})
=\ell_p(\xi)\ell_q(\eta)\ell_r(\zeta), 
\hspace{10pt} \ell_p(\xi)= \prod_{\substack{i=0 \\ i \ne p}}^{N_p} 
\frac{\xi-\xi_i}{\xi_p-\xi_i},
\label{eq:interp_poly}
\end{equation}
with equivalent definitions for the other two directions. The components of the contravariant fluxes are approximated by interpolation with the employment of a similar process used for obtaining the solution. The contravariant flux from the $\xi$-component is given by
\begin{equation}
\mathcal{F} \left( \mathbf{U}(\boldsymbol{\xi}) \right) \approx 
\sum_{p,q,r=0}^N \mathcal{F}_{\xi,h}
(\xi_p,\eta_q,\zeta_r,t)\phi_{pqr}(\boldsymbol{\xi}),
\label{eq:cont_flux}
\end{equation}
with analogous definitions for the other two directions.

The discontinuous Galerkin formulation is obtained when the mapped Navier-Stokes equations, Eq.\ (\ref{eq:refNS}), are multiplied by the test function, $\psi=\psi(\xi)$, and integrated over the reference element, $\Omega$. The strong formulation is obtained when Eq.\ (\ref{eq:refNS}) is integrated by parts twice in sequence. The discontinuous Galerkin formulation in strong form is given by 
\begin{multline}
\int_\Omega J \frac{\partial \mathbf{U}}{\partial t} \psi d 
\boldsymbol{\xi}
+ \int_{\partial \Omega} \left( \left( \mathcal{F} \cdot \vec{N} \right)^{\star}
- \mathcal{F}^{int} \right) \psi d \boldsymbol{S}_\xi
+ \int_{\partial \Omega} \left( \nabla_\xi \cdot 
\mathcal{F}\right) \psi d \boldsymbol{\xi}=0.
\label{eq:DGstrongmapfilterNS}
\end{multline}
The boundaries of $\Omega$ element are defined by $\partial \Omega$, and $\vec{N}$ is the unit normal vector of the faces of the elements in the reference space. The $(\cdot)^*$ superscript indicates the need to handle the properties at the interface of the elements, and the $(\cdot)^{int}$ represents the interior contribution to the surface fluxes at the interface. The discrete form of the scheme is obtained by substituting the integrals with Gaussian quadratures and collocating the solution and integration points.

A split formulation is employed to improve the stability of the numerical method. The formulation is based on the work of \citet{Pirozzoli2011} and, subsequently, it was adapted for the DGSEM formulation \cite{Gassner2016}. The split formulation was used in the solution of the Taylor-Green vortex problem \cite{Gassner2016}, the channel flow \cite{DauricioAzevedo2021}, and a 2-D aerodynamic profile \cite{DauricioAzevedo2023}. It was capable of stabilizing the numerical schemes in the three problems. The Roe Riemann solver \cite{Roe1981} with the entropy fix of \citet{Harten1983} is used. The lifting scheme of \citet{BassiRebay1997}, BR2, is employed for the spatial discretization of viscous fluxes. The temporal evolution of the simulations is performed with a low-storage Runge-Kutta scheme with three stages and third-order accuracy \cite{Kopriva2009}. The finite-volume sub-cell shock capturing method of \citet{Sonntag2017} handles the shock waves in the flow field. The switching function of Jameson-Schmidt-Turkel \cite{JST1981} is the base for the shock wave indicator. The shock-capturing method operation switches the formulation of the cells from the DGSEM to a second-order reconstruction finite volume (FV) scheme with a minmod slope limiter \cite{Hirsch1990b}. The shock-capturing method preserves the number of degrees of freedom independently of the formulation employed.

\section{Supersonic Free Jet Flow}
\label{sec:supjet}
\subsection{Problem Description and Modeling Approach}
\label{sec:modeldesc}

The study investigates a supersonic free round jet flow with isothermal, perfectly expanded operating conditions. The jet flow has an inlet Mach number of $1.4$ and a Reynolds number based on the inlet diameter of $1.58 \times 10^6$. The inflow pressure matches the ambient pressure in the perfectly expanded condition. The isothermal jet flow indicates that the jet temperature is the same as the ambient temperature. The jet operating condition is defined based on the experiments performed by \citet{BridgesWernet2008}.

The physical domain comprises the region where the jet flow interacts with the ambient air. The nozzle geometry is not included in the physical domain in order to assess the influence of inviscid, viscous, and unsteady viscous inflow conditions in the jet flow field. Additionally, when the nozzle geometry is not represented in the computational domain, the computational costs are reduced. Therefore, it is of primary interest to investigate options for imposing realistic inflow conditions. The jet inlet section is represented by a circular face positioned at station $x/D_j=0.0$, with the center positioned in $y/D_j=z/D_j=0.0$. The reference for the non-dimensional spatial variables is the jet inlet diameter, $D_j$. The $x$ position vector component is associated with the jet axial direction, and the $y$ and $z$ position vector components are associated with the jet axis perpendicular directions. The variable $r$ is the radial position from the domain center axis. It is calculated by $r = (y^2 + z^2)^{0.5}$. The external surfaces of the domain have an axisymmetric divergent shape.

\begin{figure}[htb!]
\centering
	\includegraphics[trim = 40mm 12mm 25mm 20mm, clip, width=0.8\linewidth]{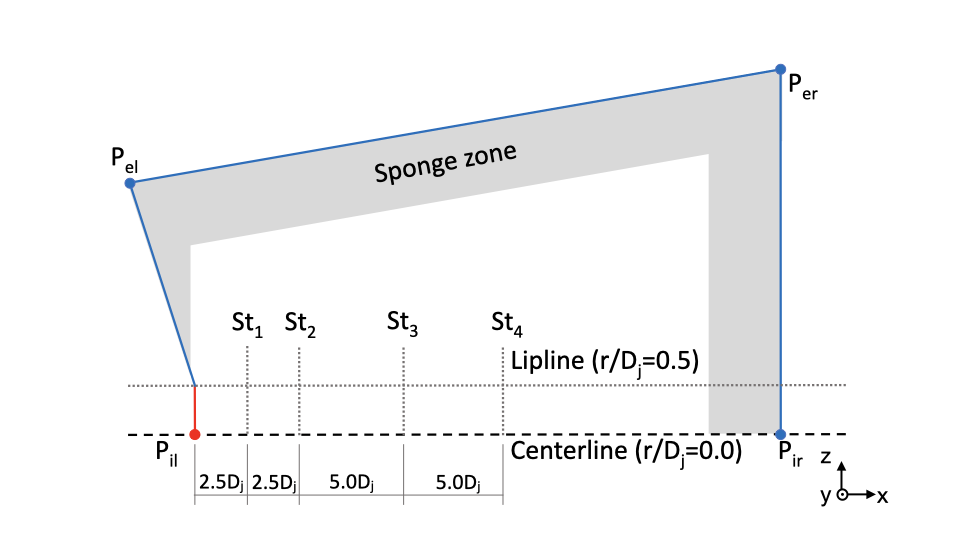}
    \caption{2-D schematic representation of the computational domain employed for the jet flow simulation.}
    \label{fig:geo}
\end{figure}

A sketch of the computational domain is presented in Fig.\ \ref{fig:geo}. The figure shows 
a visual description of the computational geometry and the definition of some important 
regions of the flow field where the numerical data is extensively investigated. Simulation 
data from those regions are available in a public repository \cite{database12024, database22024, database32024, database42024, database52024, database62024} 
and the reader can find more details in Sect.\ \ref{sec:database}. The computational domain extends from the jet inlet section, centered at $P_{il}$ point, to a distance of $40D_j$ along the centerline until the $P_{ir}$ point. The divergent shape starts at $x/D_j = -0.95$ and $r/D_j = 8.0$, $P_{el}$ point, and ends at $x/D_j = 40.0$ and $r/D_j = 12.5$, $P_{er}$ point. The red boundary in Fig.\ \ref{fig:geo} represents the nozzle inlet section, and the blue boundaries represent the region where the far field condition is imposed. A dashed black line illustrates the jet flow centerline, $r/D_j=0.0$. The gray area is the region where a sponge zone is used. The gray dot lines represent the jet lipline, $r/D_j=0.5$, and stations, $St_1$ through $St_4$, which are positioned at $x/D_j=2.5$, $x/D_j=5.0$, $x/D_j=10.0$ and $x/D_j=15.0$, respectively. The computational domain used in the present simulations is the same used for the finest mesh in the resolution study \cite{Abreuetal2024}.

The numerical mesh is generated using the multiblock strategy. In such a strategy, the computational domain is divided into multiple non-overlapping hexahedral elements, which is necessary to satisfy the requirements of the present computational tool \cite{Kopriva2010, Hindenlang2012}. The numerical grid was created using the GMSH mesh generator \cite{Geuzaine2009}. The mesh topology and refinement level were investigated in the resolution study previously performed \cite{Abreuetal2024}. Such study has identified that the use of a numerical mesh with $15.4 \times 10^6$ elements together with a third-order accurate spatial discretization provided, in general, a good agreement with experimental data. The numerical data from the resolution study are validated with two other numerical datasets in Sect.\ \ref{sec:validation}.

\begin{figure*}[htb!]
\centering
  \subfloat[Edges of the block structure from the multiblock mesh.]{
	\includegraphics[trim = 0mm 0mm 20mm 0mm, clip, width=0.48\linewidth]{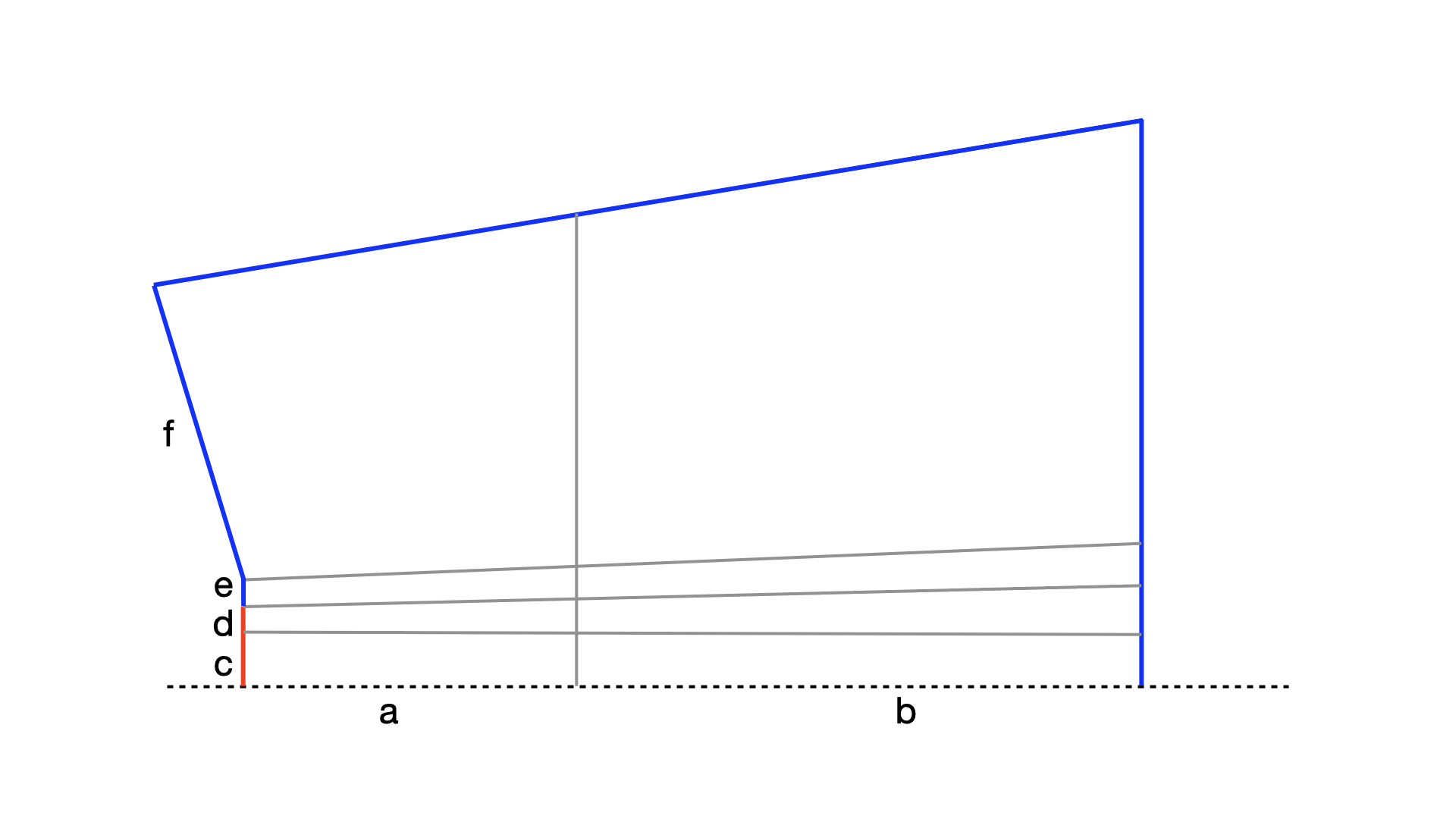}
    \label{fig:2dblock}
    }
  \subfloat[Numerical mesh used in the simulations.]{
  	\includegraphics[trim = 0mm 0mm 20mm 0mm, clip, width=0.48\linewidth]{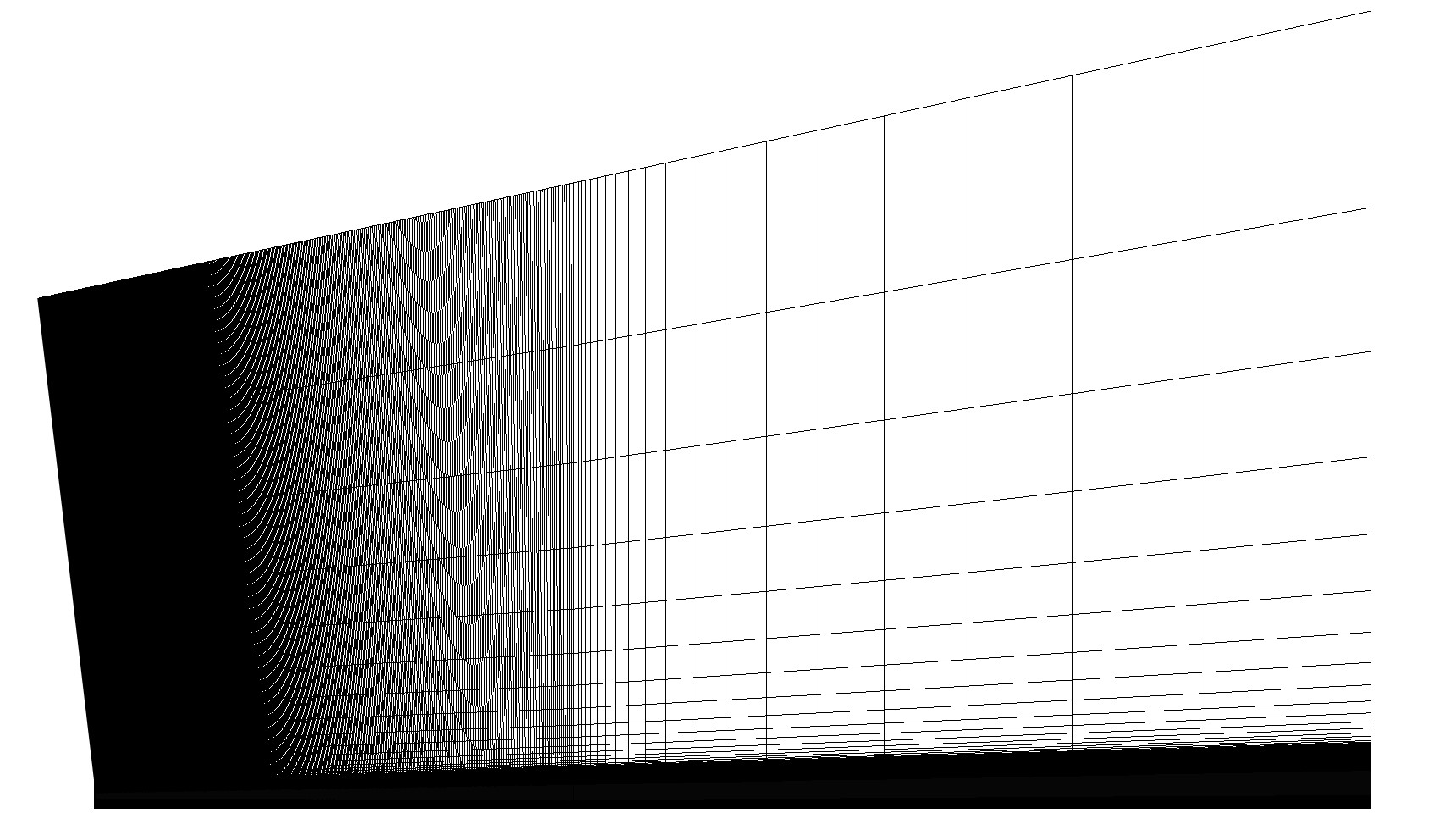}
    \label{fig:2dmesh}
  }
  \caption{2-D $xr$-plane cut from the numerical mesh representing the multiblock structure and the mesh generated.}
\end{figure*}

The multiblock structure employs a central block around the axis of the computational domain, which is connected to four other blocks in the faces that are parallel to the geometry axis. After the central block, the subsequent blocks have an annular shape. The 2-D block structure in the $xr$-plane is presented in Fig.\ \ref{fig:2dblock}. The computational domain is divided into two blocks in the axial direction of the flow, identified by $a$ and $b$ edges, and four blocks in the radial direction, which are identified by $c$, $d$, $e$, and $f$ edges. The first set of blocks in the axial direction, $a$ edge, has $651$ elements, and it extends up to $x/D_j=15.0$. The second set of blocks in the axial direction has $21$ elements and a large stretching ratio to produce large elements close to the boundaries of the domain to dampen any wave that could reach the far field boundaries and potentially could be reflected back into the jet flow.
\begin{figure*}[htb!]
\centering
  \subfloat[2-D \textit{xr}-plane cut close to the inlet section.]{
	\includegraphics[trim = 0mm 0mm 0mm 0mm, clip, width=0.48\linewidth]{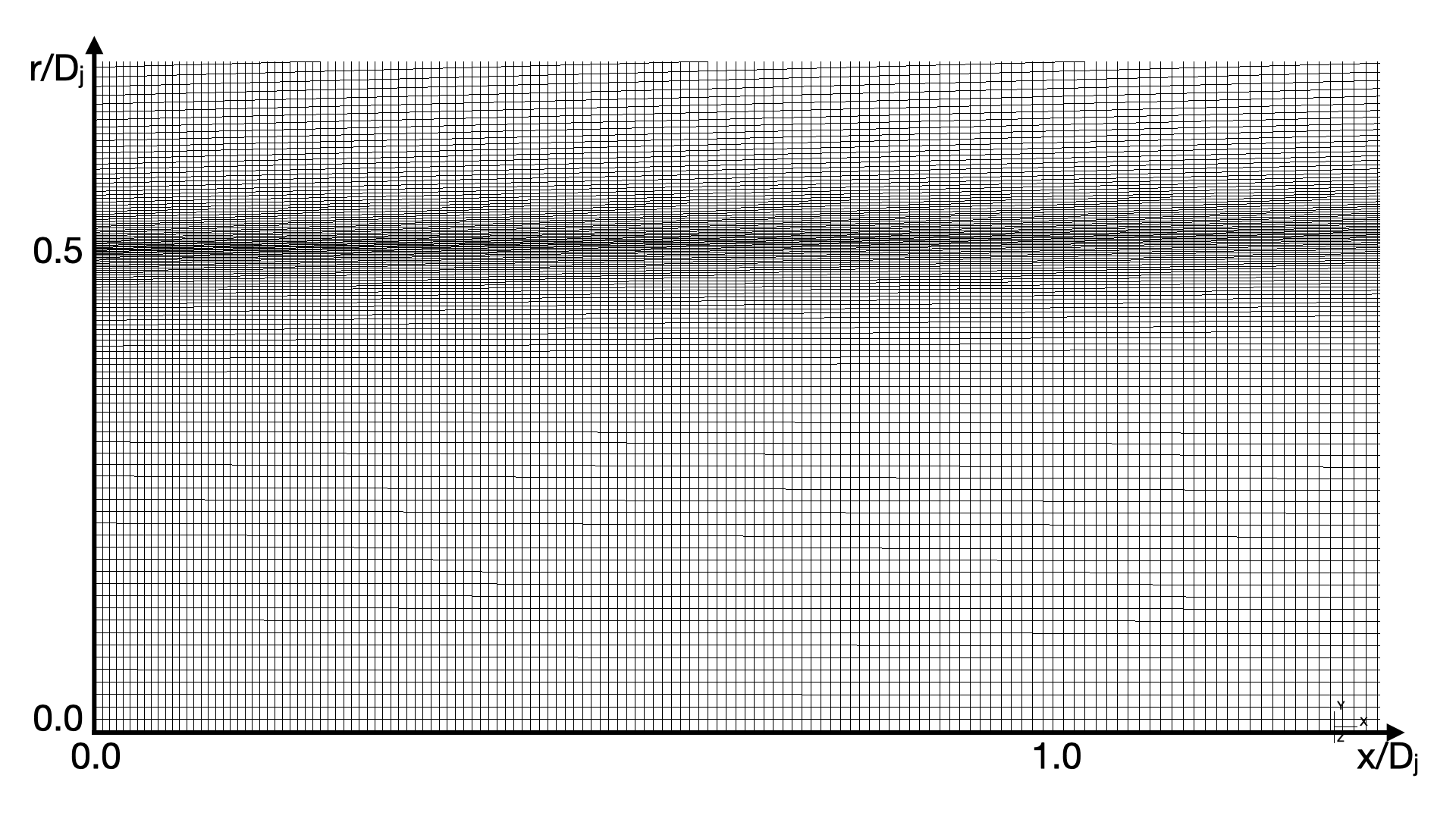}
    \label{fig:2dzoom}
    }
  \subfloat[2-D \textit{yz}-plane cut from the inlet section.]{
  	\includegraphics[trim = 0mm 0mm 0mm 0mm, clip, width=0.48\linewidth]{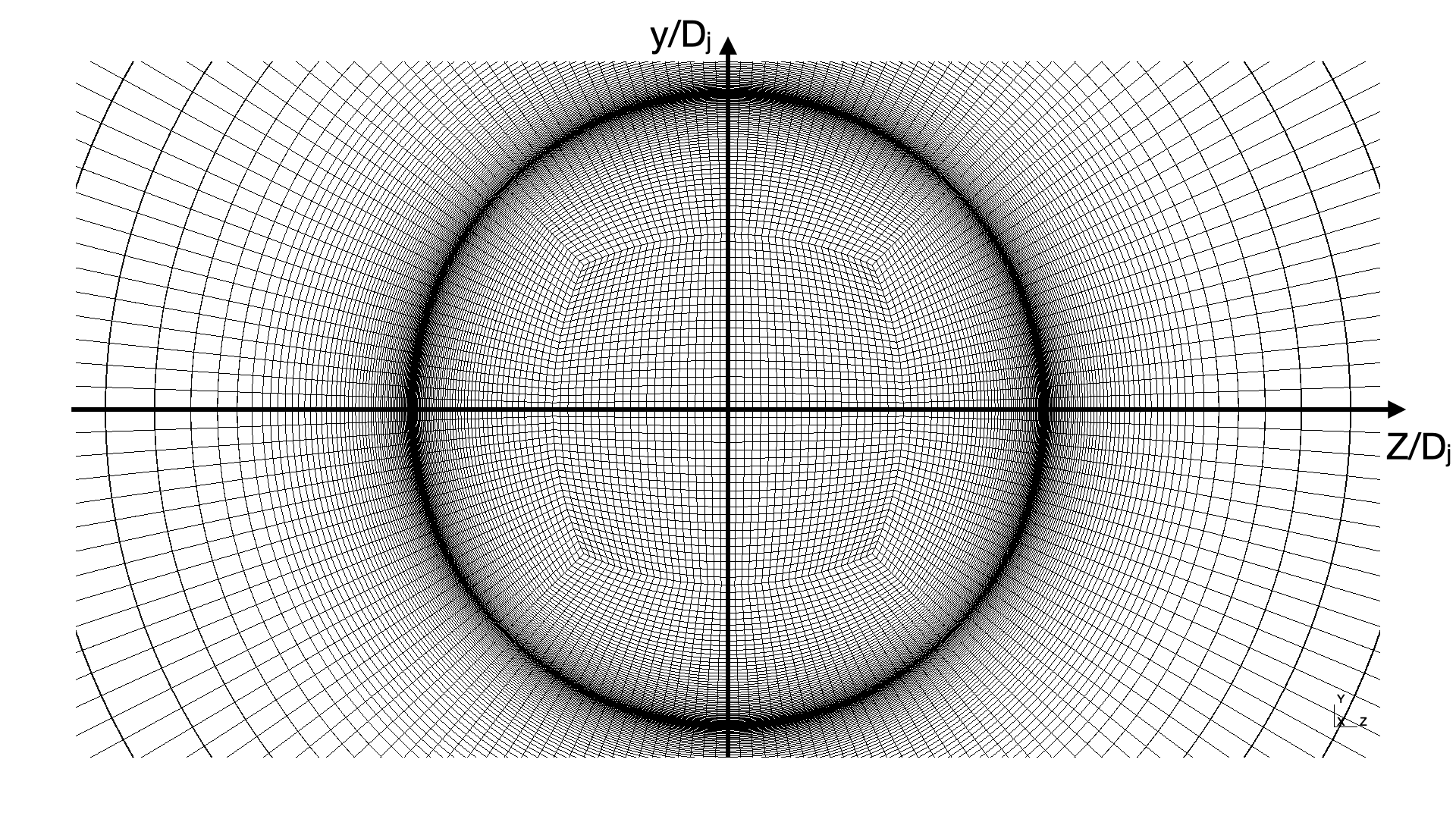}
    \label{fig:inletzoom}
  }
  \caption{Details from the inlet section of the numerical mesh.}
\end{figure*}

In the radial direction, the first block, $c$ edge, is the central block, which presents a uniform element distribution with $45$ elements. The topology used in the numerical mesh imposes the same number of elements in this region as the number of elements in the azimuthal direction due to the mesh structure. The second and third layers of blocks, $d$ and $e$ edges in Fig.\ \ref{fig:2dblock}, are the ones where the development of the jet mixing layer effectively occurs. They contain small elements close to their adjacent face at the inlet section. Such small elements are needed to capture the small eddies in the region. A 2-D longitudinal plane view of the numerical mesh used in the simulations is presented in Fig.\ \ref{fig:2dmesh}. The inlet section of the mesh is presented in detail in Fig.\ \ref{fig:2dzoom}. It is possible to observe the small element size in the axial direction and a highly refined region in the jet mixing layer. The $d$ and $e$ edges present $51$ elements. The opening angle of $1^\circ$ in the region of the jet mixing layer was identified in the mesh resolution study \cite{Abreuetal2024}. A uniform cell size distribution across $c$, $d$, and $e$ edges at $x/D_j=15.0$ is imposed. The crossflow plane of the mesh at in the inlet section is presented in Fig.\ \ref{fig:inletzoom}. It illustrates the interior block structure. 

\begin{figure*}[htb!]
\centering
	\subfloat[Radial mesh cell size distribution.] {
	\includegraphics[width=0.46\linewidth]{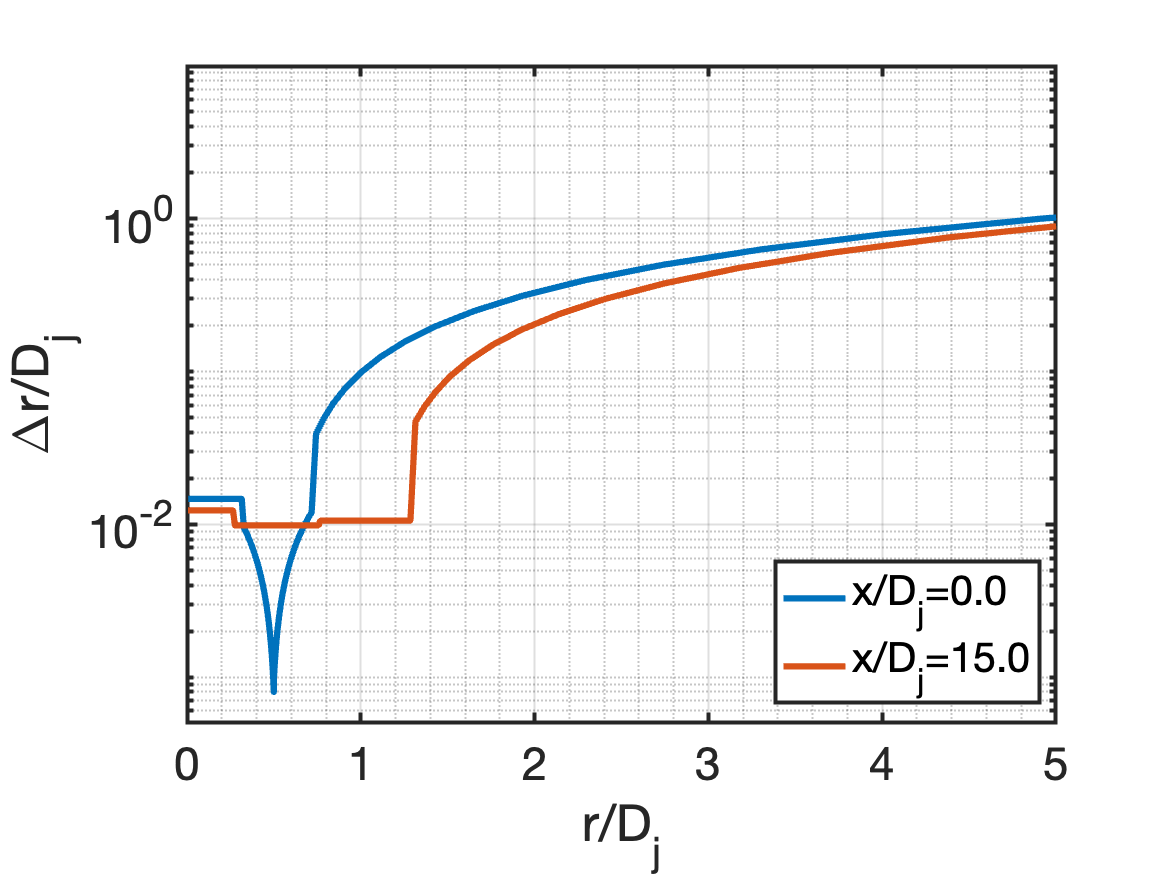}
    \label{fig:radref}
	}
	\subfloat[Axial mesh cell size distribution.]{
	\includegraphics[width=0.46\linewidth]{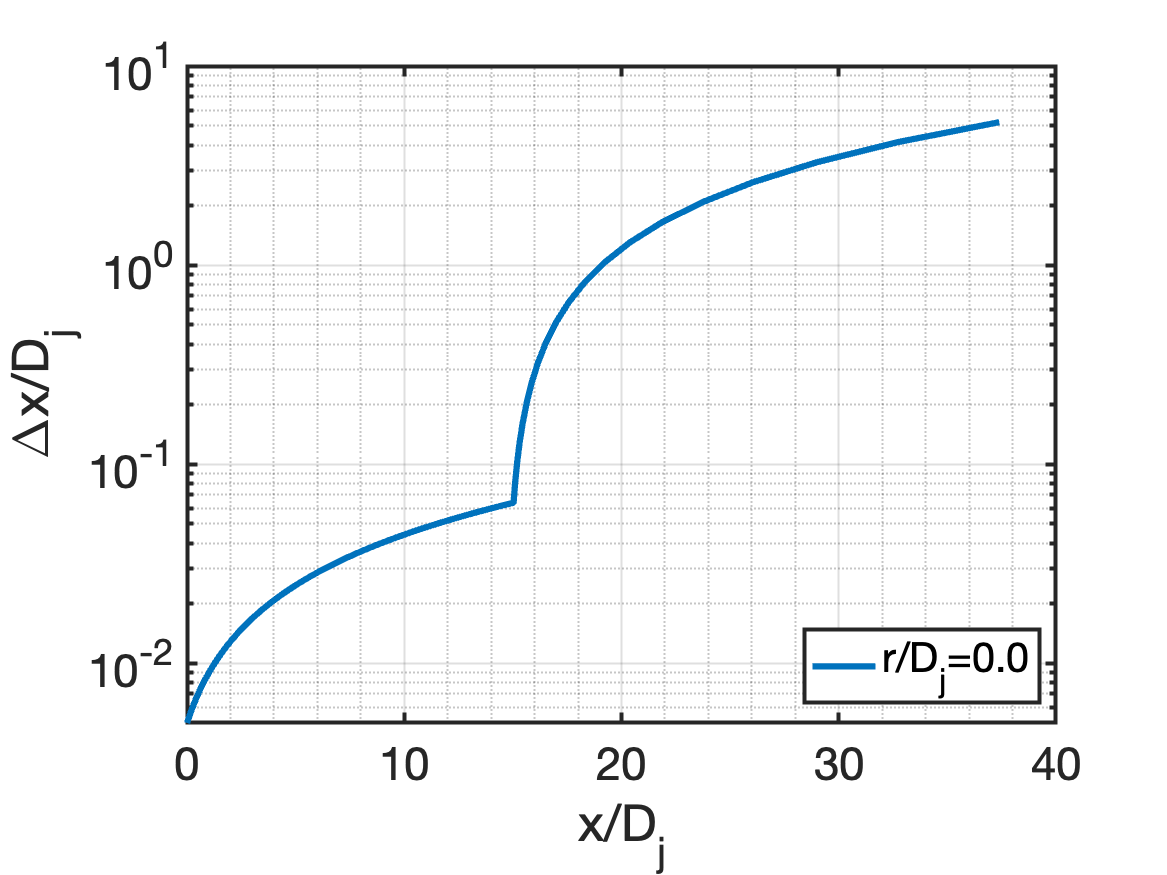}
    \label{fig:axialref}
	} 
    \caption{Mesh cell size distributions at two longitudinal positions, $x/D_j=0.0$ and $x/D_j=15.0$, and one radial position, $r/D_j=0.0$.}
    \label{fig:meshref}
\end{figure*}

The mesh cell size distributions are presented in Fig.\ \ref{fig:meshref}. The radial mesh cell size distributions are presented in two longitudinal stations to illustrate the different cell size distributions from the inlet section and at a position that is distant from the inlet section. The radial cell size distributions are presented in Fig.\ \ref{fig:radref} for $x/D_j=0.0$ and $x/D_j=15.0$. The axial mesh size distributions are presented in Fig.\ \ref{fig:axialref} at $r/D_j=0.0$.

It is important to mention that, following the large-eddy simulation guidelines of \citet{Larssonetal2016}, the authors estimated the mesh requirements for explicitly including the nozzle geometry. Applying the recommended near-wall and streamwise resolution criteria to a nozzle of length $L = 2.75D_j$ with an estimated boundary-layer thickness $\delta \approx 0.05D_j$ yields a mesh on the order of $160 \times 10^6$ grid points when using the same multiblock strategy adopted for the jet domain. This mesh size is approximately one order of magnitude larger than the $15.4 \times 10^6$ elements employed in the present jet simulations. Furthermore, such a substantial increase in mesh size is typically associated with smaller grid spacings, which impose numerical stability limitations due to CFL constraints. This reaffirms the decision to exclude the nozzle geometry from the simulation domain and study the effects of different input boundary layers on the flow.

\subsection{Boundary Conditions}
\label{sec:boundcond}

Weakly enforced Dirichlet-type boundary conditions \cite{Arnoldetal2002, BazilevsHughes2007} are applied to all boundaries of the computational domain. Far field boundaries further implement sponge zone conditions \cite{Flad2014} with the objective of guaranteeing that no signals are reflected back into the domain, as indicated in Fig.\ \ref{fig:geo}. In the present work, the main objective is precisely to assess the influence of different inflow conditions on the development of the jet flow field. The difference between the numerical simulations is the Dirichlet condition employed, which is referred to as the reference state. The numerical simulations use two reference states: one for the jet inflow condition, which is applied to the red surface in Fig.\ \ref{fig:geo}, and the second for the ambient flow condition applied to the blue surfaces. The jet inflow condition is referred to by $(\cdot)_{jet}$, while the ambient flow is referred to by $(\cdot)_{ff}$. Three reference states are investigated for the inflow condition: an inviscid profile, a steady viscous profile, and an unsteady viscous profile. The detailed description of each reference state is presented in sequence in this section. The ambient air reference state has a small longitudinal velocity component with a Mach number of $M_{ff}=0.01$ in order to avoid any spurious recirculation close to the boundaries \cite{Mendezetal2012}. The ambient air pressure, $p_{ff}$, and temperature, $T_{ff}$, are based on the Reynolds number obtained in the reference experiments \cite{BridgesWernet2008}.

\subsubsection{Inviscid Profile}
\label{sec:invprof}

The inviscid inflow profile is the simplest inflow condition investigated in the present work. Uniform properties in the radial and azimuthal directions characterize the inviscid inflow profile. The values of the properties of the inviscid profile are based on the experimental data \cite{BridgesWernet2008}, Mach number of $1.4$, and the Reynolds number based on the jet inlet diameter of $1.58 \times 10^6$. The flow is characterized as perfectly expanded and isothermal, which leads to $p_{jet}/p_{ff} = T_{jet}/T_{ff} = 1$.

\subsubsection{Steady Viscous Profile}
\label{sec:ransprof}

The inviscid profile is a simplistic model of the experimental test case due to the absence of the correct boundary layer and the turbulent characteristics of the flow. The addition of these features is treated in the present work as an attempt to improve the physical representation of the jet inflow boundary conditions. In the present section, an \textit{a priori} RANS simulation of the nozzle flow is employed to generate the primitive variable profiles of the nozzle-exit section, which are imposed as the inflow condition for the LES simulations. This strategy introduces the boundary layer to the mean inflow condition.

\begin{figure*}[htb!]
\centering
	\subfloat[2-D representation of the computational domain of the nozzle.] {
	\includegraphics[width=0.47\linewidth]{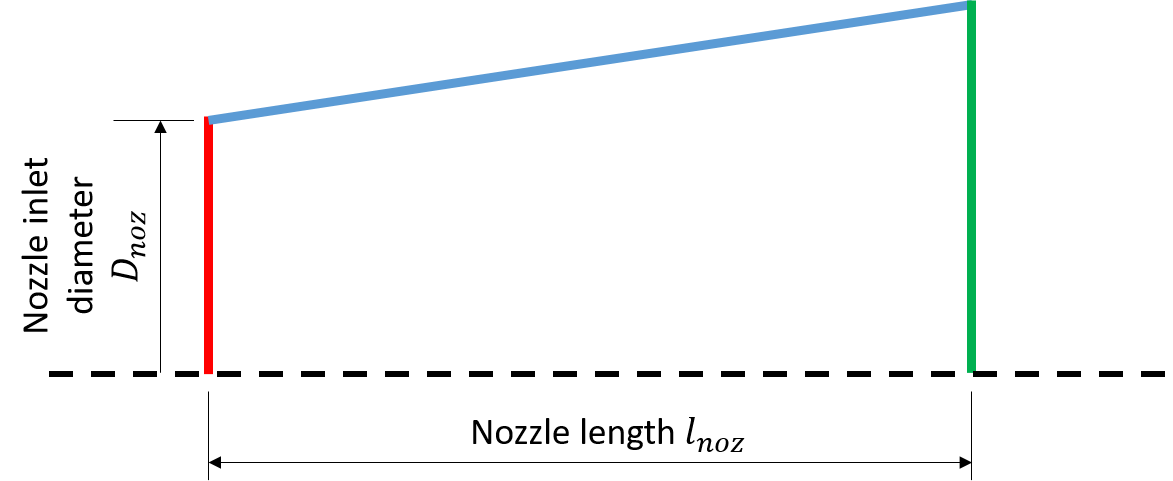}
    \label{fig:nozz_sketch}
	}
	\subfloat[Cut plane of the nozzle mesh used in the simulations.]{
	\includegraphics[width=0.47\linewidth]{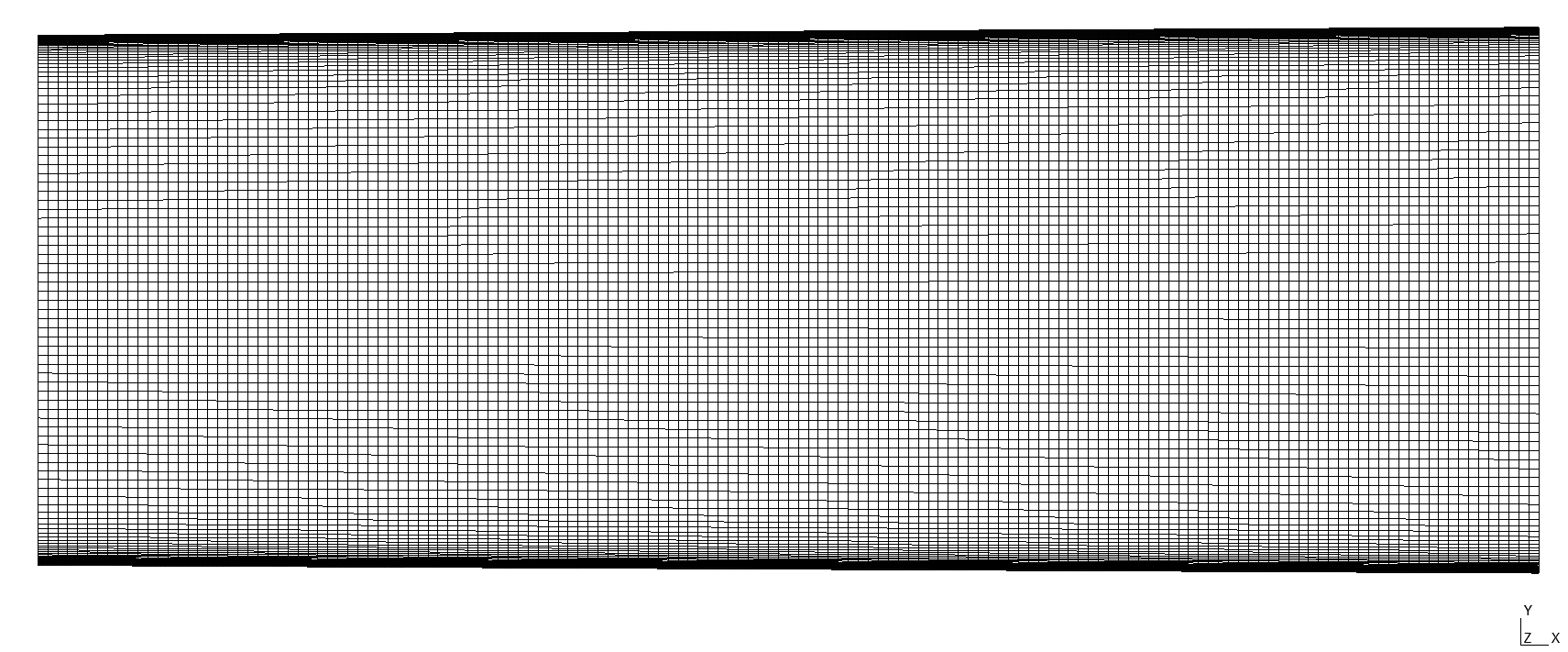}
    \label{fig:mesh_rans}
	}
    \caption{2-D schematic representation and cut plane of the grid used 
    for the RANS calculation of the nozzle flow.}
    \label{fig:rans_sketch}
\end{figure*}

The RANS simulations are performed with the CFD++ solver \cite{Chakravarthy1999}, which employs a second-order accurate finite volume formulation with the one-equation Spalart-Allmaras turbulence model \cite{SpalartAllmaras94}. A point implicit scheme, which is based on an implicit Euler scheme, is used for the relaxation method. The nozzle geometry is represented by a slightly divergent shape, which avoids choking the supersonic flow inside the nozzle duct. The strategy of simulating the nozzle by an almost constant section is adapted from the simulations that used constant section nozzle ducts for subsonic flows \cite{BogeyBailly2010, BogeyMarsdenBailly2011, BogeyMarsden2013, BogeyMarsden2016, Bresetal2018}. The nozzle design led to a nozzle with an inlet diameter of $D_{noz}=0.9686D_j$ and a nozzle length of $L_{noz}=2.75D_j$. A 2-D sketch of the physical domain used in the simulations of the nozzle flow is presented in Fig.\ \ref{fig:nozz_sketch}.

The meshes generated for the nozzle simulations also use the multiblock topology to generate hexahedral elements. The meshes are generated with the GMSH mesh generator \cite{Geuzaine2009}. The near-wall distance is defined to provide a $Y^+$ of one, in order to provide adequate near-wall resolution for the RANS turbulence model. The mesh has approximately $50$ elements per nozzle-exit diameter and one element for each $2$ degree in the azimuthal direction. The nozzle mesh has $1.6 \times 10^6$ elements. Figure\ \ref{fig:mesh_rans} presents the 2-D cut plane of mesh used in the numerical simulations.

The boundary conditions applied in the RANS simulations are supersonic inlet and outlet, and no-slip wall. A linear CFL ramp is employed to avoid any numerical instabilities in the first steps of the simulations, reaching a maximum value of $100$. The simulation is performed for $2000$ iterations and presents a residual drop of $6$ orders of magnitude compared to the initial residue value for each variable.

\begin{figure*}[htb!]
\centering
	\subfloat[2-D Cut plane of the nozzle colored by Mach number contours.]{
	\includegraphics[width=0.5\linewidth]{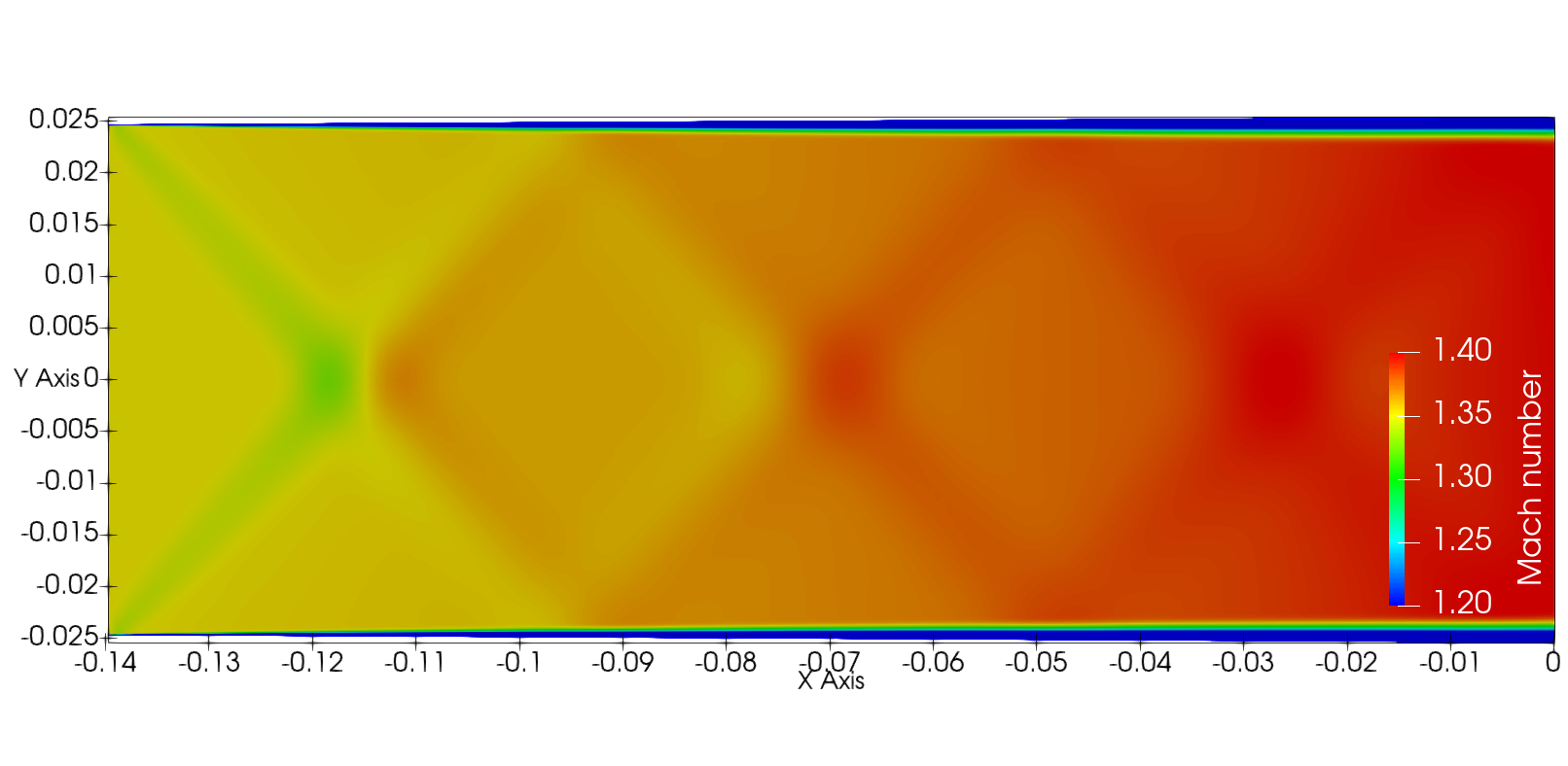}
    \label{fig:noz_contours}
	}
    \subfloat[Longitudinal velocity profile at the output of the nozzle.]{
	\includegraphics[width=0.38\linewidth]{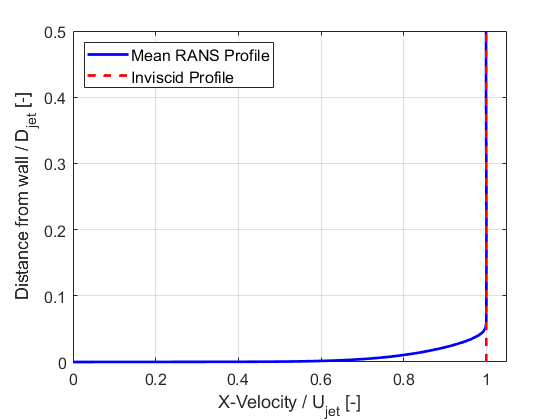}
	\label{fig:noz_velx}
	}
\caption{Numerical data obtained from the RANS simulation of the nozzle flow.}
\label{fig:noz_results}
\end{figure*}

The 2-D cut plane of the nozzle flow, Fig.\ \ref{fig:noz_contours}, is colored by Mach number contours, and it allows the visualization of the flow inside the nozzle and highlights the shock wave structure generated in the interior of the nozzle flow. The flow inside the nozzle presents small velocity increase in the jet centerline. The flow Mach number increases from $1.35$ to $1.40$. The nozzle opening angle and length were defined to produce a small velocity increment. The velocity profile in the nozzle-exit section is presented in Fig.\ \ref{fig:noz_velx}. The velocity profile shows the development of a turbulent boundary layer with thickness  $\delta_{BL}=0.05D_j$. A small velocity deviation is observed out of the boundary layer when compared to the constant value from the inviscid profile.

The three velocity components, density, and pressure profiles are extracted from the RANS simulations and stored to be used in the LES calculations. The profiles are stored on a radial basis. The interpolation between the data stored from the RANS simulations and the LES inflow surface is performed through a piece-wise cubic interpolation \cite{VanLoan1999} of the primitive variables, which are, then, converted to the conserved variables used as the reference state for the inlet boundary condition.

\subsubsection{Unsteady Viscous Profile}
\label{sec:trip}

The two conditions presented in the previous sections result in steady flows in the inlet section. Steady flows represent mean flow characteristics. However, they are not capable of reproducing the high-frequency unsteady data from real flows. Tripping techniques are an option to enhance the turbulence intensity close to the region where it is applied. The tripping methods are numerical features that introduce unsteadiness to flow simulations that can trigger flow turbulence. There are several tripping techniques available in the literature \cite{BogeyMarsdenBailly2011,BogeyMarsdenBailly2011, Langenaisetal2019, GandHuet2021, Dhamankaretal2017}.

\begin{figure}[htb!]
\centering
	\includegraphics[trim = 245mm 100mm 180mm 65mm, clip, width=0.35\linewidth]{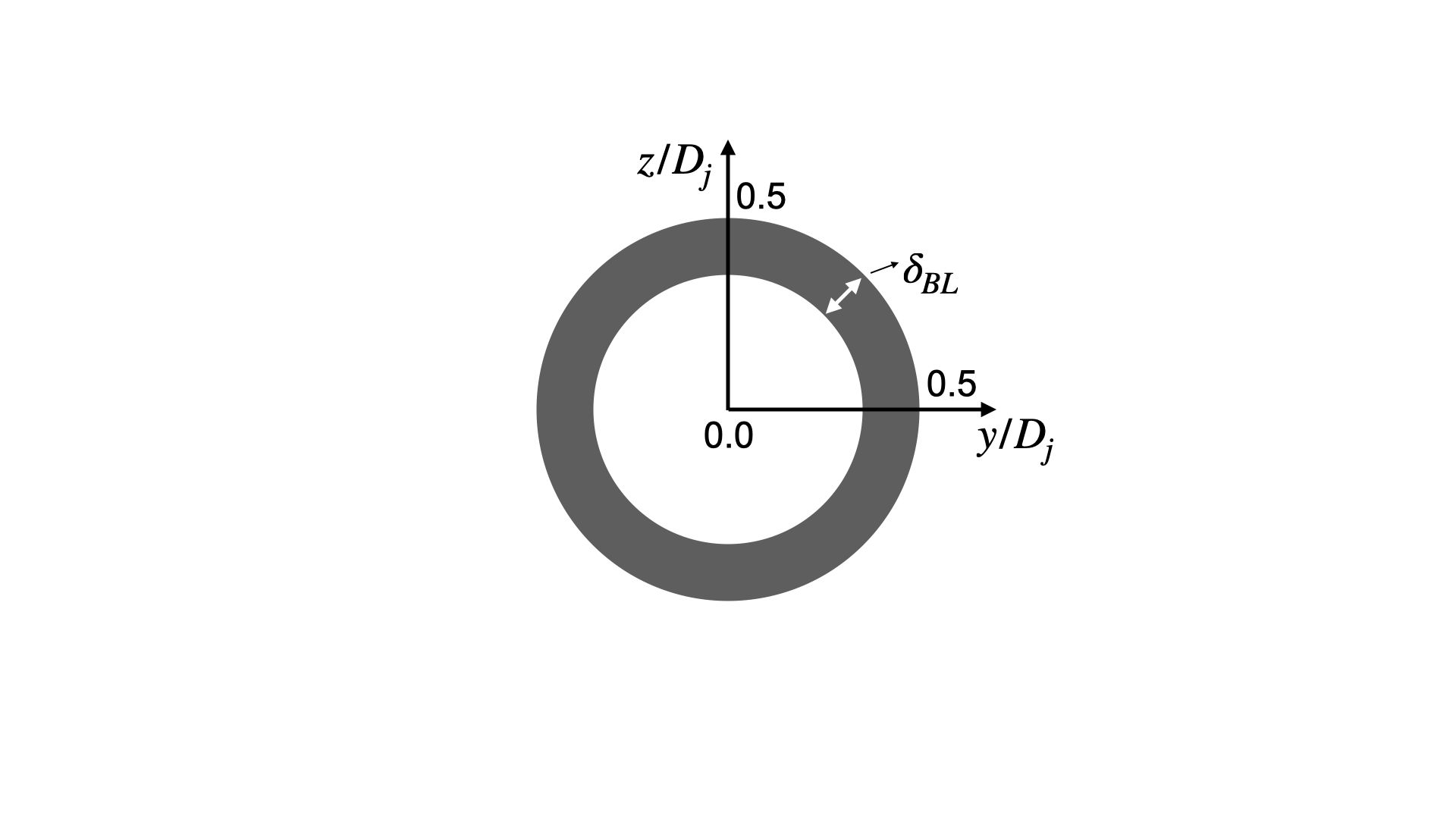}
    \caption{Annular region on the inflow surface where the tripping method is applied. The thickness of the annular region is equal to the boundary-layer thickness, $\delta_{BL}$, obtained from the nozzle-exit velocity profile.}
    \label{fig:tripping-region}
\end{figure}

In the current work, the mean viscous profile presented in Sect.\ \ref{sec:ransprof} is superimposed by property fluctuations using the tripping method proposed by \citet{BogeyMarsdenBailly2011}. The approach introduces pseudo-random disturbances in the three velocity components of the inflow profiles inside the boundary layer. In Fig.\ \ref{fig:tripping-region}, the region of the inflow condition where the tripping method is applied is highlighted in gray. The three velocity components in the cylindrical coordinate system are calculated by
\begin{equation}
\left[ \begin{array}{c} 
             \tilde{u}_x \\ \tilde{u}_r \\ \tilde{u}_{\theta}
       \end{array} \right] = 
       \left[ \begin{array}{c} 
              \tilde{u}_{x,RANS} \\ \tilde{u}_{r,RANS} \\ 
              \tilde{u}_{\theta,RANS}
       \end{array} \right]
       + \alpha U_j
       \left[ \begin{array}{c} 
              3\epsilon_x(x,r,\theta,t) \\ \epsilon_r(x,r,\theta,t) \\ 
              2\epsilon_{\theta}(x,r,\theta,t)
       \end{array} \right].                 
\end{equation}
The vector $\tilde{\textbf{u}}_c = [\tilde{u}_x, \tilde{u}_r, \tilde{u}_{\theta}]^T$ is the filtered velocity vector in cylindrical coordinate system. The three velocity components with the subscript $RANS$ are the velocity components from the RANS calculation of the nozzle flow. The $\alpha$ constant is used to adjust the intensity of the velocity fluctuations and has the value proposed in \citet{BogeyMarsdenBailly2011}. The $\alpha$ constant multiplies the jet velocity, $U_j$, and three random numbers, $\epsilon_x(x,r,\theta,t)$, $\epsilon_r(x,r,\theta,t)$, and $\epsilon_{\theta}(x,r,\theta,t)$, which have values between $-1$ and $1$.

The three inflow reference states described above are implemented in the LES calculation as weakly enforced Dirichlet boundary conditions at the jet inlet plane. For each case, the corresponding primitive-variable profiles are prescribed as the reference state of the numerical fluxes at the inflow boundary, outside the computational domain, and an approximate Riemann problem is solved to find a common flux at the boundary interface \cite{Abreuetal2024}. This approach ensures a consistent and controlled comparison of the impact of the inflow modeling on the downstream jet development.

\subsection{Simulation Settings}
\label{sec:simset}

\begin{figure*}[tbp]
\centering
\subfloat[Density profiles.]{
	\includegraphics[width=0.45\linewidth]{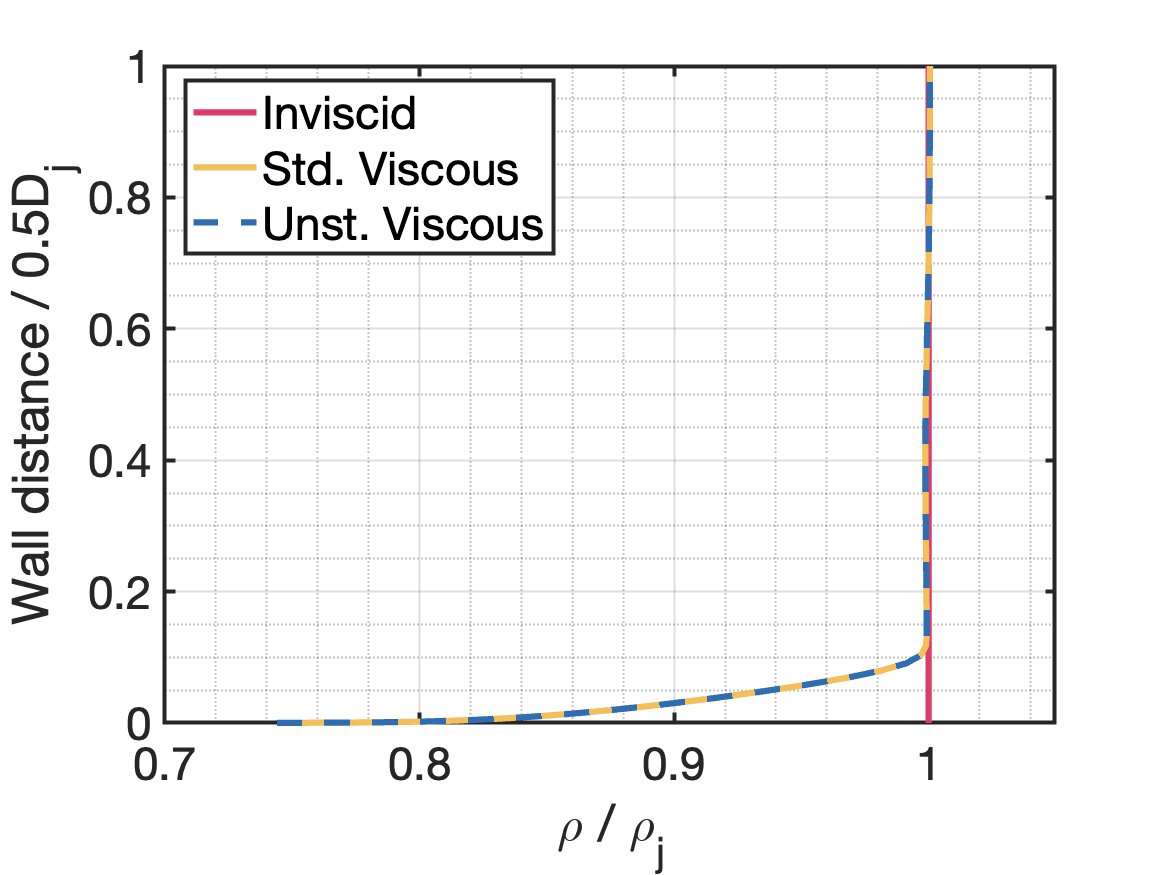}
	\label{res:inflow_profile1}	
	}
\subfloat[Pressure profiles.]{
	\includegraphics[width=0.46\linewidth]{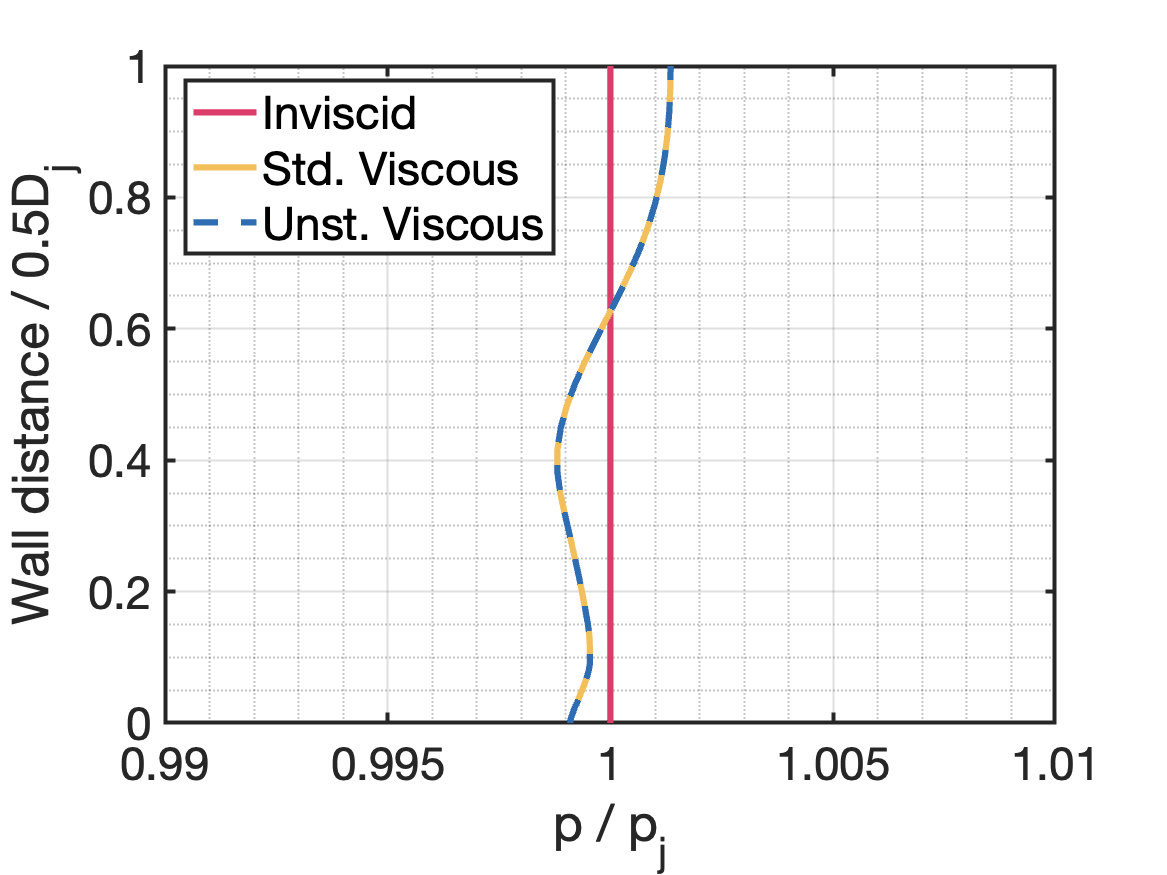}
	\label{res:inflow_profile2}	
	}
\\
\subfloat[Longitudinal velocity component profiles.]{
	\includegraphics[width=0.45\linewidth]{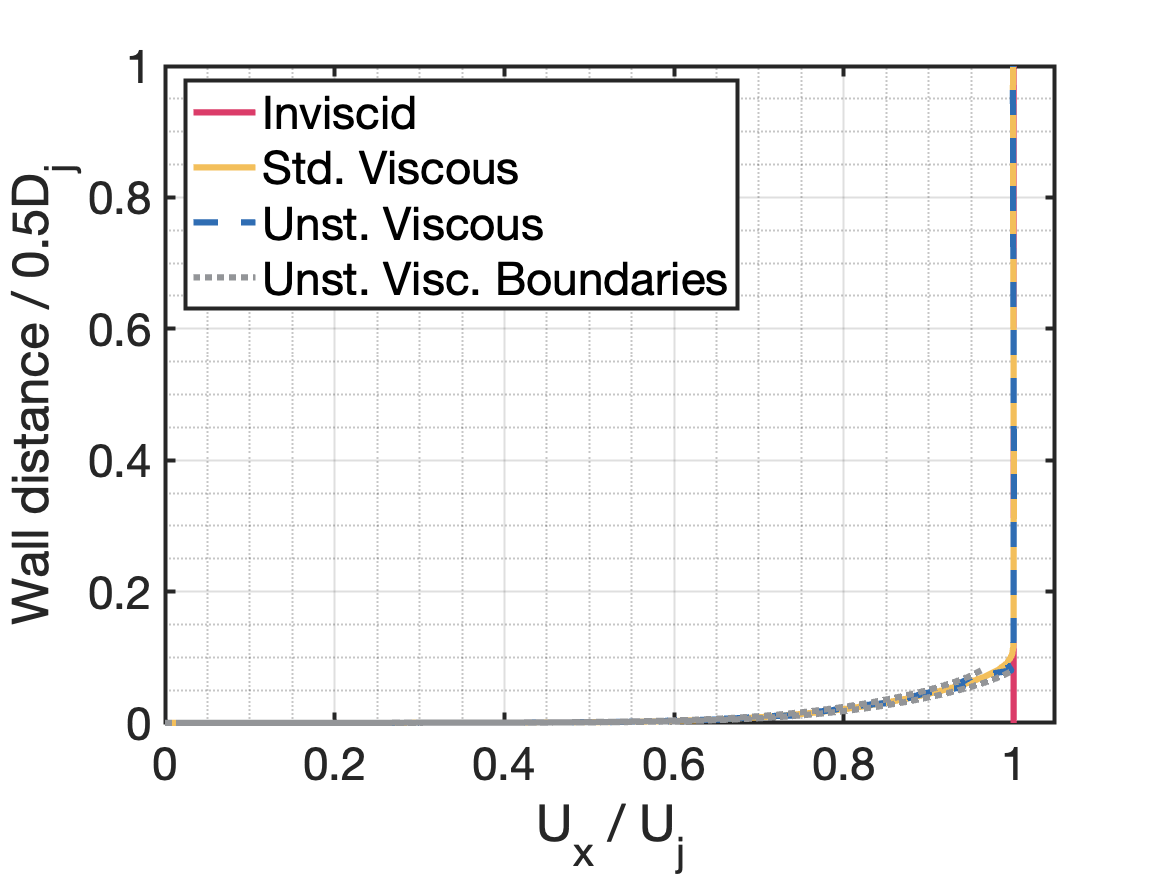}
	\label{res:inflow_profile3}
	}
\subfloat[Detail on the boundary layer of the longitudinal velocity component profiles.]{
	\includegraphics[width=0.45\linewidth]{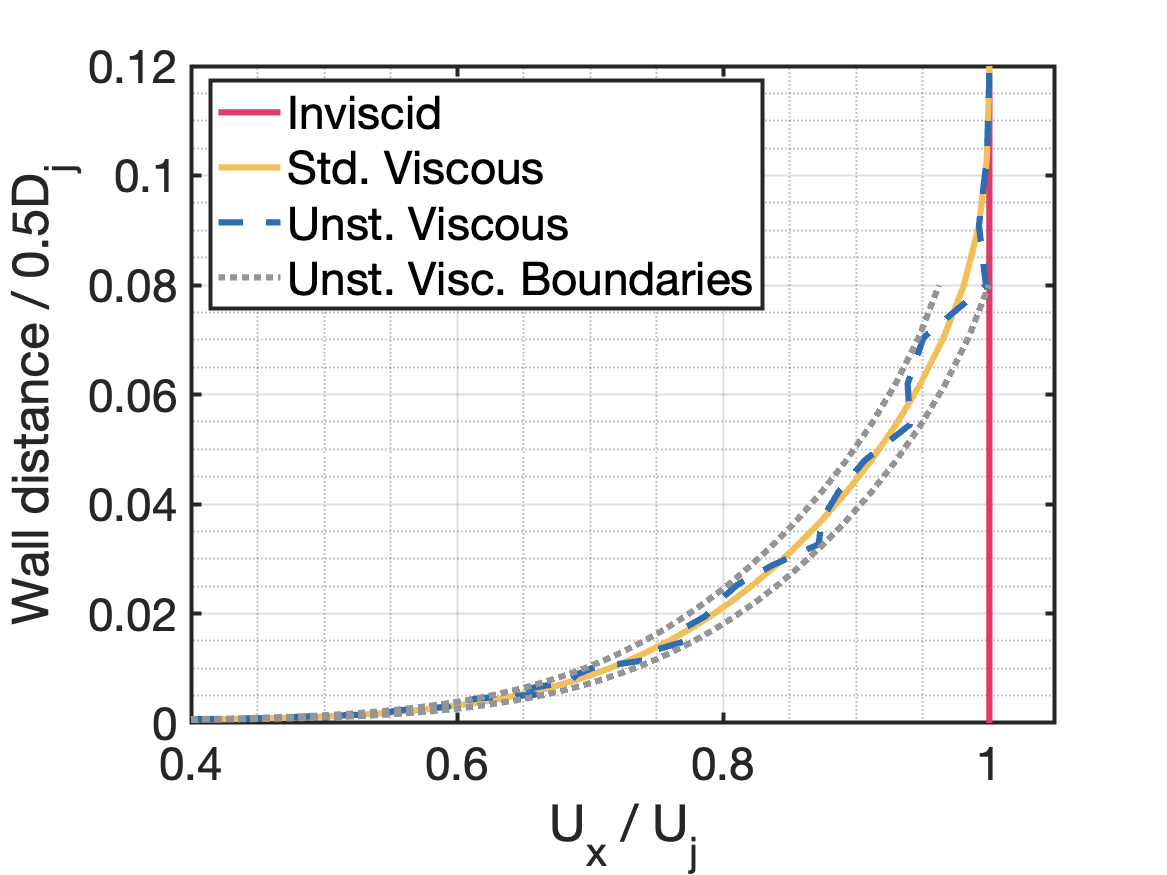}
	\label{res:inflow_profile4}	
	}
\\
\subfloat[Radial velocity component profiles.]{
	\includegraphics[width=0.45\linewidth]{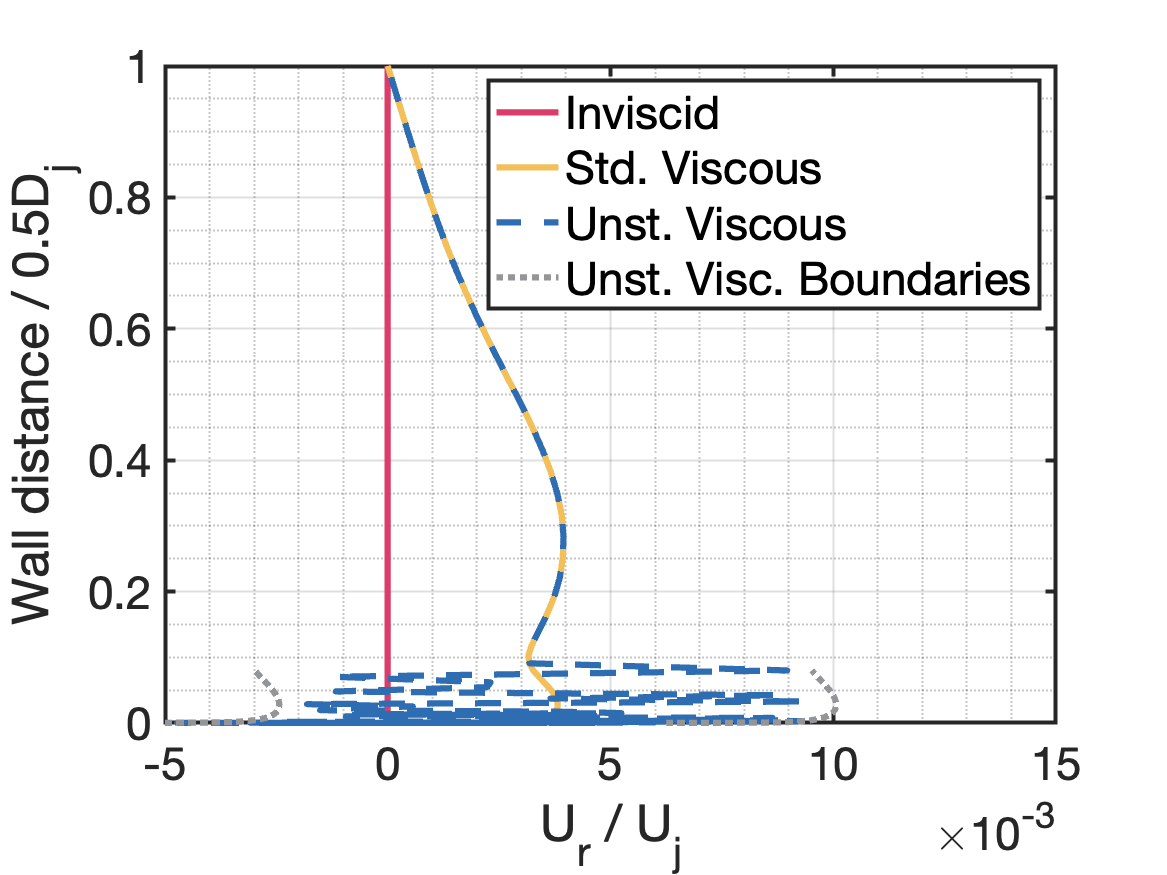}
	\label{res:inflow_profile5}
	}%
\subfloat[Detail on the boundary layer of the radial velocity component profiles.]{
	\includegraphics[width=0.45\linewidth]{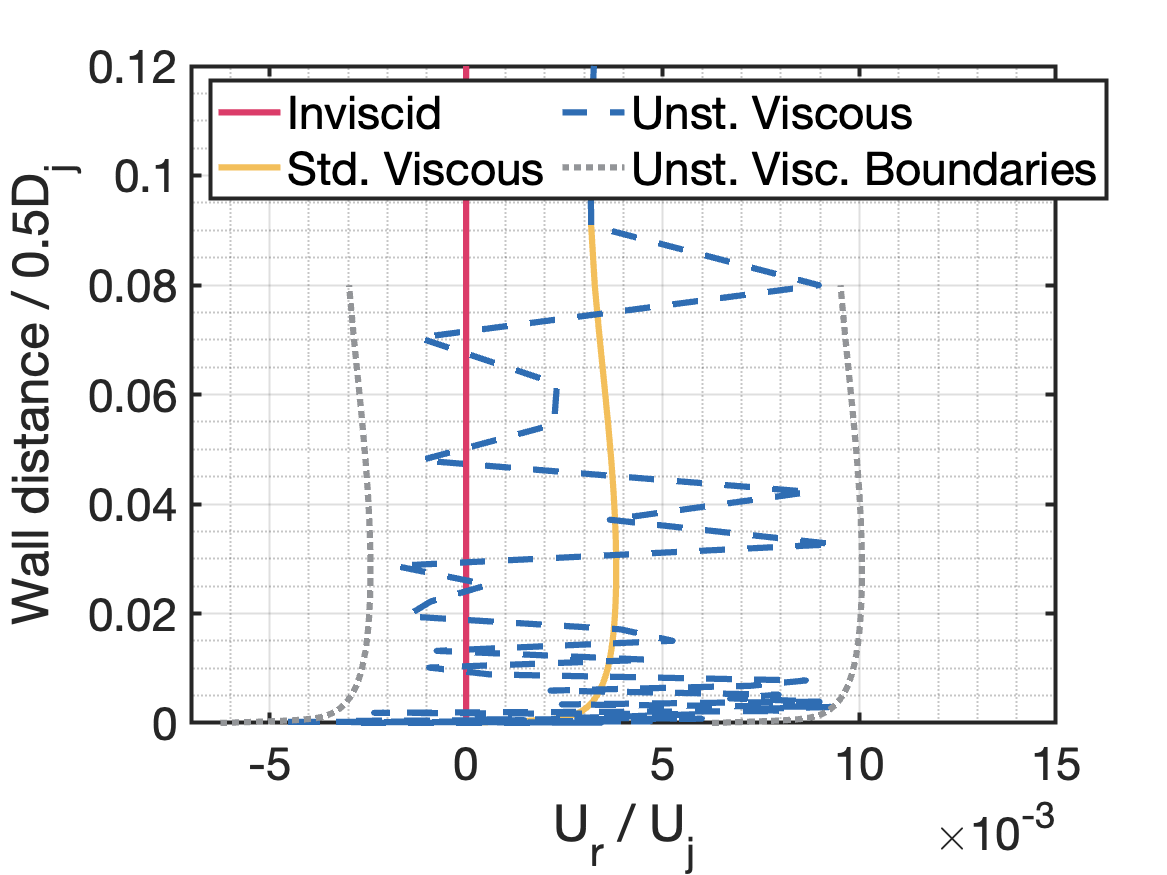}
	\label{res:inflow_profile6}	
	}
\caption{Inflow profiles from the three boundary conditions investigated 
in the work.}
\label{res:inflow_profile}
\end{figure*}

The simulations are performed with a computational mesh of $15.4 \times 10^6$ elements and a third-order accurate spatial discretization, which results in approximately $410 \times 10^6$ degrees of freedom per equation. Three numerical simulations are performed in the current work, and each one presents a different jet inflow boundary condition. The three inflow conditions imposed are inviscid, steady viscous, and unsteady viscous profiles. Figure\ \ref{res:inflow_profile} compares the three inflow conditions used in the jet flow simulations. They represent the profiles imposed as the reference state at the jet inflow section. The profiles from density and pressure are presented in Figs.\ \ref{res:inflow_profile1} and \ref{res:inflow_profile2}. The viscous profile shows non-uniform variables.

The longitudinal velocity component profiles are presented in Figs.\ \ref{res:inflow_profile3} and \ref{res:inflow_profile4}, and the radial velocity profiles are presented in Figs.\ \ref{res:inflow_profile5} and \ref{res:inflow_profile6}. The velocity profiles are plotted in two charts to allow a complete visualization of the inflow profiles and a detailed view of the boundary layer region. The flow is nearly uniform outside the boundary layer for the longitudinal velocity component. However, it presents a non-uniform behavior for the radial velocity component. The steady viscous profile presents a boundary layer in the axial velocity component. The two velocity profiles are snapshots indicating one realization of the velocity profile that can be imposed in the inlet section when imposing the unsteady viscous profile. The two gray dotted profiles are the boundaries of the unsteady viscous profile that can be obtained with the defined $\alpha$ constant from the tripping method.

\subsection{Calculation of Statistical Properties}
\label{sec:stats}

The numerical simulations are performed in two steps. In the first step, the jet flow develops from its initial state, which can be a restart from a previous simulation, until it reaches a statistically steady flow condition. The first step requires approximately $100$ flow-through times (FTT). The flow-through time (FTT) is the basis for comparison used by many authors in the literature, and it is defined based on the jet inlet velocity and the nozzle exit diameter. It can represent the necessary time for the fluid to cross one nozzle outlet diameter with the imposed velocity on the input boundary. In the second step, once the statistically steady flow condition is achieved, the data is recorded for the statistical analysis of the flow. The second step is performed for approximately $114$ FTT. The non-dimensional time acquisition frequency is $21.124$. The time sample is composed of $2401$ flow snapshots. The definition of the sampling time and the acquisition frequency are associated with the Strouhal number range in which a spectral analysis can be performed. It also comprises the interval available from experimental tests \cite{BridgesWernet2008}.

The statistical analysis of the flow field is performed via the investigation of the mean and root mean square (RMS) values of the flow properties. The data is studied through the axial direction in the jet centerline, $r/D_j=0.0$, and the lipline, $r/D_j=0.5$. The radial data profiles are also studied in four longitudinal stations: $x/D_j=2.5$, $x/D_j=5.0$, $x/D_j=10.0$, and $x/D_j=15.0$. The regions of interest from the flow field are presented in the 2-D sketch of the domain in Fig.\ \ref{fig:geo}. The data for the jet lipline and the radial stations are obtained by averaging $120$ evenly distributed azimuthal positions.

A spectral analysis is performed through the investigation of the power spectral density values of the longitudinal velocity component fluctuations. The power spectral density values are obtained from the four stations where radial profiles are investigated at the jet centerline and lipline. The power spectral density is calculated by applying the Fast Fourier Transform algorithm to the longitudinal velocity component fluctuations temporal signal in each position. The magnitude of the resulting values is squared and divided by frequency and time sample.

\section{Database From the Numerical Simulations}
\label{sec:database}

One strategy to enhance the current numerical techniques, particularly with regard to turbulence modeling, is the application of machine learning and neural networks. Additionally, for well-trained configurations, such an approach can reduce calculation costs \cite{Amarloo23, DEZORDOBANLIAT24, Bruntonetal2020, Mauliketal2019, Kurzetal2022, Kurzetal2023}. In the model training from these methods, a large amount of data is needed. The numerical data generated by the large-eddy simulations of supersonic jet flow configurations performed for the present work are available in a public repository \cite{database12024, database22024, database32024, database42024, database52024, database62024}. The repository includes data from four numerical simulations \cite{database12024, database22024, database32024, database42024} investigated in the resolution study \cite{Abreuetal2024}, as well as data from the present work, which addresses viscous and unsteady boundary layer profiles \cite{database52024, database62024}. The supersonic jet flow database provides time-resolved samples collected along a set of probes distributed in lines and planes, as illustrated in Fig.\ \ref{fig:geo}.
The simulation data generated in the current work are extracted from \( 121 \) axial and \( 240 \) radial probe lines, along with probe planes aligned and perpendicular to the jet flow. The probes contain \( 2401 \) time samples of conserved variables of the flow, extracted at a fixed non-dimensional frequency of \(21.124\). The conserved variables are the components of the $\mathbf{U}$ algebraic vector, which is defined immediately after Eq.\ (\ref{eq:filteredNS})\@.

\begin{figure*}[htb!]
\centering
	\subfloat[Axial probe lines.]{
	\includegraphics[width=0.7\linewidth]{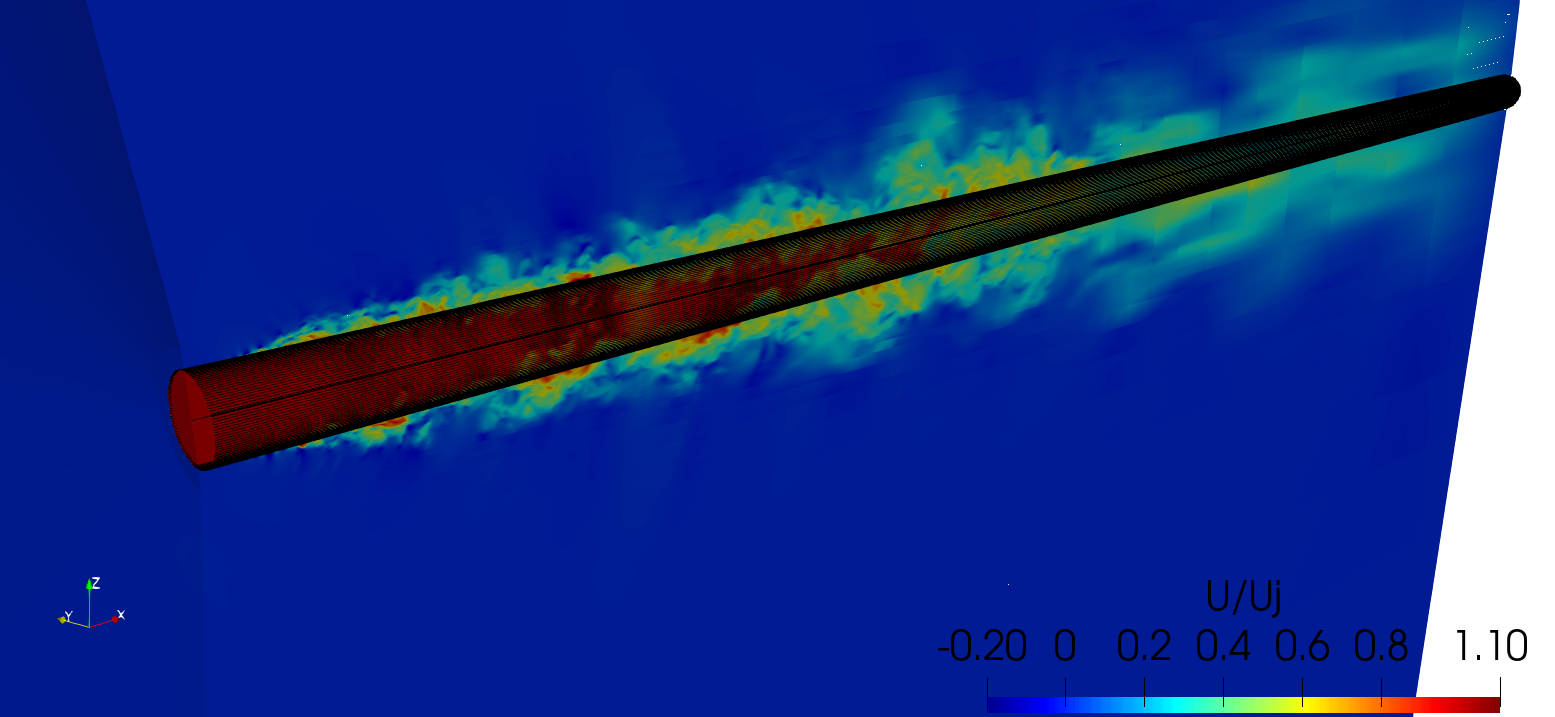}
    \label{fig:axial-probes}
	}
    \\
    \subfloat[Radial probe lines.]{
	\includegraphics[width=0.7\linewidth]{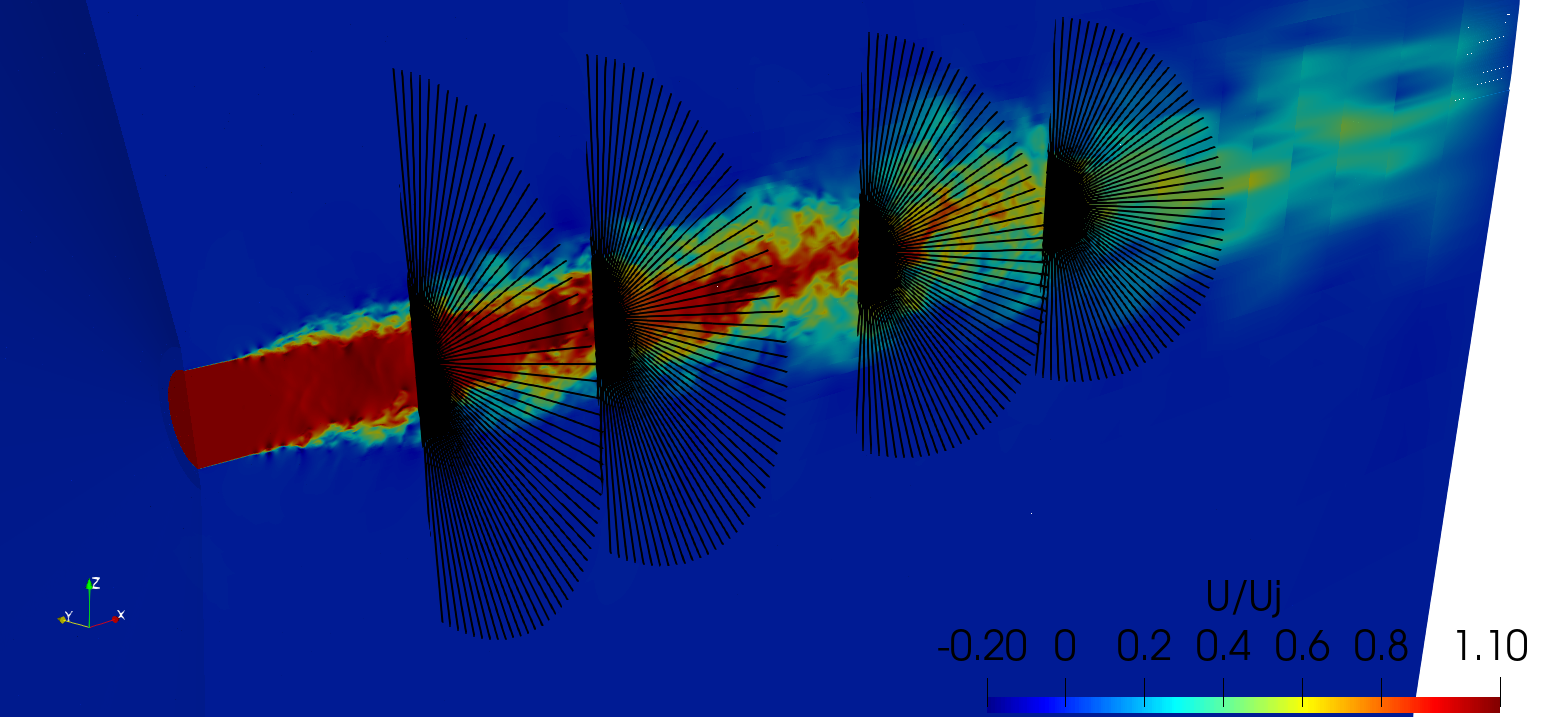}
	\label{fig:radial-probes}
	}
\caption{A set of probes arranged along axial and radial lines, where unsteady data is collected from numerical simulations.}
\label{fig:line-probes}
\end{figure*}

Figure \ref{fig:line-probes} illustrates the positioning of the probes. Each probe line consists of a group of measurement points where the flow properties are recorded. The axial probe lines, Fig.\ \ref{fig:axial-probes}, contain \( 401 \) probes each, while the radial probe lines, Fig.\ \ref{fig:radial-probes}, consist of \( 101 \) probes each. The probes are uniformly distributed along both the axial and the radial probe lines.

The axial probe lines span from \( x/D_j = 0.0 \) to approximately \( x/D_j = 39.0 \), covering two key regions of the jet flow. The \textit{line000} axial probe line is positioned along the jet centerline at \( r/D_j = 0.0 \). The remaining \( 120 \) axial probe lines are azimuthally distributed along the jet lipline at \( r/D_j = 0.5 \), with an angular separation of \( 3^\circ \). These probe lines are labeled sequentially from \textit{line001} to \textit{line120}. The \( 240 \) radial probe lines are located in four distinct axial positions: \( x/D_j = 2.5 \), \( x/D_j = 5.0 \), \( x/D_j = 10.0 \), and \( x/D_j = 15.0 \). At each axial location, \( 60 \) radial probe lines are available, uniformly spaced in the azimuthal direction with \( 3^\circ \) separation. These axial positions are depicted in Fig.\ \ref{fig:geo}. 
The radial probe lines extend symmetrically along $0 \leq r/D_j \lessapprox 3.15$ for the 
azimuthal positions between $ 0 \leq \theta \leq \pi $. The corresponding labels are as 
follows:  
\begin{itemize}
    \item \( x/D_j = 2.5 \): \textit{line201} to \textit{line260},  
    \item \( x/D_j = 5.0 \): \textit{line301} to \textit{line360},  
    \item \( x/D_j = 10.0 \): \textit{line401} to \textit{line460},  
    \item \( x/D_j = 15.0 \): \textit{line501} to \textit{line560}.  
\end{itemize}  

\begin{figure*}[htb!]
\centering
	\subfloat[Probe planes positioned perpendicular to the jet flow.]{
	\includegraphics[width=0.7\linewidth]{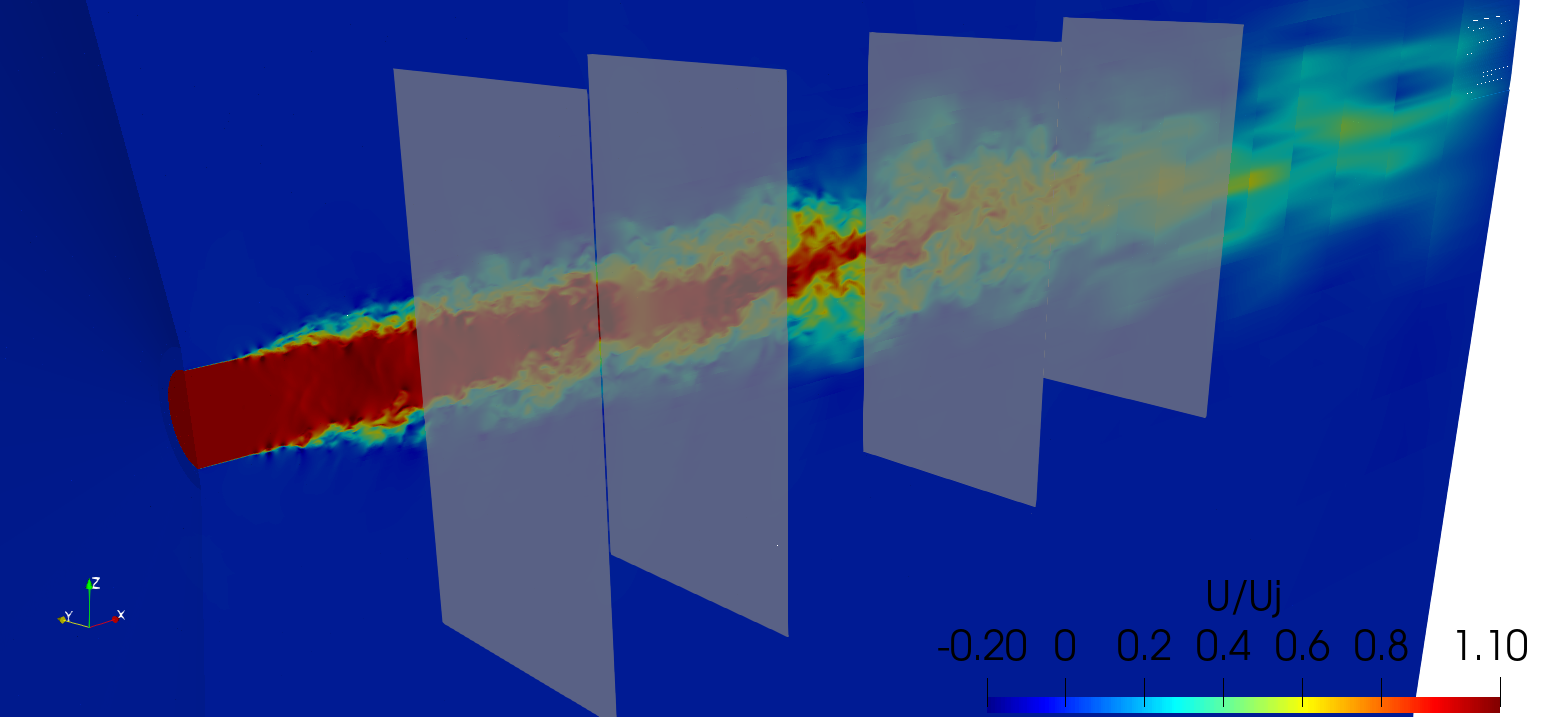}
    \label{fig:probe-planes-perp}
	}
    \\
    \subfloat[Probe plane aligned with the jet flow]{
	\includegraphics[width=0.7\linewidth]{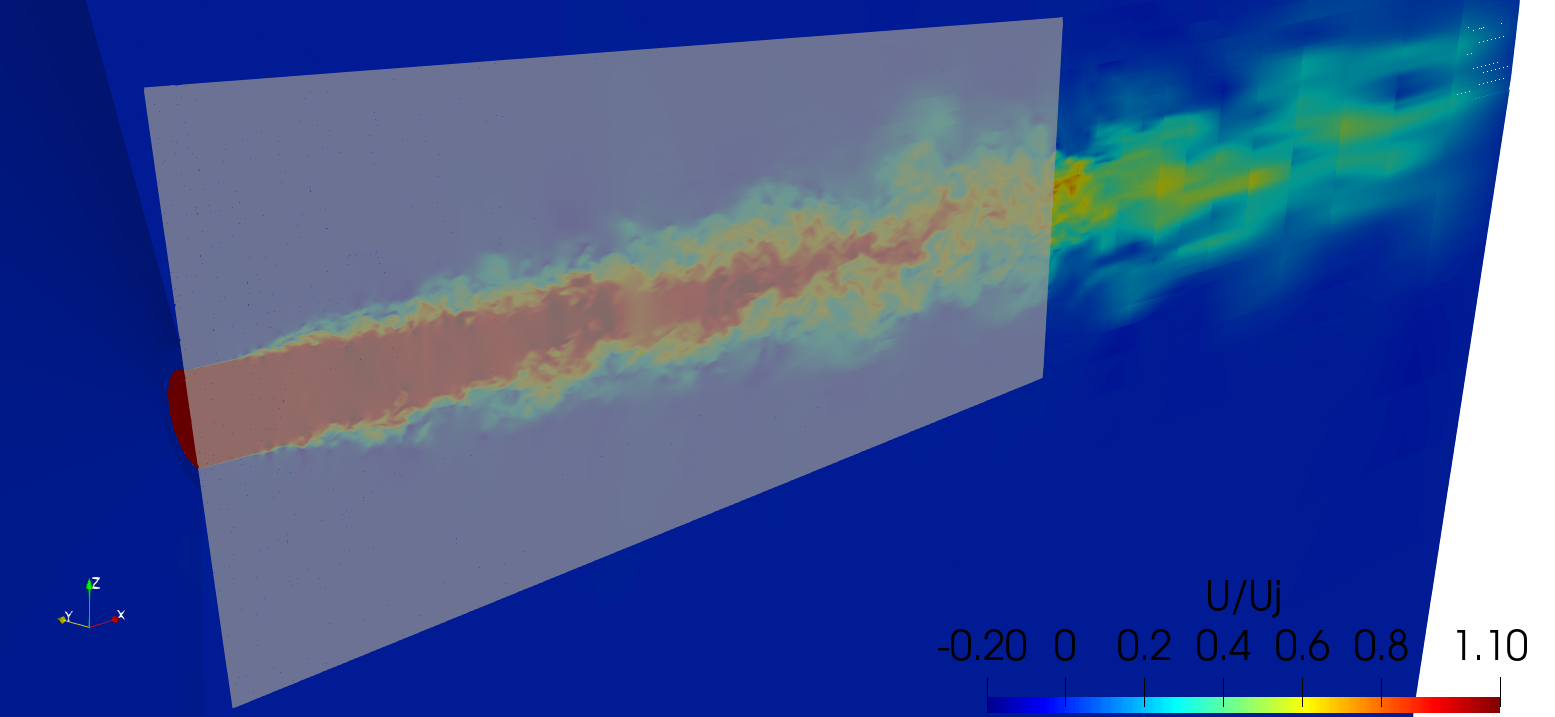}
	\label{fig:probe-plane-para}
	}
\caption{A set of probes is arranged along four planes perpendicular to the jet flow and one plane aligned with it, from which unsteady data is collected in the numerical simulations.}
\label{fig:probe-planes}
\end{figure*}

Five probe planes are also available in the database, as shown in Fig.\ \ref{fig:probe-planes}. Four of these planes, Fig.\ 
\ref{fig:probe-planes-perp}, are positioned perpendicular to the jet axis at the same axial locations as the radial 
probe lines: \( x/D_j = 2.5 \), \( x/D_j = 5.0 \), \( x/D_j = 10.0 \), and \( x/D_j = 15.0 \). Each of these four 
probe planes consists of \( 201 \) probes in both directions, extending from \( y/D_j \approx -3.15 \) and 
\( z/D_j \approx -3.15 \) to \( y/D_j \approx 3.15 \) and \( z/D_j \approx 3.15 \). These planes are labeled 
\textit{plane001}, \textit{plane002}, \textit{plane003}, and \textit{plane004}, according to their distance from 
the jet inlet section.
The fifth probe plane, labeled as \textit{plane005}, is aligned with the jet axis. This plane is composed by 
$0.0 \leq x/D_j \leq 15.0$  and $ -3.15 \lessapprox z/D_j \lessapprox 3.15 $, as illustrated in Fig.\ 
\ref{fig:probe-plane-para}. It consists of \( 800 \) probes in the radial direction and \( 1200 \) probes in 
the axial direction.
The probe planes can be used to investigate features of the flow field, which can be useful for machine learning and/or stability analysis. The probe planes present a finer refinement level than probe lines, and, therefore, they could provide additional information for algorithm training purposes. 

The data collected from each probe line and plane are stored in PVD (ParaView Data) file format. The files are named based on the numerical simulation and the corresponding probe line or plane. The filename consists of a prefix related to the numerical simulation, which is the number of the database preceded by the \textit{s} letter, followed by the term ``\_RP\_'', the name of the probe line or plane, and additional identifying information. For example, a file might be named:
\begin{center}
\texttt{s5\_RP\_line450\_Line\_000291.pvd}
\end{center}
The PVD file serves as an index that links to multiple VTS files (structured VTK format \cite{schroeder98} ) corresponding to different time steps. These VTS files are stored inside the \textit{timeseries} folder and can be accessed independently of the PVD file. For each simulation, data from multiple probe lines and planes are archived in \( 15 \) compressed tarball files (\texttt{.tar.gz}). The probe lines are divided into groups of approximately \( 30 \) lines per file. The four perpendicular planes are stored in a single file, while the axial plane data is split into two files due to its large size.

\section{Numerical Results}
\label{sec:numres}

The present section discusses the results obtained from numerical simulations of the jet flow. First, the validation of the solver, together with a mesh and polynomial resolution study, for the perfectly expanded supersonic jet is summarized. This is followed by an assessment of the three inlet boundary conditions, {\em i.e.}, inviscid, steady viscous, and unsteady viscous profiles. The inviscid inflow profile prescribes constant values for all flow variables at the inlet plane. The steady viscous profile imposes velocity components, pressure, and density distributions obtained \textit{a priori} from a RANS simulation of a supersonic nozzle. The unsteady viscous profile extends the steady viscous case by introducing a tripping method to the boundary-layer, following the approach of \citet{BogeyMarsdenBailly2011}. Further details of the inflow profiles are provided in Sect.\ \ref{sec:boundcond}.

\subsection{Validation of the Numerical Framework}
\label{sec:validation}

The perfectly expanded supersonic jet flow is simulated using a nodal discontinuous Galerkin method \cite{Kopriva2010, Hindenlang2012}. A previous resolution study \cite{Abreuetal2024} validated the numerical framework and the numerical setup using four simulations with varying polynomial orders and meshes, ranging from $50 \times 10^6$ to $410 \times 10^6$ degrees of freedom per equation. The simulations were performed under the operating conditions of the experiments from \citet{BridgesWernet2008}, which evaluate an isothermal, perfectly expanded supersonic flow with a Mach number of 1.4 and a Reynolds number based on the nozzle exit diameter of $1.58\times 10^6$. This configuration enables direct comparisons of velocity fields along the jet centerline, lipline, and four axial planes, as well as spectral analysis of the velocity signal.

The validation study \cite{Abreuetal2024} assessed the combined effects of mesh resolution and DG polynomial degree on the accuracy of supersonic free-jet simulations. By comparing progressively refined meshes and increasing polynomial orders, it demonstrated that polynomial enrichment can yield accuracy improvements comparable to mesh refinement, particularly in the near-field region of the jet. Conversely, insufficient spatial resolution or low-order discretizations were shown to introduce excessive numerical dissipation, leading to an underprediction of turbulence levels and premature decay of the jet potential core.
The reader can find the four databases that evaluate mesh sizes, point distributions, and polynomial order for LES of the perfectly expanded supersonic jet flow using the FLEXI framework in the Zenodo repository \cite{database12024, database22024, database32024, database42024}.

The results showed that increasing the resolution improved agreement with experimental data, particularly by reducing numerical dissipation and extending the jet potential core, defined where $U \geq 0.95U_j$. The lowest-resolution case underestimated the potential core length, while higher-resolution cases provided a more accurate representation of both mean and fluctuating velocity profiles.
Based on quantitative comparisons with experimental measurements, a baseline DGSEM configuration was identified that achieves grid-independent predictions while maintaining a manageable computational cost. In particular, simulations using a third-order accurate spatial discretization and a mesh of $15.4 \times 10^6$ elements, corresponding to approximately $410 \times 10^6$ degrees of freedom per equation, were found to provide robust agreement with experimental data. This calibrated numerical setup, therefore, serves as a reliable reference configuration for the present investigation of inflow boundary condition effects.
Notably, the largest deviations were observed in regions strongly influenced by the inflow profile, which directly motivated the present study on the impact of different inflow boundary condition formulations.

In addition to exploring these inflow effects, the next subsection includes further validation through comparisons with both experimental data \cite{BridgesWernet2008} and prior numerical results from \citet{Mendezetal2012}\@. These later authors have simulated similar flow conditions, but at a significantly lower Reynolds number. In particular, \citet{Mendezetal2012} have performed computations at two different Reynolds numbers. For the isothermal jet configuration, which is precisely the same jet configuration addressed in the present work, \citet{Mendezetal2012} performed their calculations at a Reynolds number of $1.5\times 10^{5}$, which is one order of magnitude lower than the Reynolds number used for the present calculations. Most of the comparisons with the data of \citet{Mendezetal2012} have used their isothermal jet data. However, there were some comparisons of power spectral densities (PSDs) for which there are no data in the work of \citet{Mendezetal2012} at $1.5\times 10^{5}$ Reynolds number. For these cases, our comparisons used the data of \citet{Mendezetal2012} generated for a heated jet test case, which was calculated at a Reynolds number of $7.6 \times 10^{4}$. Therefore, this is an even lower Reynolds number, and the analyses should take into account the considerably large differences on the flow Reynolds number.

\subsection{Numerical Investigation of the Jet Inflow Conditions}
\label{sec:jetinflow}

The resolution study conducted by the authors \cite{Abreuetal2024} employed an inviscid 
inflow profile, which did not fully capture the inlet flow physics observed in the
experiments \cite{BridgesWernet2008}. That study applied four different resolutions, and the
corresponding data are available in the first four databases of the Zenodo repository 
\cite{database12024,database22024,database32024,database42024}, which contains six 
databases in total. The most refined calculation \cite{database42024} using the 
inviscid inlet boundary condition employed approximately $410 \times 10^6$ degrees of freedom per equation 
and revealed deviations between the numerical and experimental data near the inlet, 
particularly around the lip\-line region. In the present work, two additional inlet boundary 
conditions incorporating more realistic flow physics, steady viscous and unsteady 
viscous profiles, are used in the simulations. The data from calculations using the 
steady and unsteady viscous inlet boundary layers are, respectively, available in the 
last two databases of the repository \cite{database52024,database62024}.
The impact of these three inflow conditions is assessed by analyzing the mean flow 
field, axial profiles of the streamwise velocity component, radial profiles of the 
streamwise and radial velocity components, shear stress tensor components, and 
power spectral densities. 
Results from simulations using all three inflow conditions are compared with both the experimental data of \citet{BridgesWernet2008} and the numerical data of \citet{Mendezetal2012}\@. Please, observe the previous comments, in the last paragraph of the previous subsection, regarding the Reynolds number for the simulations of \citet{Mendezetal2012}.

\subsubsection{Mean Flow Field}

In the current subsection, a qualitative analysis of the velocity and pressure flow fields is performed. The contours of the mean longitudinal velocity component, RMS values of the longitudinal velocity component fluctuations, and mean pressure from the three numerical simulations are compared to understand the influence of the inflow condition on the jet development. The mean longitudinal velocity component contours additionally present the boundaries of the potential core, identified in black. The potential core region is where the jet velocity is higher than $0.95U_j$. The borders identify where $<U_x>/U_j=0.95$.

\begin{figure}[htb!]
\centering
\subfloat[Inviscid profile.]{
	\includegraphics[trim = 0mm 0mm 0mm 0mm, clip, width=0.72\linewidth]{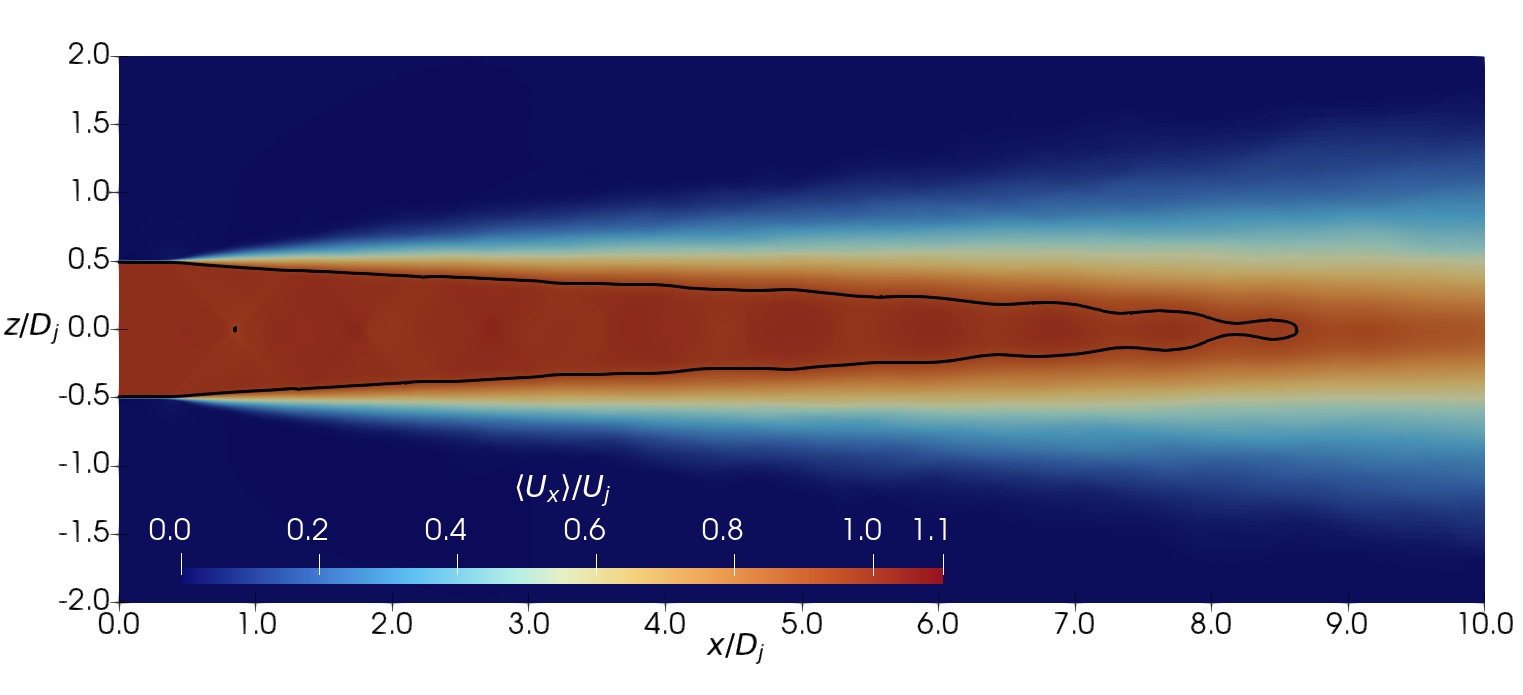}
	\label{res:mean_velx1}
	}
\\
\subfloat[Steady viscous profile.]{
	\includegraphics[trim = 0mm 0mm 0mm 0mm, clip, width=0.72\linewidth]{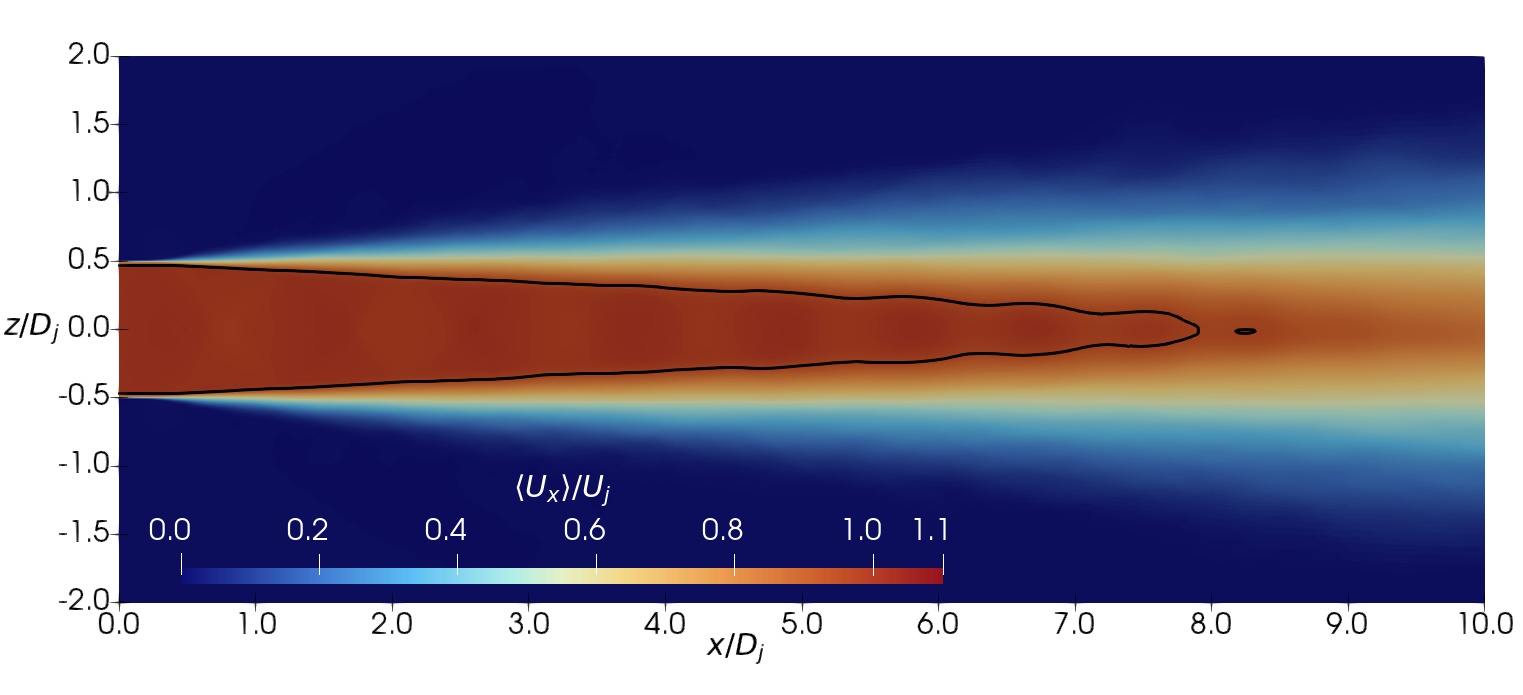}
	\label{res:mean_velx2}
	}
\\
\subfloat[Unsteady viscous profile.]{
	\includegraphics[trim = 0mm 0mm 0mm 0mm, clip, width=0.72\linewidth]{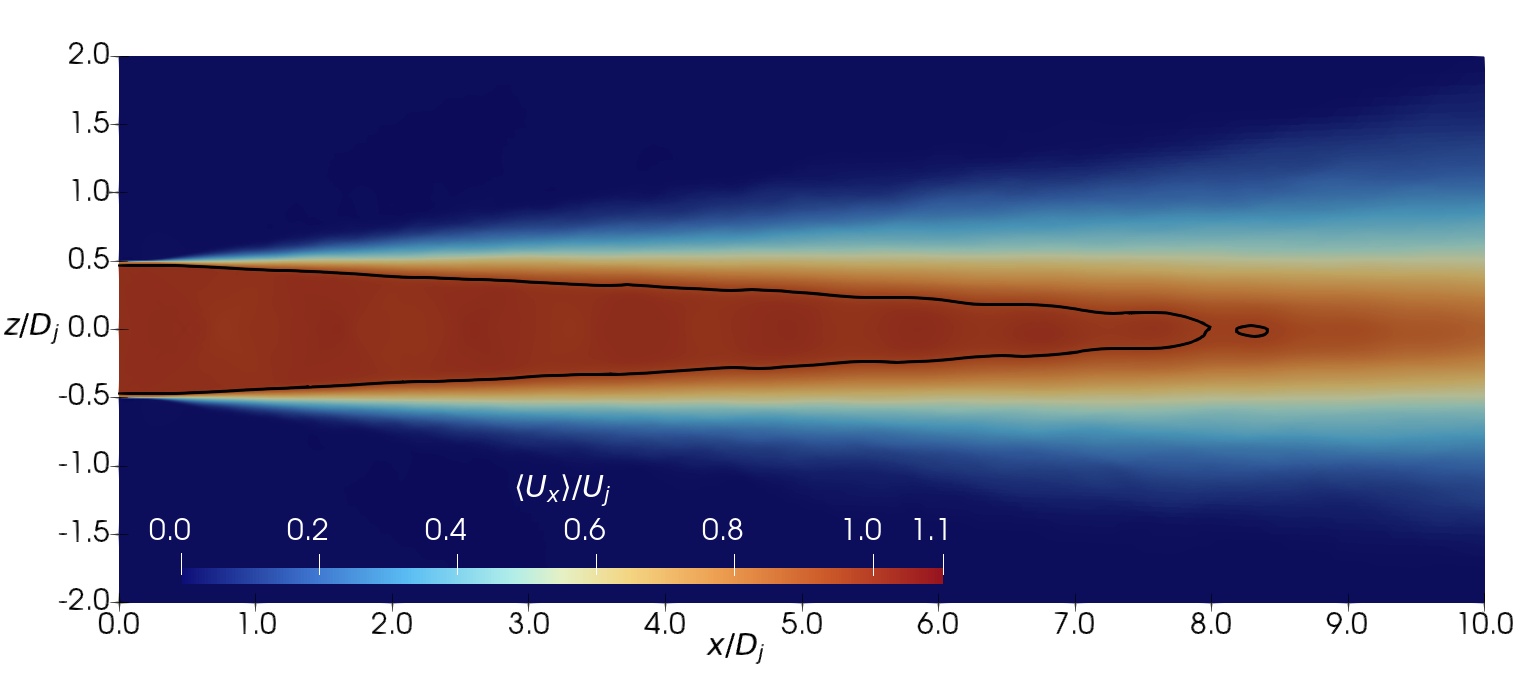}
	\label{res:mean_velx3}
	}
\caption{Mean longitudinal velocity component, $\langle U_x \rangle/U_j$, contours on a plane through the centerline of the jet with a black contour delimiting the jet potential core, $\langle U_x \rangle/U_j=0.95$. The 2-D cut-plane is composed of $1200$ probes in the x-direction and $800$ probes in the z-direction. The statistical analysis is performed with a $2401$ time sample gathered with a non-dimensional time frequency of $21.124$.}
\label{res:mean_velx13}
\end{figure}

Figure \ref{res:mean_velx13} presents the contours of the mean streamwise velocity component normalized by the inlet jet velocity, $U_j = 1.4$, for the three simulations using different inlet boundary conditions. In the region near $z/D_j \approx \pm 0.5$ and $x/D_j < 0.5$, the two simulations with viscous inflow profiles exhibit a velocity contour originating from the boundary layer included in the imposed profile. The potential core length obtained with the inviscid profile is longer than that obtained with both steady and unsteady viscous profiles. Both simulations using viscous inflow conditions yield similar core lengths.

\begin{figure}[htb!]
\centering
\subfloat[Inviscid profile.]{
	\includegraphics[trim = 0mm 0mm 0mm 0mm, clip, width=0.72\linewidth]{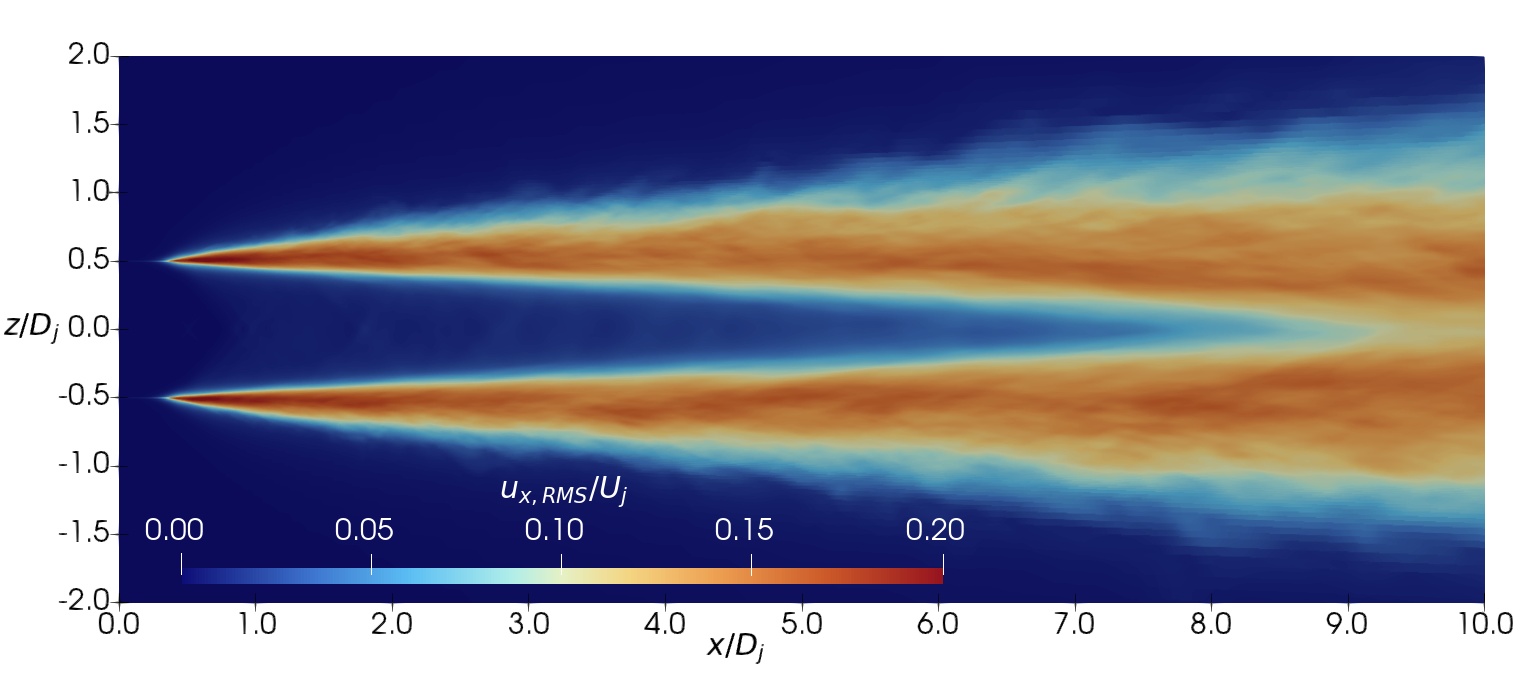}
	\label{res:rms_velx1}
	}
\\
\subfloat[Steady viscous profile.]{
	\includegraphics[trim = 0mm 0mm 0mm 0mm, clip, width=0.72\linewidth]{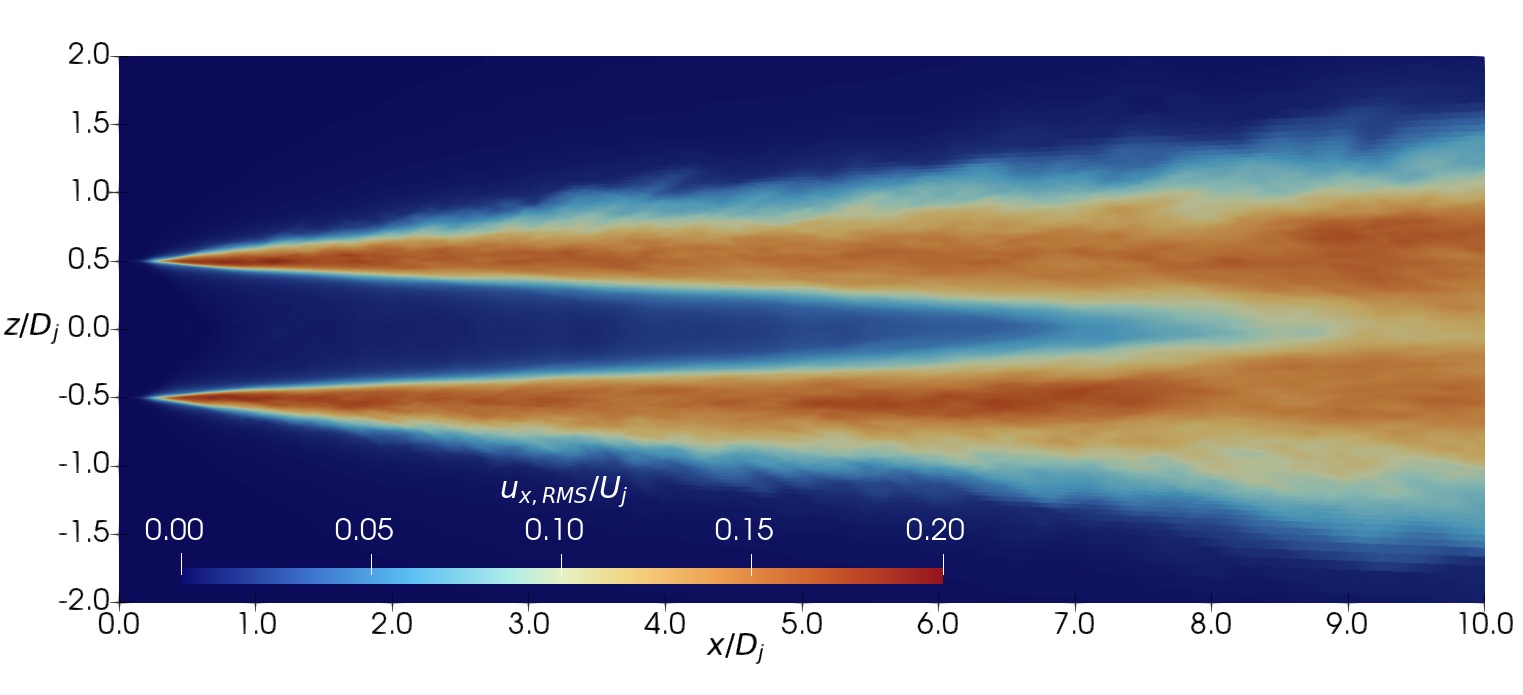}
	\label{res:rms_velx2}
	}
\\
\subfloat[Unsteady viscous profile.]{
	\includegraphics[trim = 0mm 0mm 0mm 0mm, clip, width=0.72\linewidth]{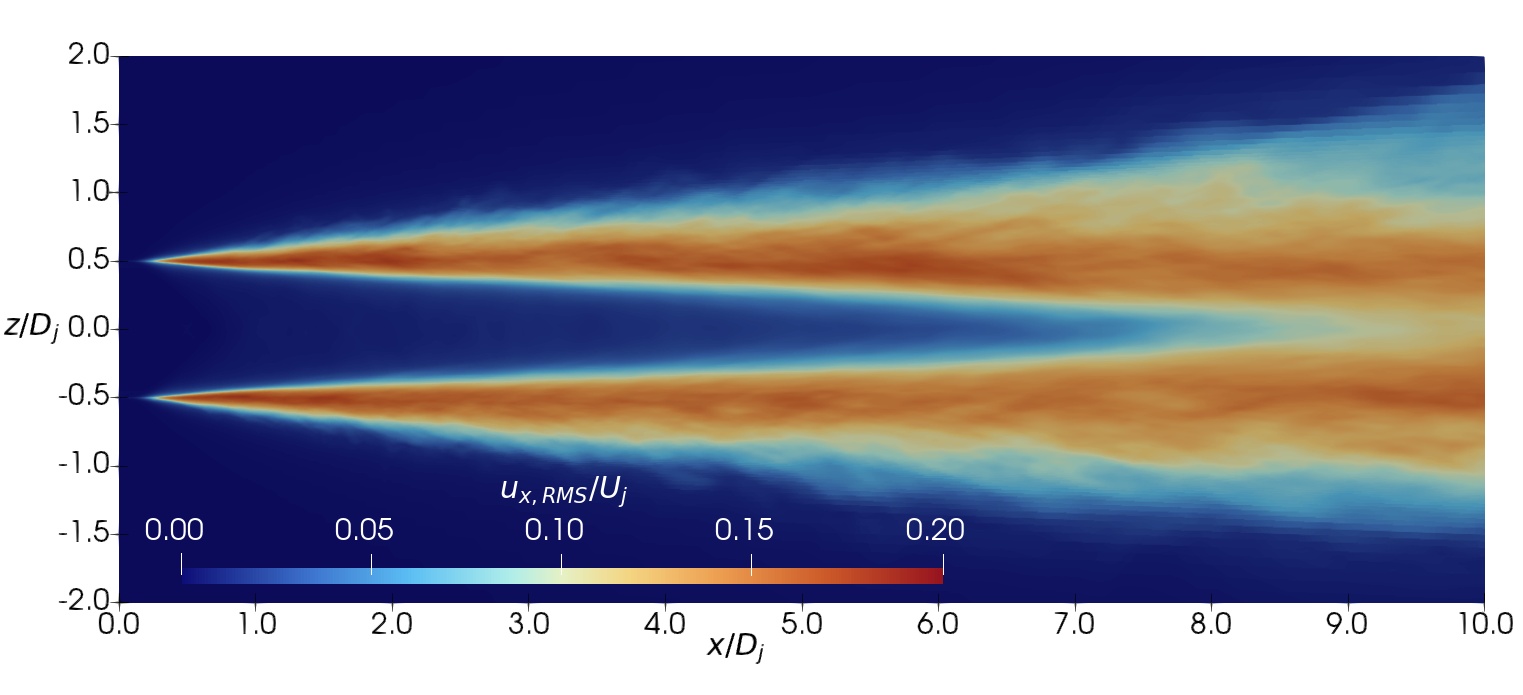}
	\label{res:rms_velx3}
	}
\caption{Contours of the RMS values of the longitudinal velocity fluctuations, $u_{x,RMS}/U_j$, on a plane through the centerline of the jet. The 2-D cut-plane is composed of $1200$ probes in the x-direction and $800$ probes in the z-direction. The statistical analysis is performed with a $2401$ time sample gathered with a non-dimensional time frequency of $21.124$.}
\label{res:rms_velx13}
\end{figure}

Figure \ref{res:rms_velx13} shows the RMS contours of the longitudinal velocity fluctuations for the three inflow conditions. The contours indicate an earlier development of the mixing layer in the simulations using steady and unsteady viscous profiles. Compared to the inviscid case, these simulations also exhibit lower peak fluctuation levels, suggesting a reduced intensity in the interaction between the jet flow and the ambient air. The two viscous inflow conditions yield very similar distributions of RMS fluctuations, with only negligible differences observed.

\begin{figure}[htb!]
\centering
\subfloat[Inviscid profile.]{
	\includegraphics[trim = 0mm 0mm 0mm 0mm, clip, width=0.72\linewidth]{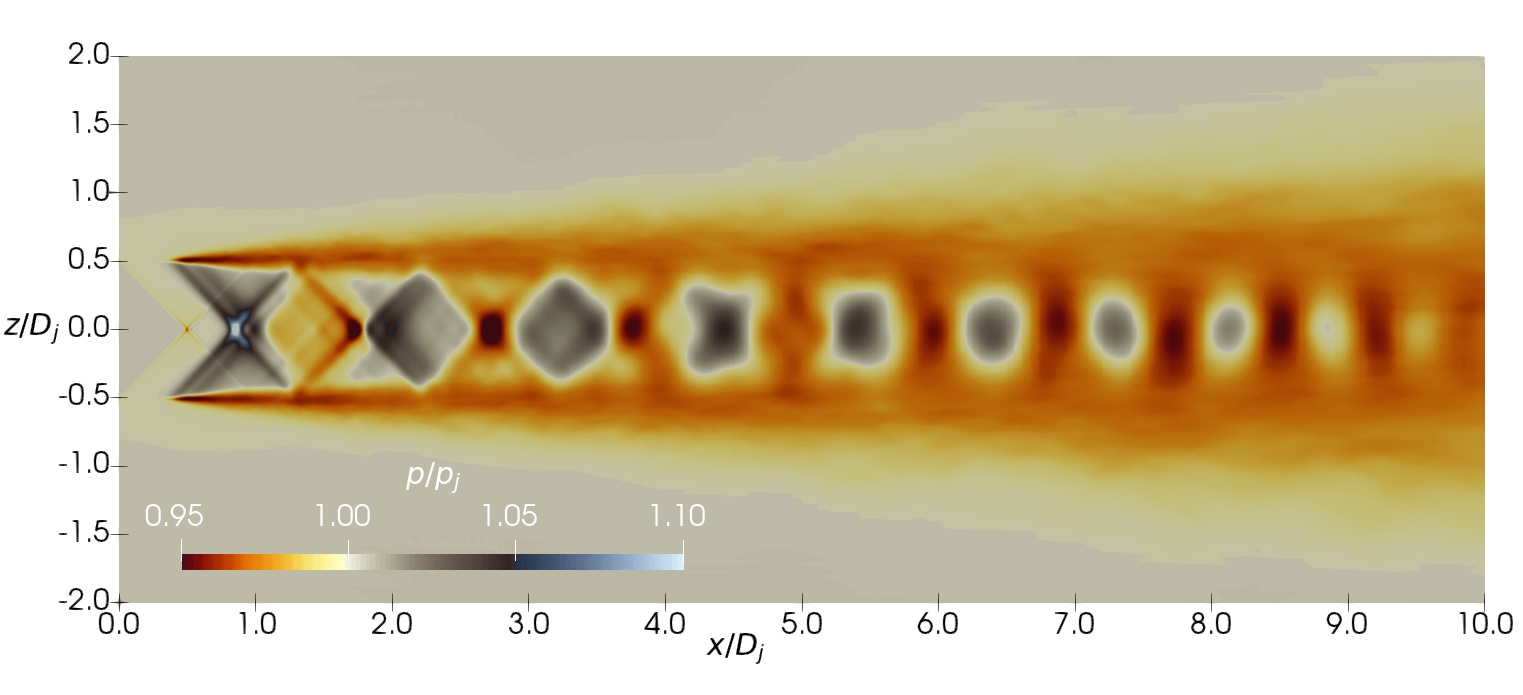}
	\label{res:mean_pres1}
	}
\\
\subfloat[Steady viscous profile.]{
	\includegraphics[trim = 0mm 0mm 0mm 0mm, clip, width=0.72\linewidth]{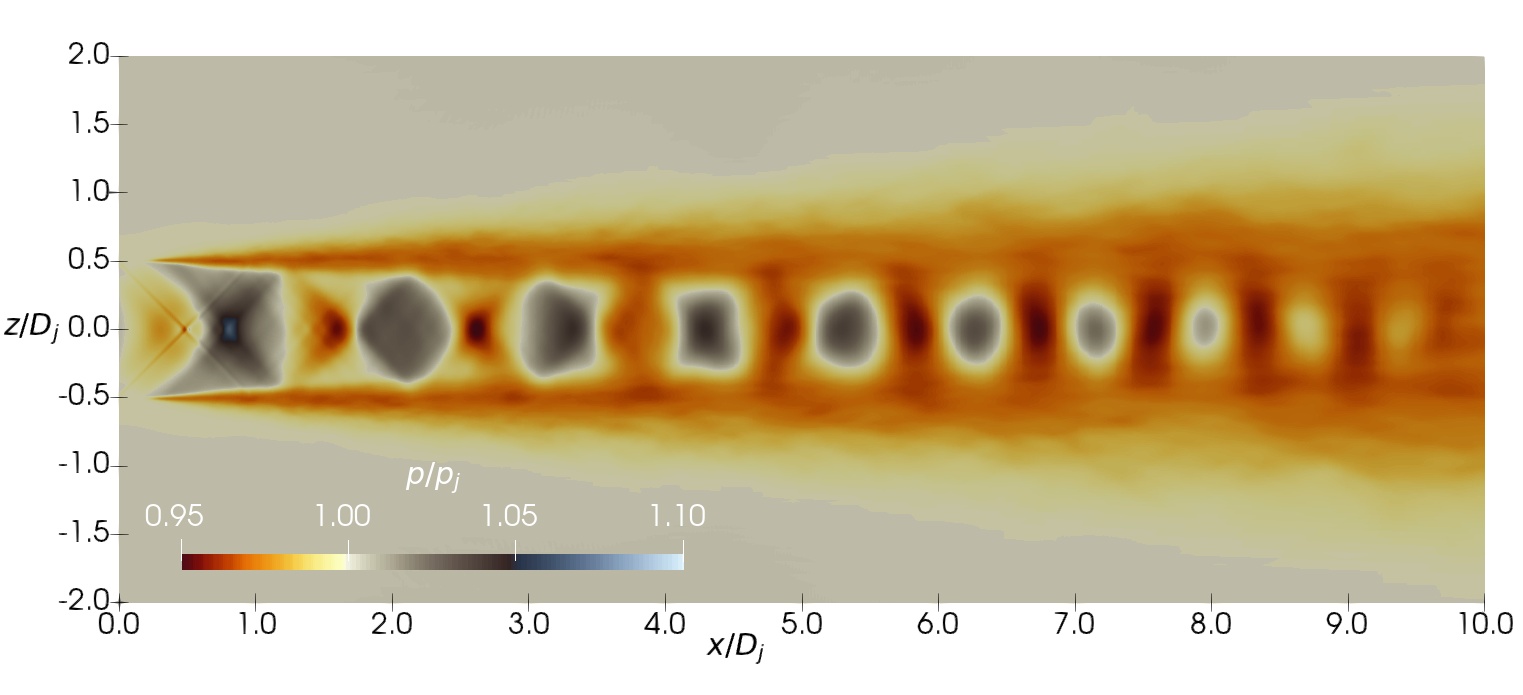}
	\label{res:mean_pres2}
	}
\\
\subfloat[Unsteady viscous profile.]{
	\includegraphics[trim = 0mm 0mm 0mm 0mm, clip, width=0.72\linewidth]{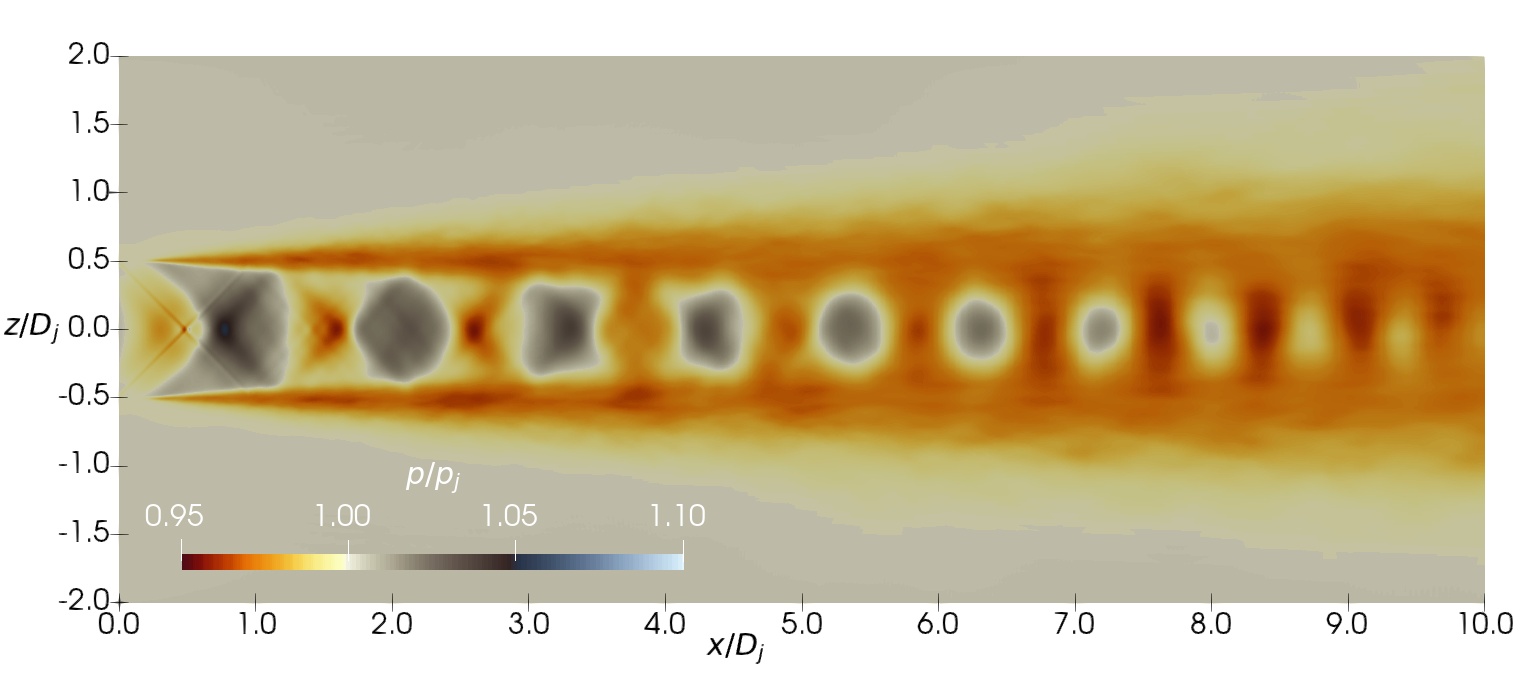}
	\label{res:mean_pres3}
	}
\caption{Mean pressure contours, $p/p_j$, on a plane through the centerline of the jet. The 2-D cut-plane is composed of $1200$ probes in the x-direction and $800$ probes in the z-direction. The statistical analysis is performed with a $2401$ time sample gathered with a non-dimensional time frequency of $21.124$.}
\label{res:mean_pres13}
\end{figure}

The mean pressure contours are shown in Fig.\ \ref{res:mean_pres13}. Results from the simulations with the inviscid and steady viscous inflow profiles, Figs.\ \ref{res:mean_pres1} and \ref{res:mean_pres2}, highlight the effects of imposing a nozzle-exit boundary layer on the jet flow. When the viscous profile is applied, the shock wave structure exhibits lower pressure amplitudes, represented by dark blue contours. In contrast, the inviscid profile results in higher pressure amplitudes, depicted in light blue. This trend is also observed in the mean streamwise velocity contours in Fig.\ \ref{res:mean_velx13}.

In the near-field region, $x/D_j < 1.5$, shock waves produced with the inviscid profile are thinner and exhibit larger pressure amplitudes compared to those generated with the steady and unsteady viscous profiles. The presence of a boundary layer at the inlet can induce shifts in the shock wave positions relative to the inviscid case, which may lead to a thickening of the shock structures and a reduction in mean pressure amplitude. The simulation using the steady viscous profile produces shock waves with pressure levels comparable to the inviscid case, whereas the unsteady viscous profile results in lower pressure amplitudes than both previous cases.


\subsubsection{Axial Profiles of Velocity Components and Shear Stress Tensor}

\begin{figure*}[tbp]
\centering
\subfloat[Averaged $U_{x}$ profile along jet centerline.]{
	\includegraphics[width=0.45\linewidth]{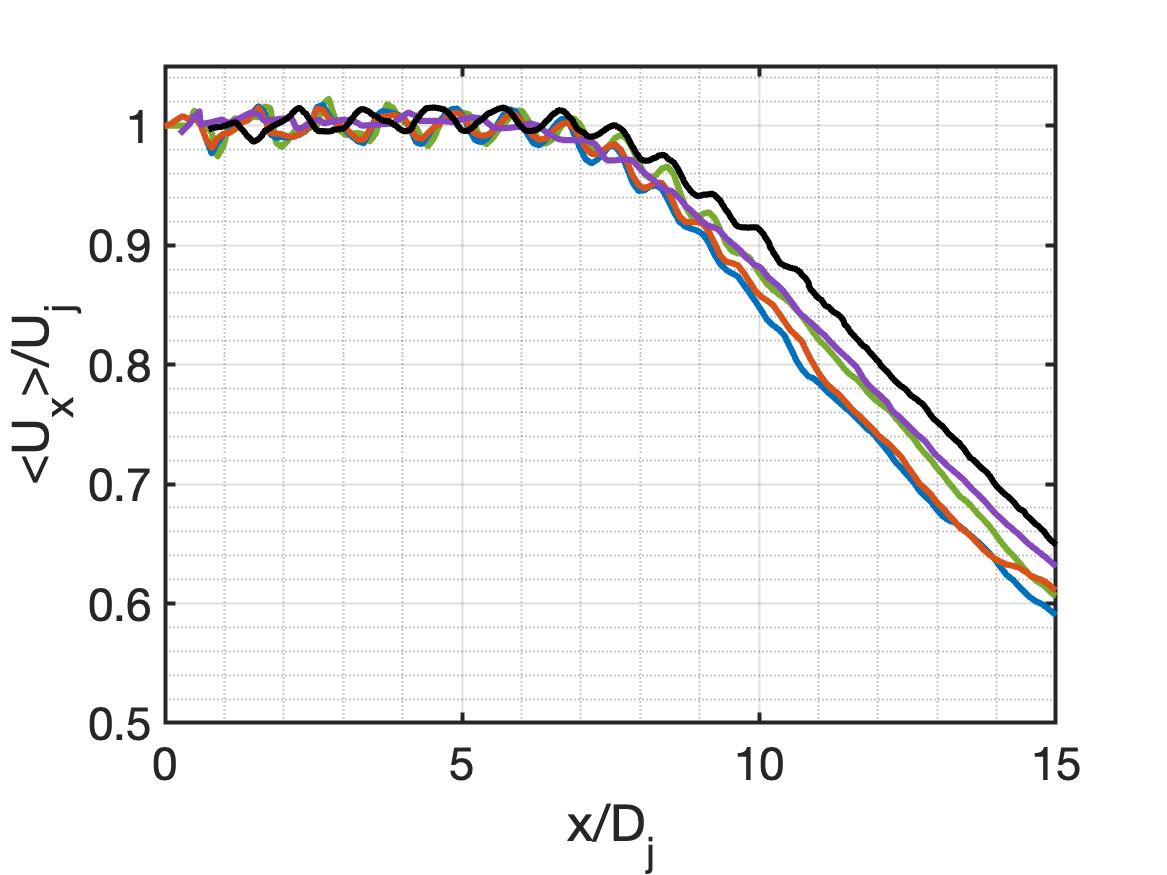}
	\label{res:ctl_mvelx_s1s3}	
	}%
\subfloat[RMS values of $u_{x}$ along jet centerline.]{
	\includegraphics[width=0.45\linewidth]{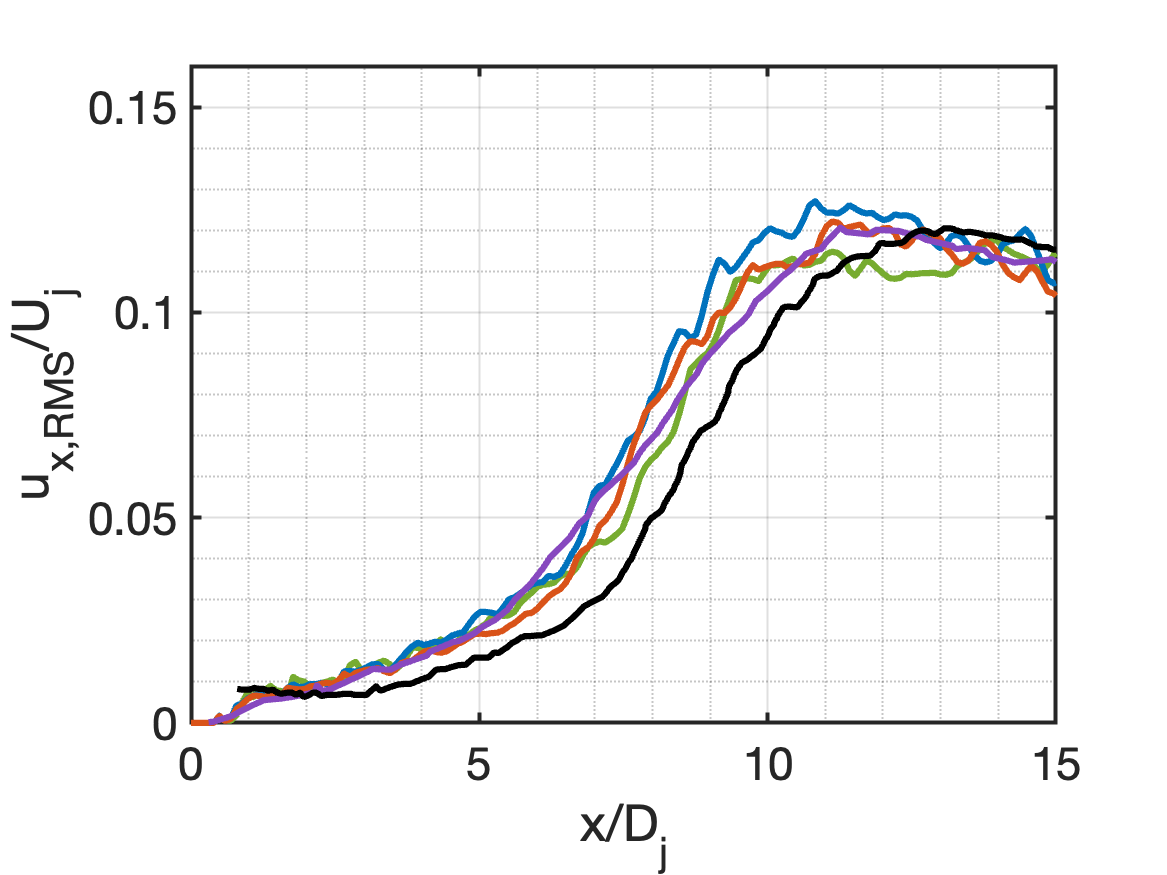}
	\label{res:ctl_rvelx_s1s3}	
	}
\\
\subfloat[RMS values of $u_{r}$ along jet centerline.]{
	\includegraphics[width=0.45\linewidth]{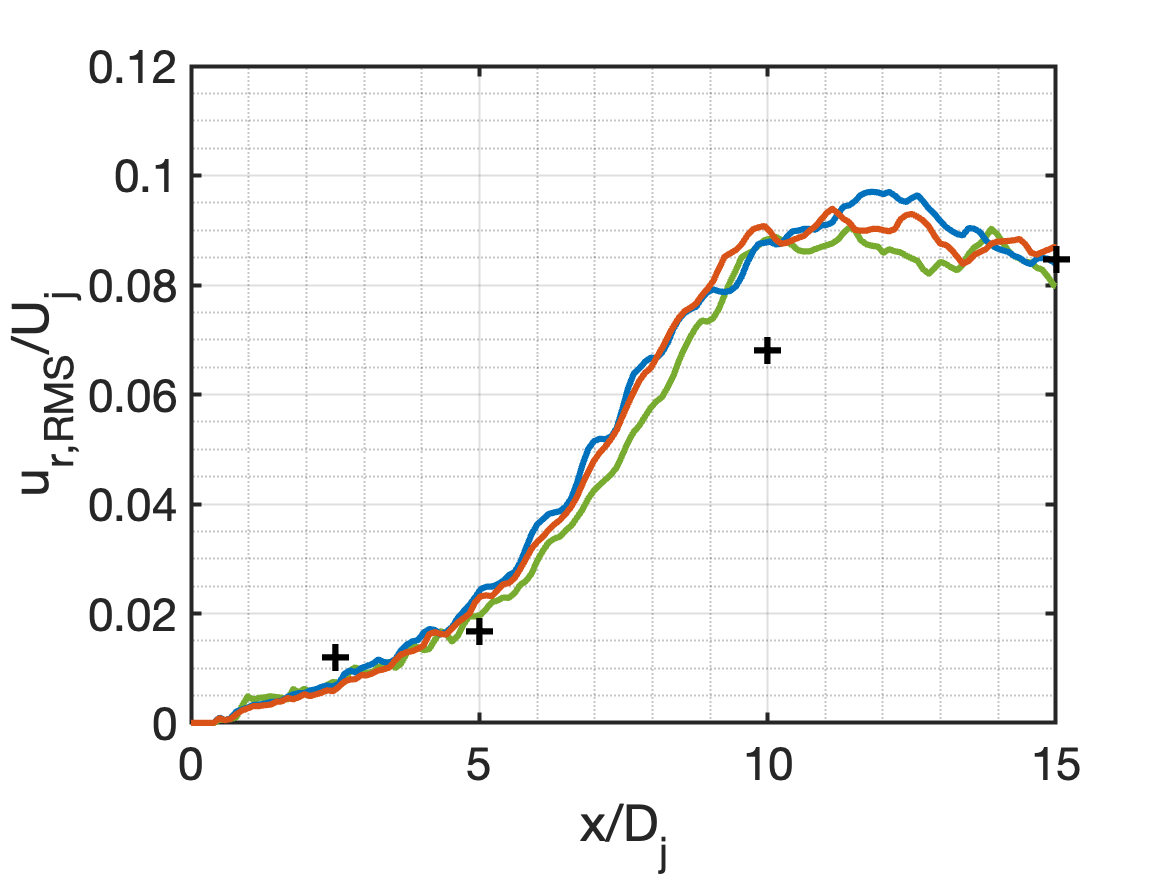}
	\label{res:ctl_rvelr_s1s3}
	}%
\subfloat[Averaged $u_{x}u_{r}$ profile along jet centerline.]{
	\includegraphics[width=0.45\linewidth]{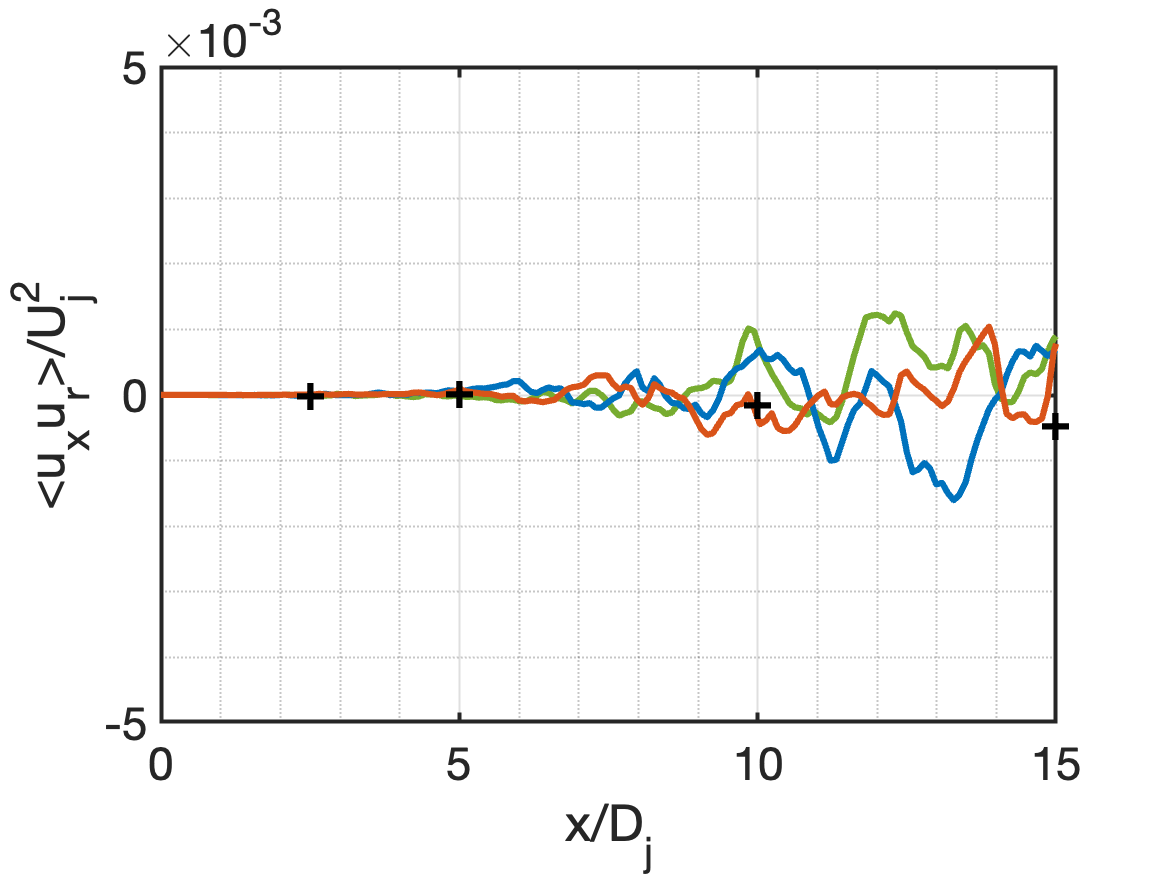}
	\label{res:ctl_mvur_s1s3}	
	}
\\
\subfloat{
	\includegraphics[trim = 9mm 8.8mm 6mm 122.5mm, clip, height=0.055\linewidth]{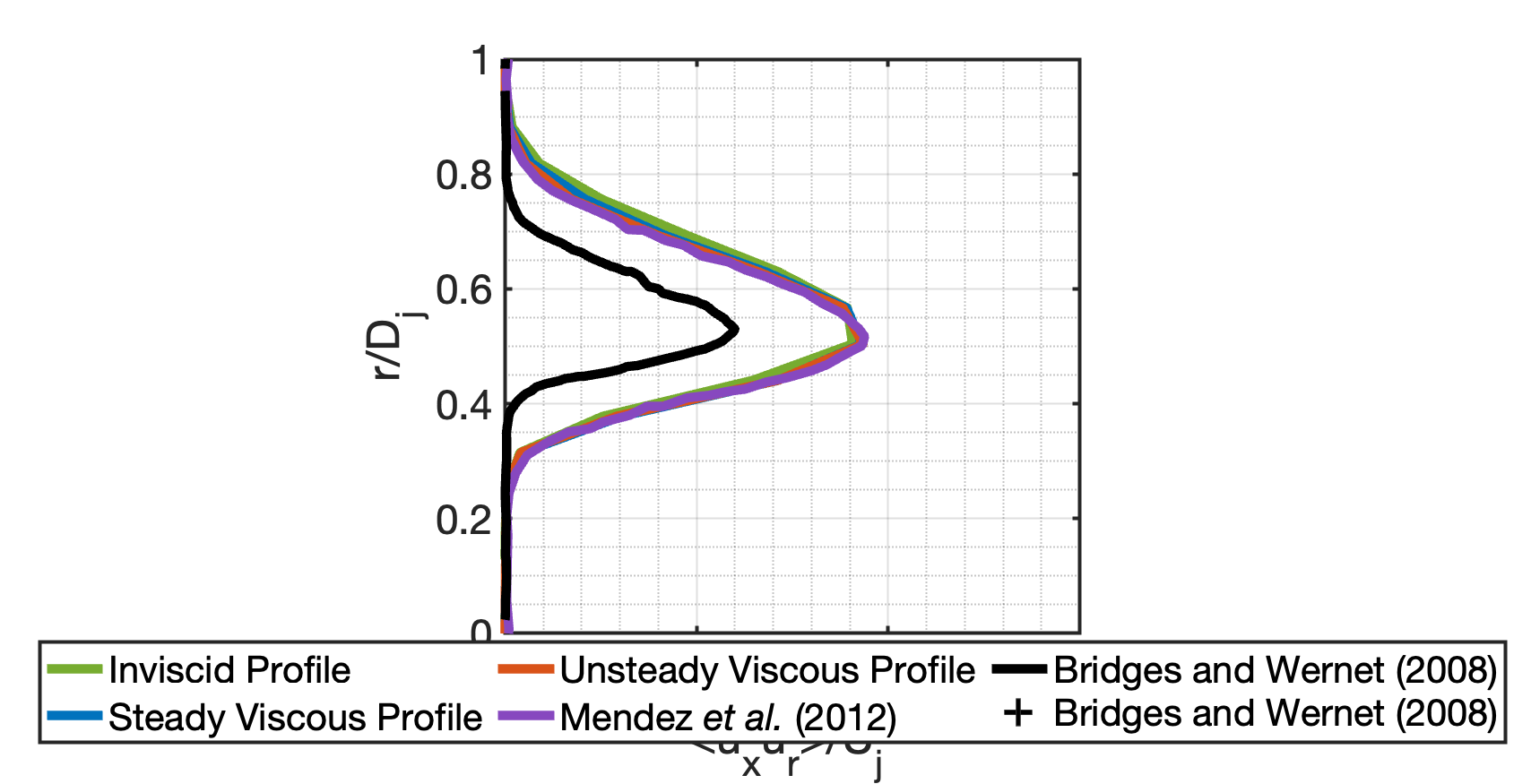}
	}
\caption{Axial profiles of velocity and stress tensor components at the centerline of the jet flow for the inflow condition studies.}
\label{res:ctl_s1s3}
\end{figure*}

The current section presents a quantitative analysis of the effects of inlet boundary conditions on the behavior of the supersonic jet flow. Figure~\ref{res:ctl_s1s3} presents axial profiles of the mean streamwise velocity component, RMS values of the streamwise and radial velocity fluctuations, and the mean shear stress tensor component along the jet centerline. The results include the present simulations using the three different inflow profiles, the numerical data from \citet{Mendezetal2012}, calculated at a Reynolds number of $1.5 \times 10^{5}$, and the experimental data from \citet{BridgesWernet2008}. 

The centerline mean velocity profiles shown in Fig.\ \ref{res:ctl_mvelx_s1s3} exhibit similar trends for all three inflow conditions up to $x/D_j < 7.0$. Beyond this region, $x/D_j > 7.0$, both simulations using viscous inflow profiles show lower velocity levels compared to the inviscid case. This behavior was also observed in the qualitative analysis, Fig.\ \ref{res:mean_velx13}, where the steady and unsteady viscous inlet profiles resulted in shorter potential core lengths compared to the inviscid configuration. When comparing the steady and unsteady viscous cases, the addition of the tripping method does not appear to affect the flow along the centerline significantly. The mean velocity profile obtained with the inviscid inflow closely follows the results of \citet{Mendezetal2012}. Compared to the experimental data of \citet{BridgesWernet2008}, the numerical profiles predict slightly lower velocity levels in the region $7.0 < x/D_j < 10.0$, but a similar velocity decay rate is observed for $x/D_j > 10.0$.

The axial profiles of the RMS values of the longitudinal velocity component fluctuations at the jet centerline from the three numerical simulations performed in the present work are presented in Fig.\ \ref{res:ctl_rvelx_s1s3} along with numerical \cite{Mendezetal2012} and experimental \cite{BridgesWernet2008} reference data. The velocity fluctuation profiles from the numerical simulations are nearly identical up to $x/D_j=7.0$. After that section, the RMS values of the axial velocity distributions from the simulations with the steady and unsteady viscous profiles diverge from those obtained with the inviscid profile. In the region of the peak of the axial velocity fluctuation distribution, $11.0 < x/D_j < 13.0$, it is observed that the simulation with the steady viscous profile presents the highest values of velocity fluctuation, and the simulation with the inviscid profile presents the lowest values. The profile obtained from data in \citet{Mendezetal2012} is similar to the numerical profiles from the simulations conducted in the present research effort. The comparison with the experimental data indicates that higher velocity fluctuation values from the numerical simulations are observed for $x/D_j < 12.0$. The differences between the numerical values and the experimental data reduce for $x/D_j \geq 12.0$.

The RMS values of radial velocity fluctuations and the mean shear stress component axial profiles obtained in the present work are shown in Figs.\ \ref{res:ctl_rvelr_s1s3} and \ref{res:ctl_mvur_s1s3}. They are compared with the limited experimental data available in \citet{BridgesWernet2008}. The reference simulation by \citet{Mendezetal2012} does not report data on radial velocity fluctuations or the mean shear stress component along the jet centerline, and therefore cannot be used for direct comparison with the present results. The axial profiles of the RMS values of radial velocity fluctuations exhibit similar behavior across all three inflow conditions. The peak RMS values are consistently lower than those of the streamwise velocity fluctuations along the centerline. Compared with the experimental pointwise data, the numerical results are lower at $x/D_j = 2.5$, while at $x/D_j = 5.0$ and $x/D_j = 10.0$, they exceed the experimental values. At $x/D_j = 15.0$, the numerical and experimental values are in close agreement.

The three simulations produce similar profiles of the mean shear stress component, as shown in Fig.\ \ref{res:ctl_mvur_s1s3}. All profiles remain close to zero for $x/D_j < 5.0$, after which they begin to oscillate around zero. The simulation with the steady viscous inflow exhibits the largest oscillation amplitudes, while the inviscid and unsteady viscous cases yield comparable values. Experimental data for the radial velocity component and the shear stress tensor are available only at discrete axial positions, which limits a detailed assessment of the numerical results against measurements.

\begin{figure*}[tbp]
\centering
\subfloat[Averaged $U_{x}$ profile along jet lipline.]{
	\includegraphics[width=0.45\linewidth]{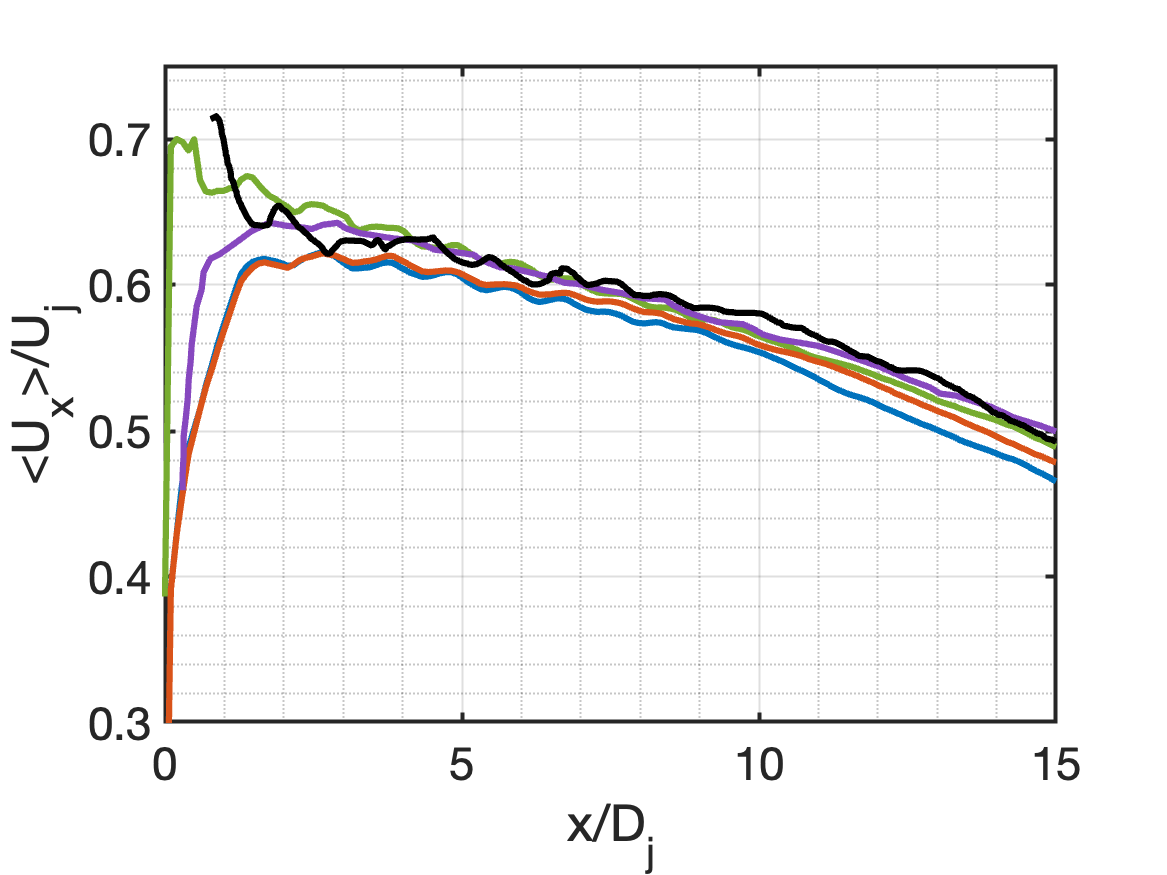}
	\label{res:lpl_mvelx_s1s3}	
	}%
\subfloat[RMS values of $u_{x}$ along jet lipline.]{
	\includegraphics[width=0.45\linewidth]{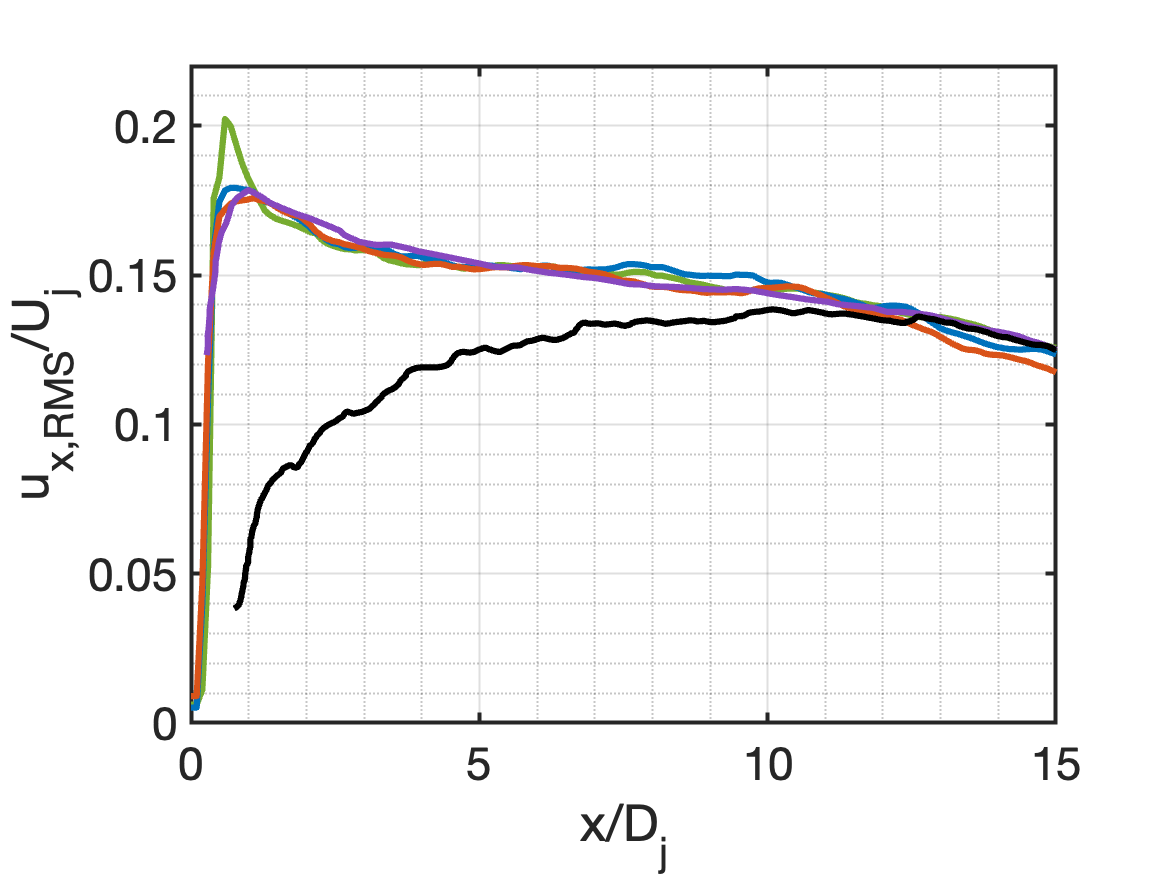}
	\label{res:lpl_rvelx_s1s3}	
	}
\\
\subfloat[RMS values of $u_{r}$ along jet lipline.]{
	\includegraphics[width=0.45\linewidth]{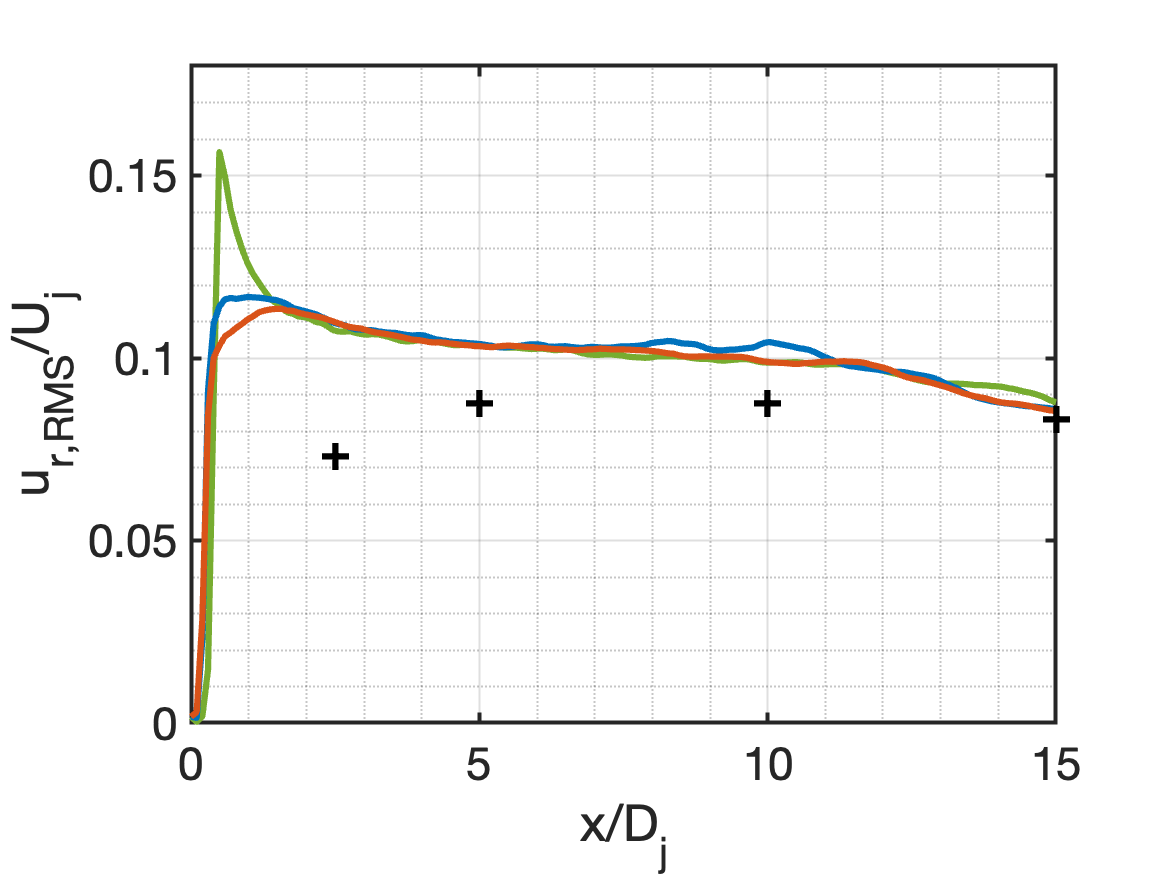}
	\label{res:lpl_rvelr_s1s3}
	}%
\subfloat[Averaged $u_{x}u_{r}$ profile along jet lipline.]{
	\includegraphics[width=0.45\linewidth]{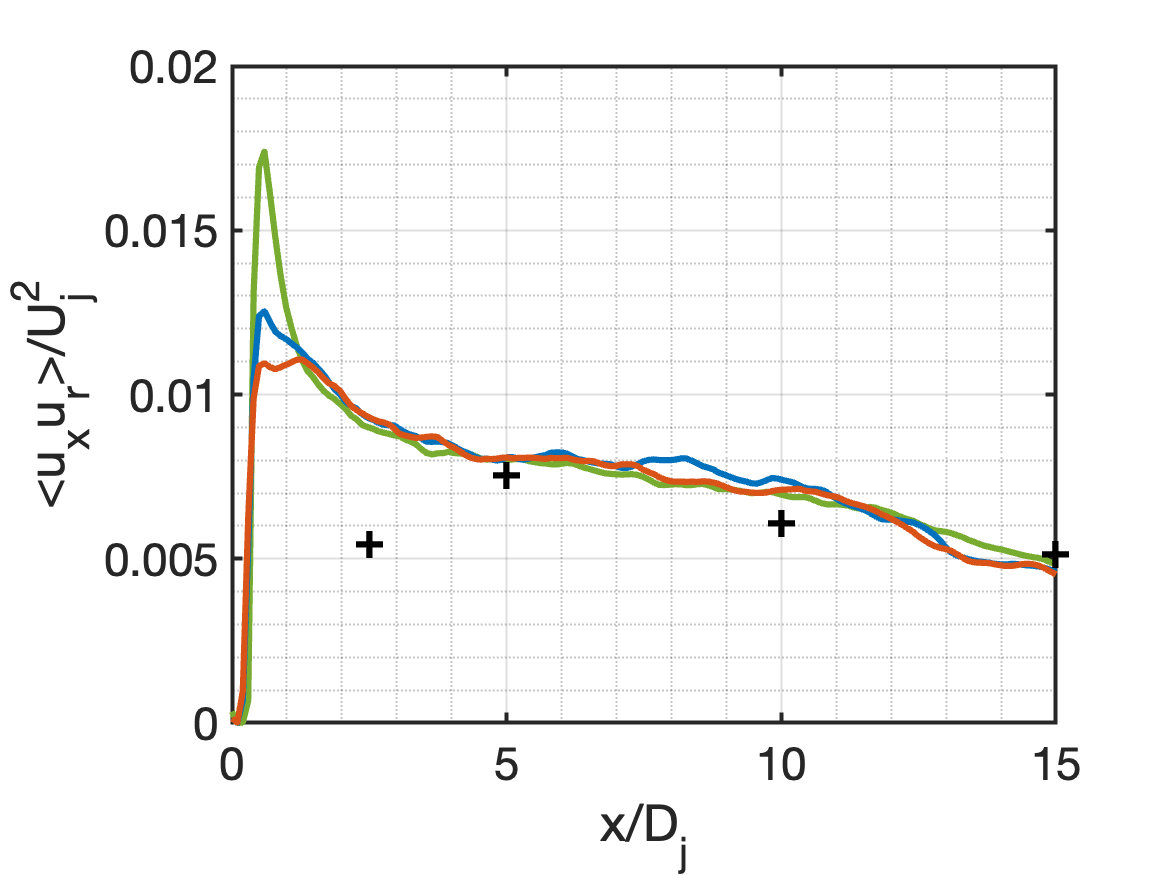}
	\label{res:lpl_mvur_s1s3}	
	}
\\
\subfloat{
	\includegraphics[trim = 9mm 8.8mm 6mm 122.5mm, clip, height=0.055\linewidth]{fig-res/legend_s1tos3_jr.png}
	}
\caption{Axial profiles of velocity and stress tensor components at the lipline of the jet flow for the inflow condition studies.}
\label{res:lpl_s1s3}
\end{figure*}

In contrast to the flow properties along the centerline, the velocity profiles at the jet lipline are significantly influenced by the imposed inflow conditions. Figure \ref{res:lpl_s1s3} shows axial profiles of the mean longitudinal velocity component, the RMS values of longitudinal and radial velocity fluctuations, and the mean shear stress tensor component along the lipline for the simulations conducted in the present study, as well as for reference data from the literature \cite{Mendezetal2012, BridgesWernet2008}. The mean longitudinal velocity profile in Fig.\ \ref{res:lpl_mvelx_s1s3}, computed with the inviscid inflow condition, exhibits good agreement with the experimental data, although it indicates higher velocity values near the jet inlet region, $x/D_j \approx 0.0$, compared to both simulations using viscous inflow conditions. The numerical reference by \citet{Mendezetal2012} shows a velocity profile shape similar to those obtained with viscous boundary layers, suggesting that the boundary layer thickness in the simulations may be larger than that in the experimental setup. For $x/D_j > 7.0$, the velocity values from the simulation using the unsteady viscous inflow profile converge toward those from the inviscid profile. The simulation with the steady viscous profile yields slightly lower values than the other two.

When evaluating the fluctuation profiles, all simulations overpredict the axial distribution of the RMS values of the longitudinal velocity component fluctuations along the lipline, Fig.\ \ref{res:lpl_rvelx_s1s3}. The imposition of a steady viscous boundary layer reduces the peak of the profile predicted using the inviscid inlet condition, occurring at $0.5 < x/D_j < 1.0$, by approximately $10\%$. A similar trend is observed in the profiles of the radial velocity component fluctuations and the Reynolds stress tensor component, Figs.\ \ref{res:lpl_rvelr_s1s3} and \ref{res:lpl_mvur_s1s3}, where the inviscid inlet condition leads to a profile peak near the jet entrance. The tripped boundary layer condition also lowers the peak around the same region, $0.0 < x/D_j < 2.0$, when compared to the case with a steady viscous boundary layer at the inlet. Aside from the peak differences, the overall shape of the profiles is similar across all simulations. The RMS values of $u_r$ along the lipline are consistently overestimated when compared to the experimental measurements at $x/D_j = 2.5$, $5.0$, and $10.0$. The time-averaged $u_x u_r$ profiles along the lipline are overpredicted at $x/D_j = 2.5$, but show good agreement with the experimental data at $x/D_j = 5.0$ and $x/D_j = 10.0$.

An important contribution of the present study is the release of a comprehensive numerical database containing the full simulation results for all three inflow conditions. This publicly available dataset \cite{database42024, database52024, database62024} is intended to support future investigations into supersonic jet flows, offering a valuable resource for validation, model development, and further analysis of inlet boundary condition effects. In particular, all the data shown in the present discussion is available in the numerical database.

\subsubsection{Radial Profiles of Velocity and Shear Stress Tensor Components}

\begin{figure*}[tbp]
\centering
\subfloat[$x/D_j=2.5$]{
	\includegraphics[trim = 20mm 0mm 32mm 8mm, clip, height=0.2\linewidth]{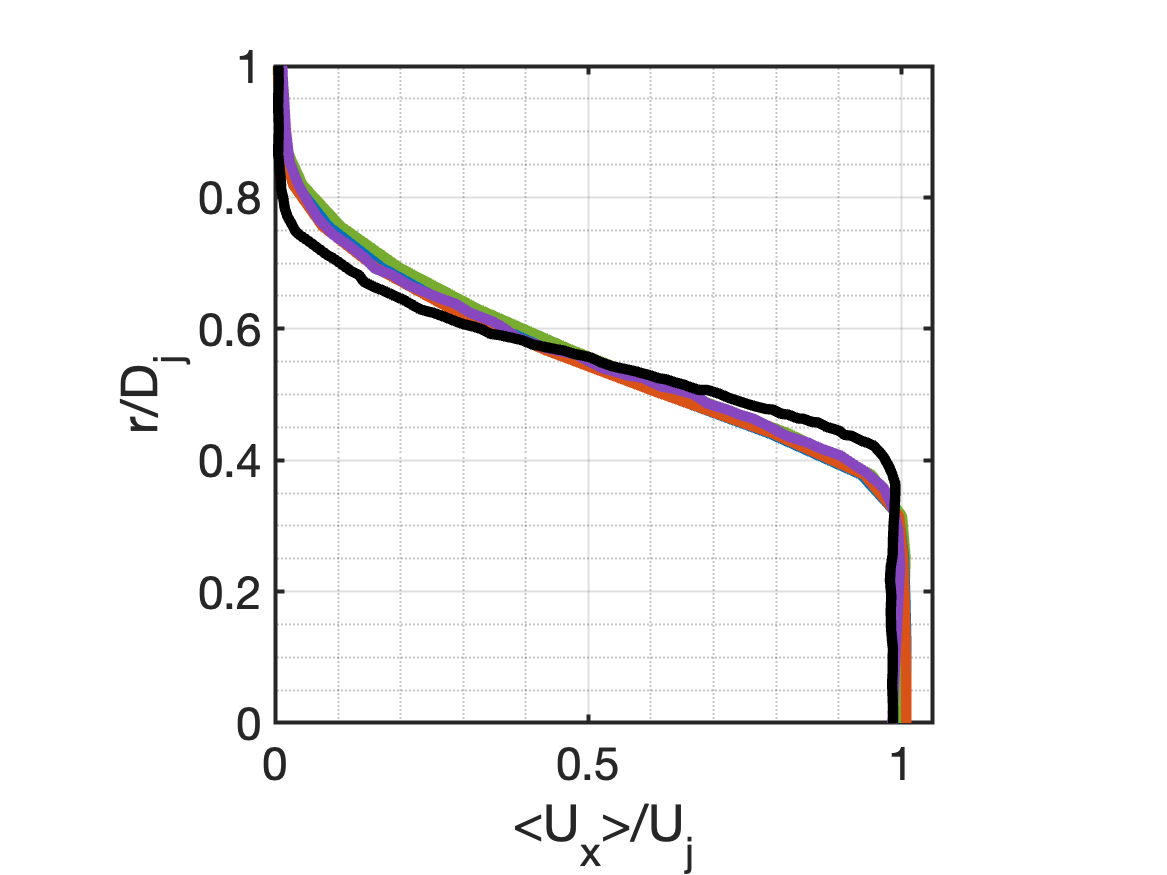}
	\label{res:s1s3_radiala}	
	}
\subfloat[$x/D_j=5.0$]{
    \includegraphics[trim = 20mm 0mm 32mm 8mm, clip, height=0.2\linewidth]{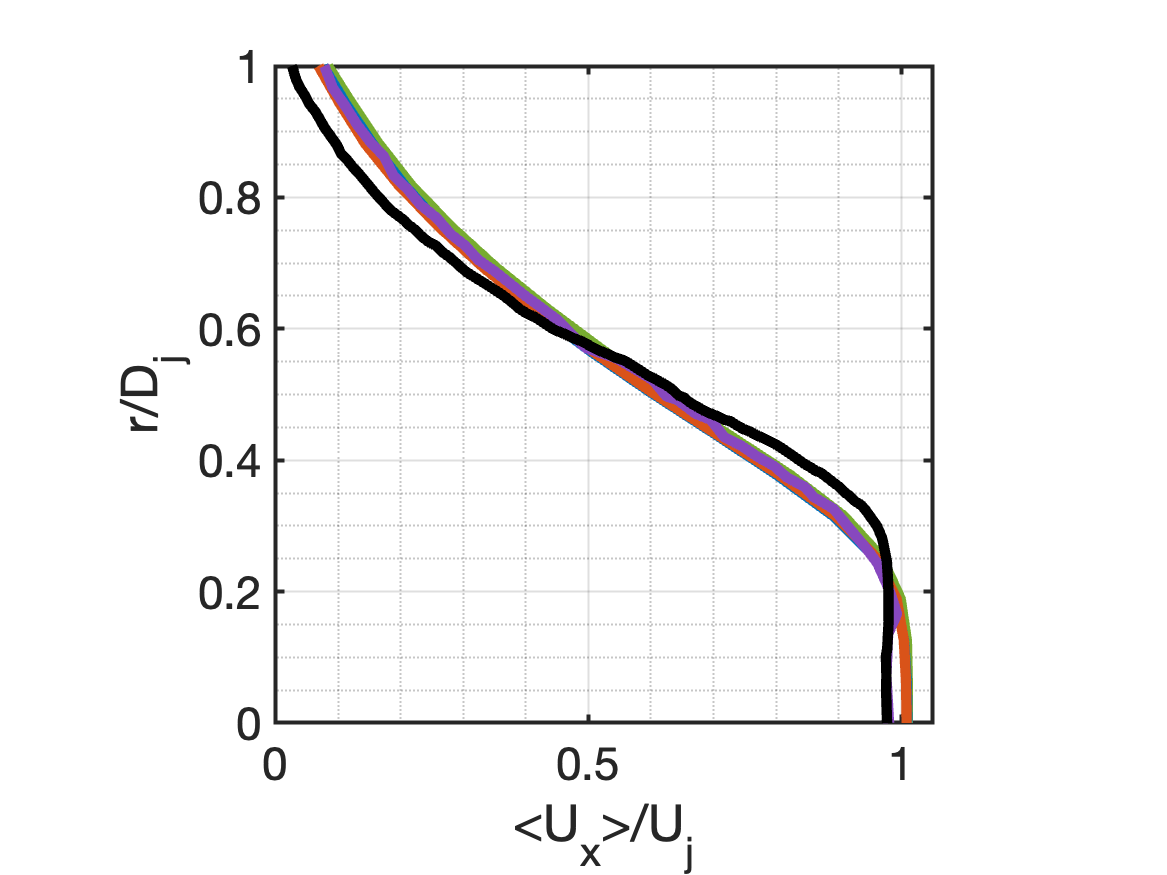}
	\label{res:s1s3_radialb}	
	}
\subfloat[$x/D_j=10.0$]{
	\includegraphics[trim = 20mm 0mm 32mm 8mm, clip, height=0.2\linewidth]{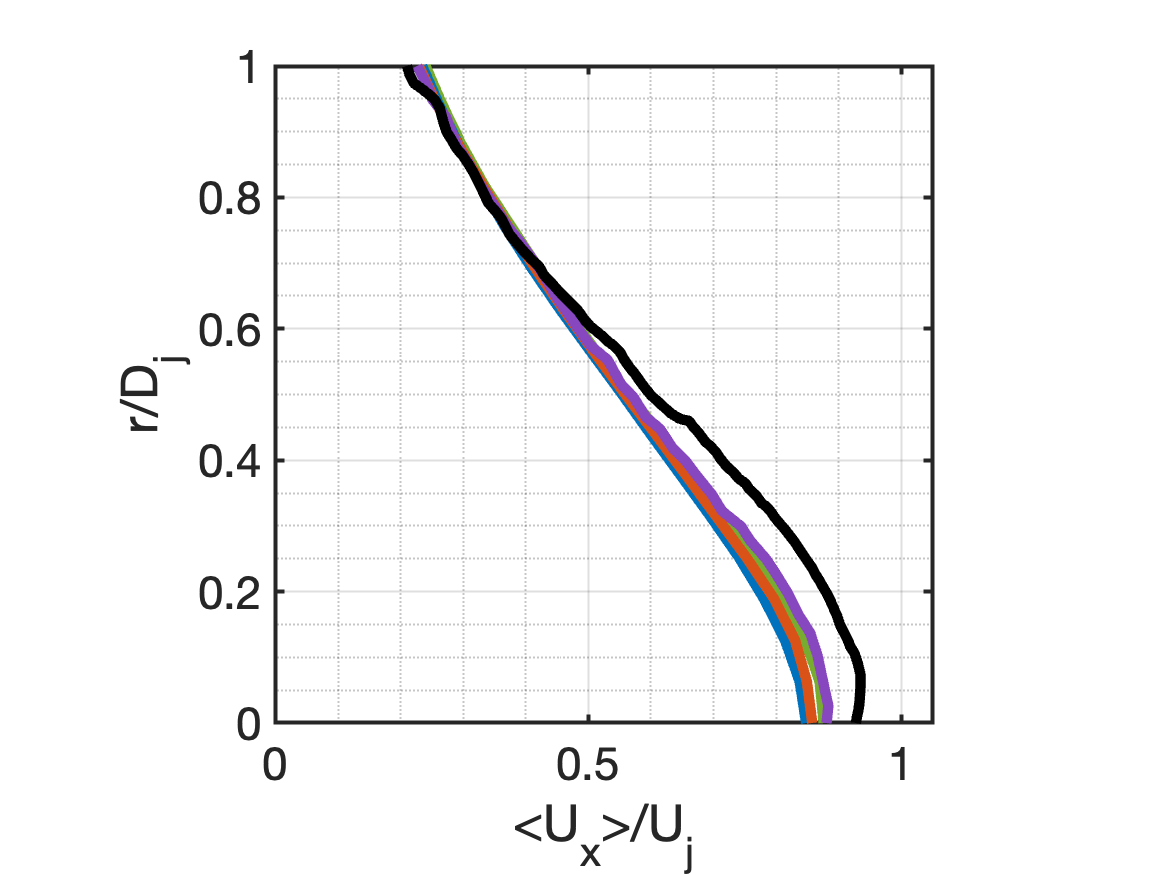}
	\label{res:s1s3_radialc}
	}
\subfloat[$x/D_j=15.0$]{
	\includegraphics[trim = 20mm 0mm 32mm 8mm, clip, height=0.2\linewidth]{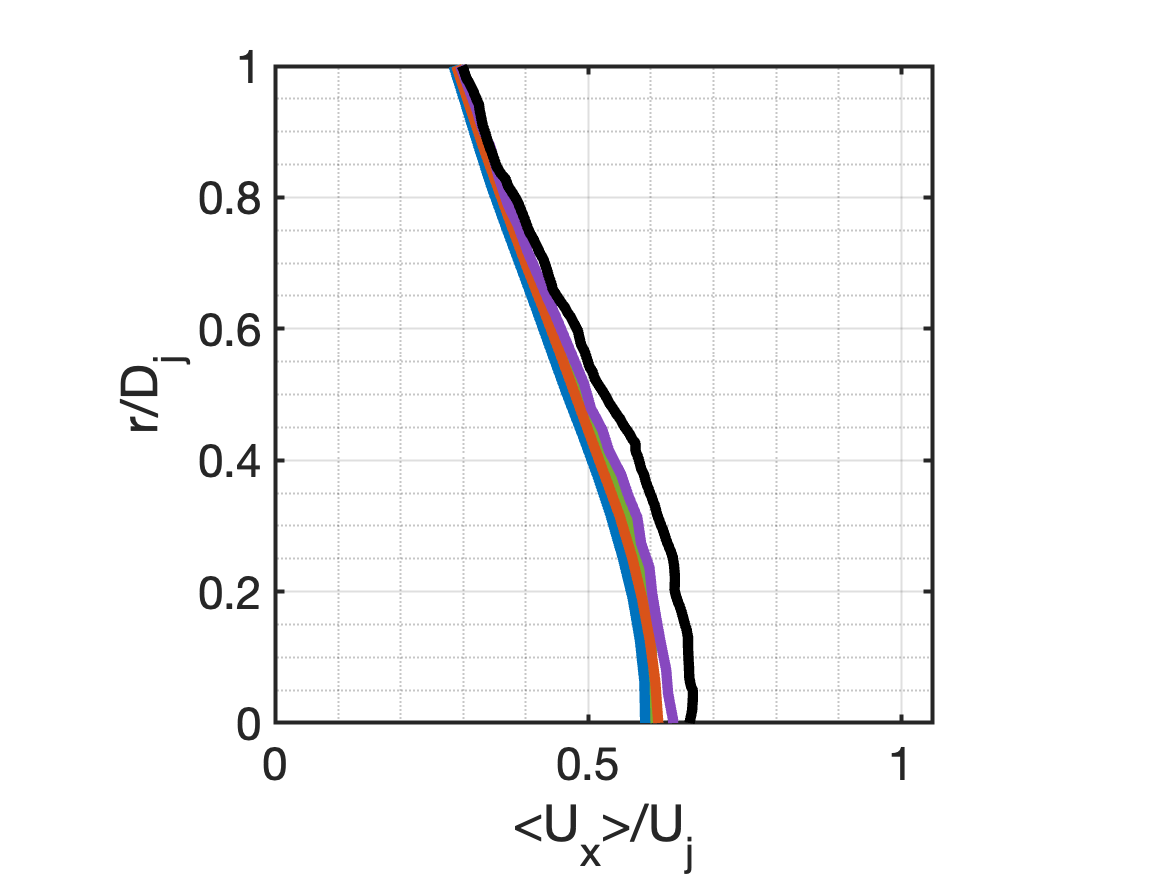}
	\label{res:s1s3_radiald}	
	}
\\
\subfloat[$x/D_j=2.5$]{
	\includegraphics[trim = 20mm 0mm 32mm 8mm, clip, height=0.2\linewidth]{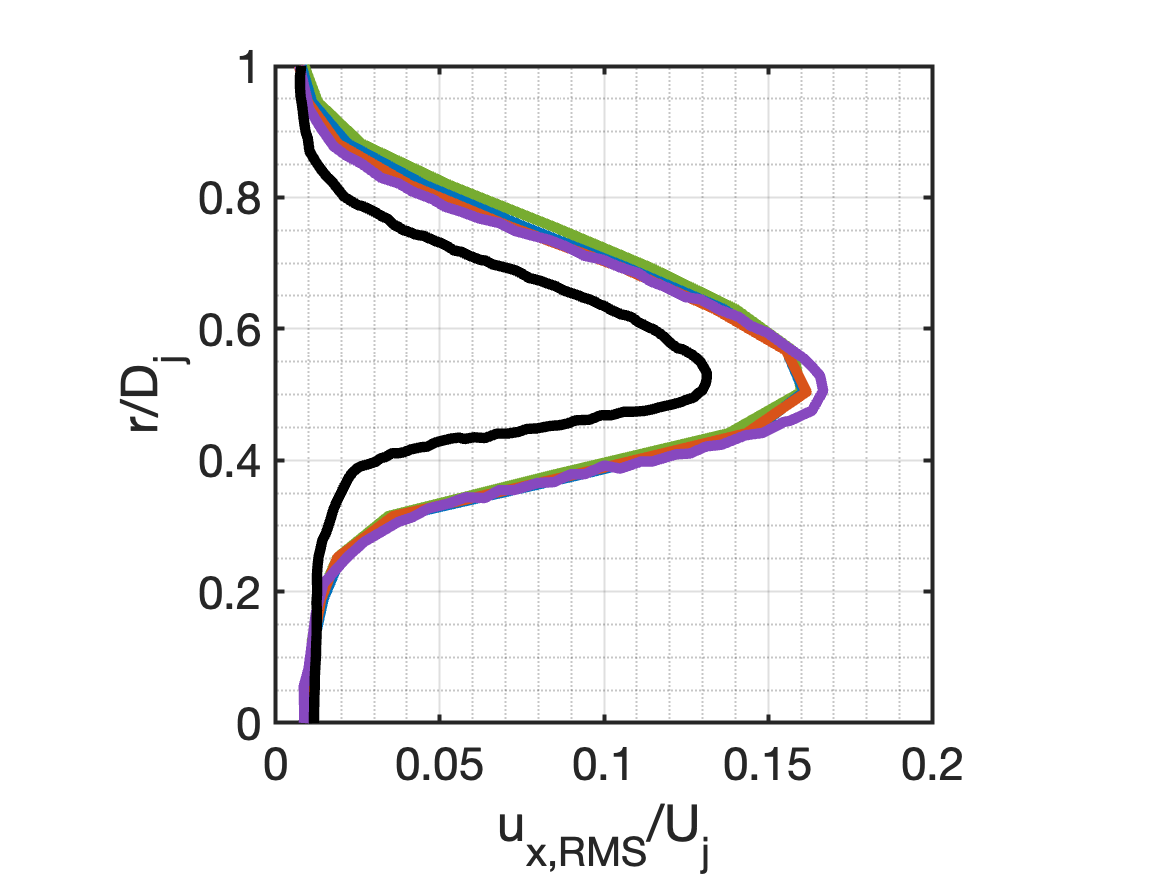}
	\label{res:s1s3_radiale}	
	}
\subfloat[$x/D_j=5.0$]{
	\includegraphics[trim = 20mm 0mm 32mm 8mm, clip, height=0.2\linewidth]{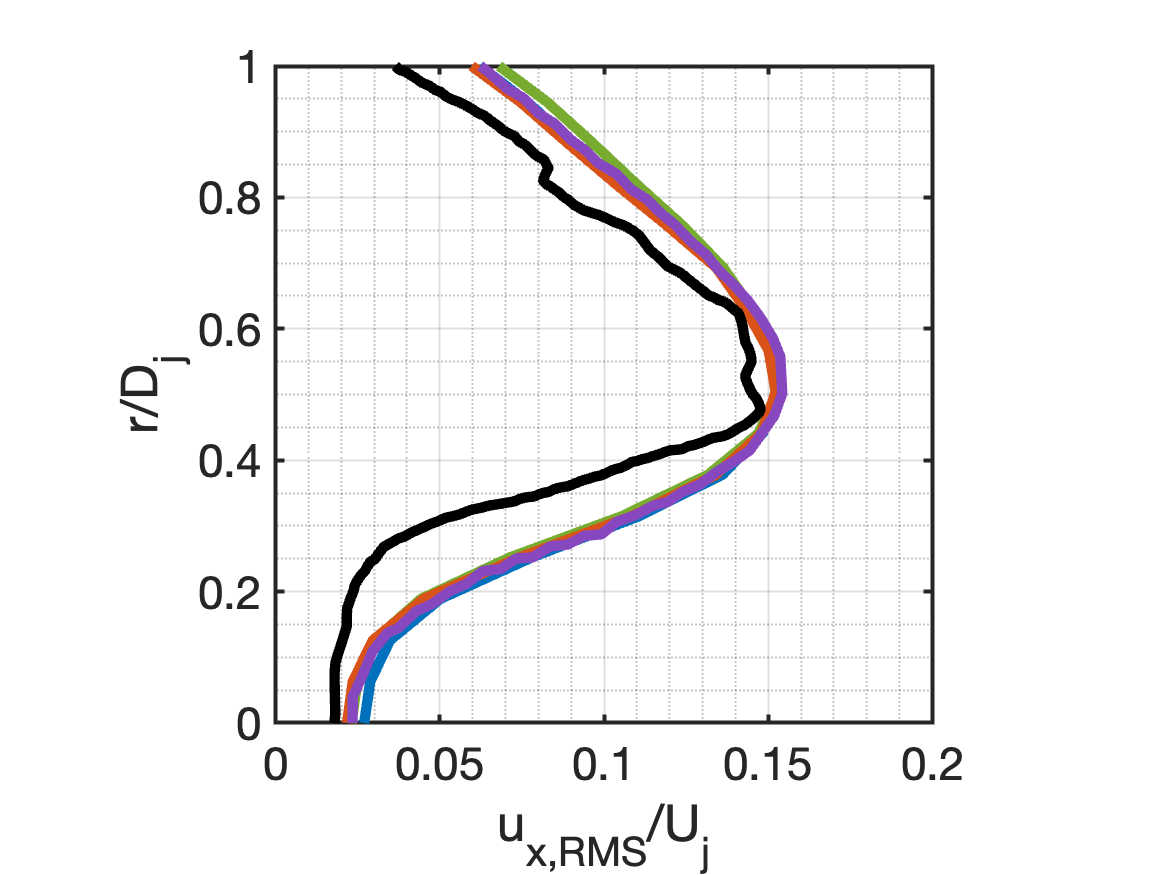}
	\label{res:s1s3_radialf}	
	}
\subfloat[$x/D_j=10.0$]{
	\includegraphics[trim = 20mm 0mm 32mm 8mm, clip, height=0.2\linewidth]{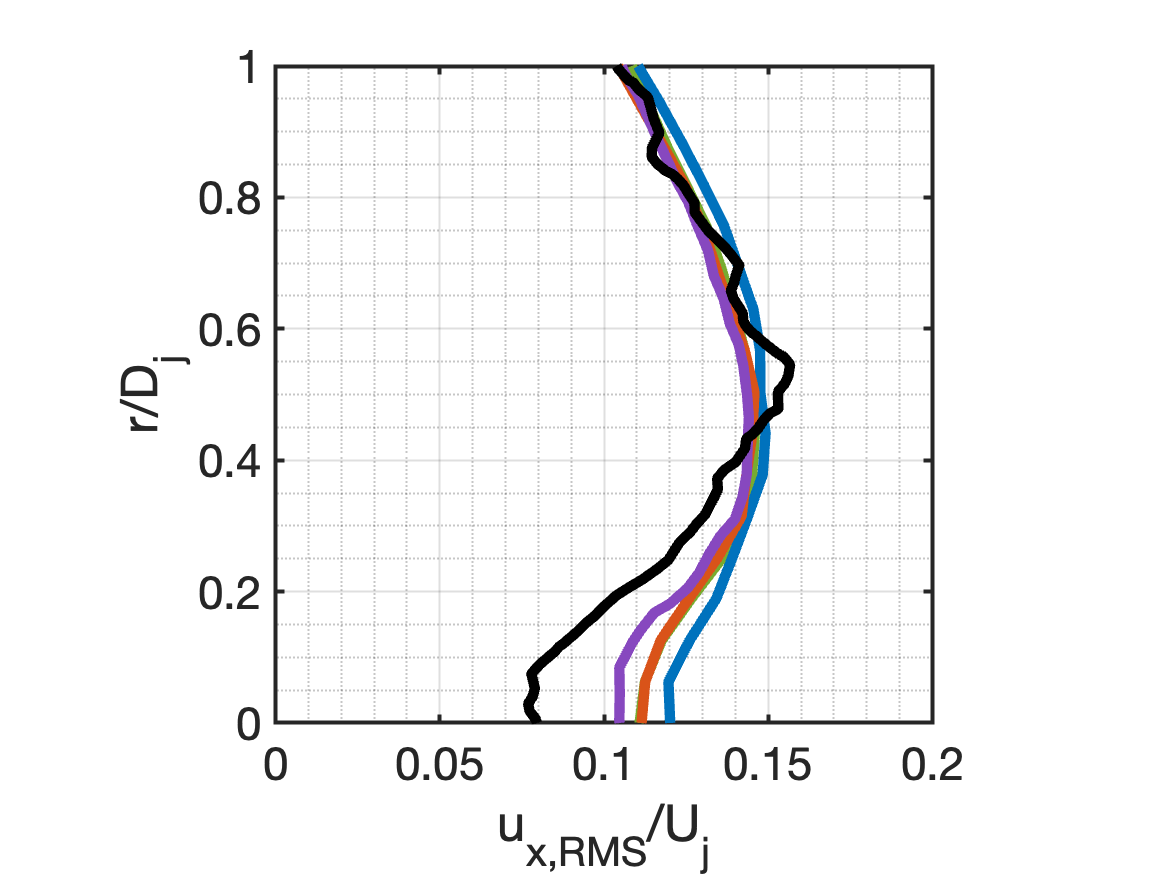}
	\label{res:s1s3_radialg}
	}
\subfloat[$x/D_j=15.0$]{
	\includegraphics[trim = 20mm 0mm 32mm 8mm, clip, height=0.2\linewidth]{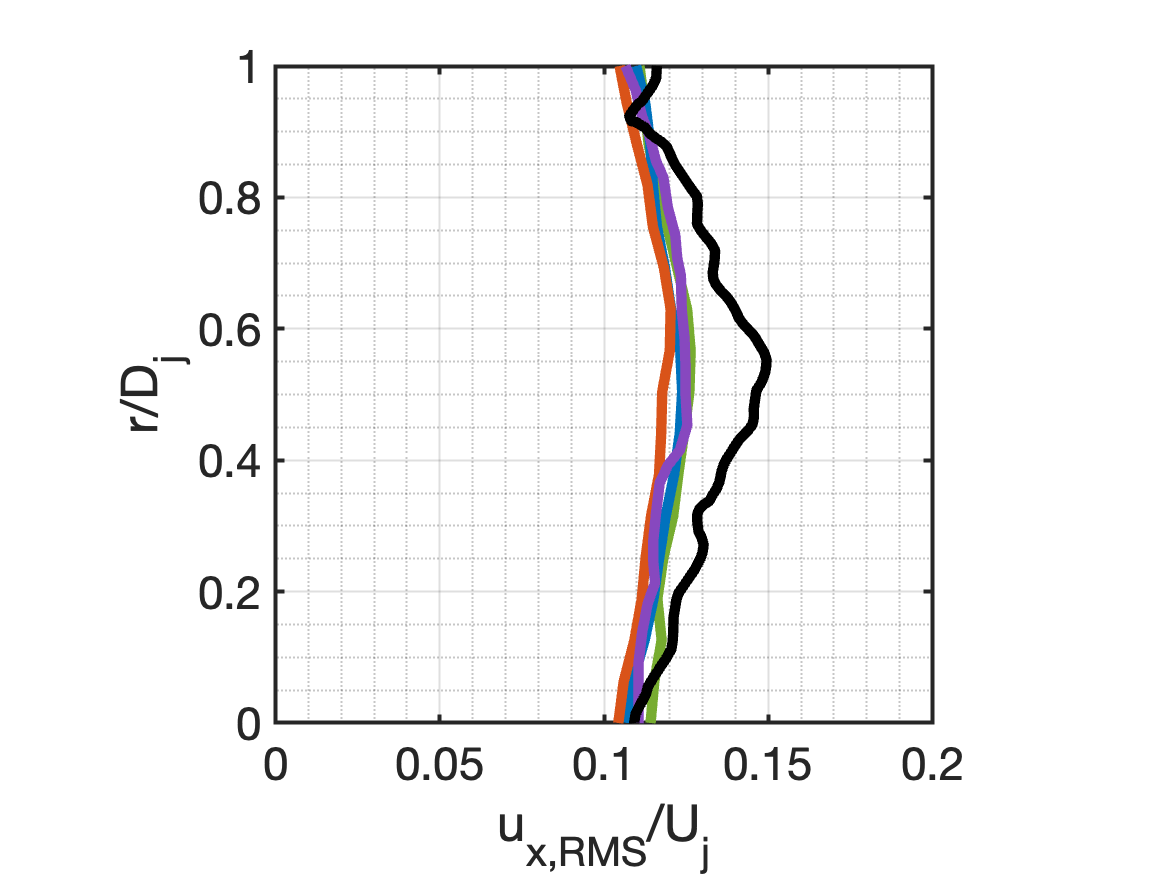}
	\label{res:s1s3_radialh}	
	}
\\
\subfloat[$x/D_j=2.5$]{
	\includegraphics[trim = 20mm 0mm 32mm 8mm, clip, height=0.2\linewidth]{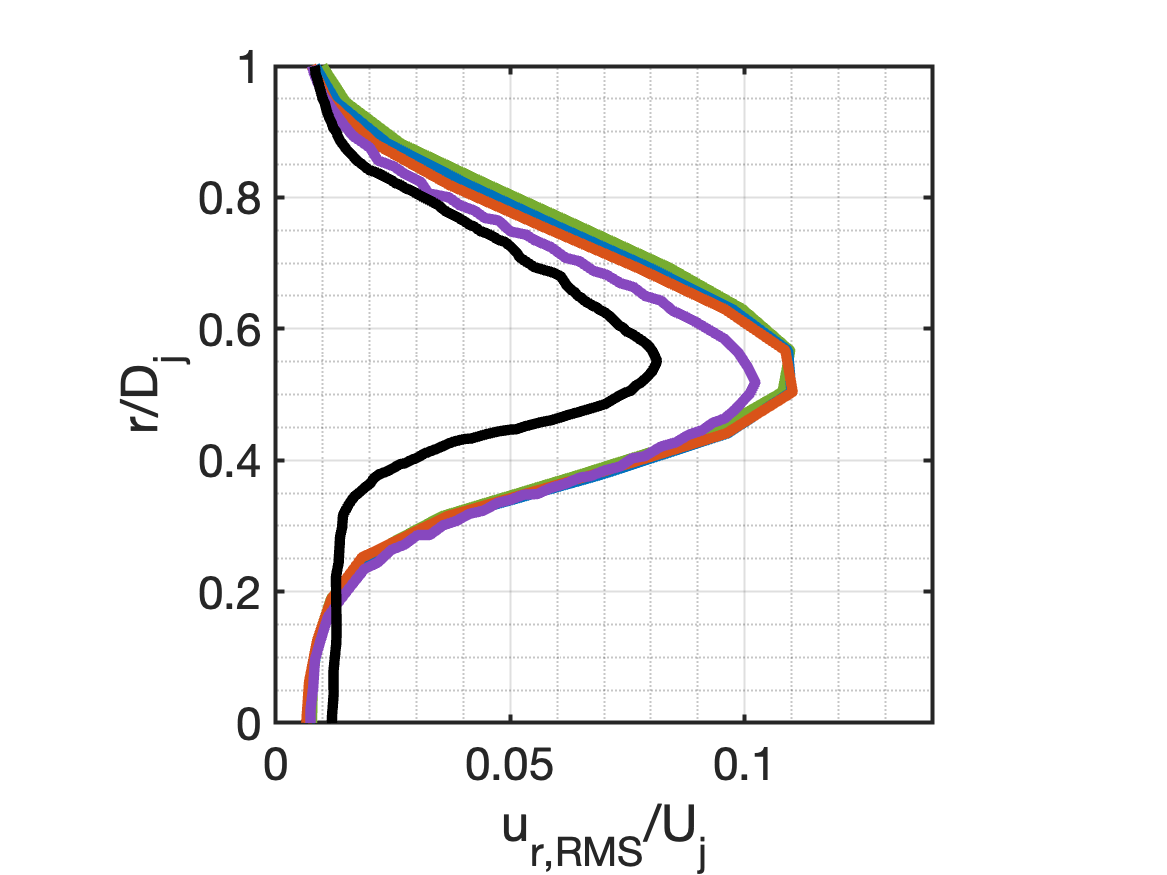}
	\label{res:s1s3_radiali}	
	}
\subfloat[$x/D_j=5.0$]{
	\includegraphics[trim = 20mm 0mm 32mm 8mm, clip, height=0.2\linewidth]{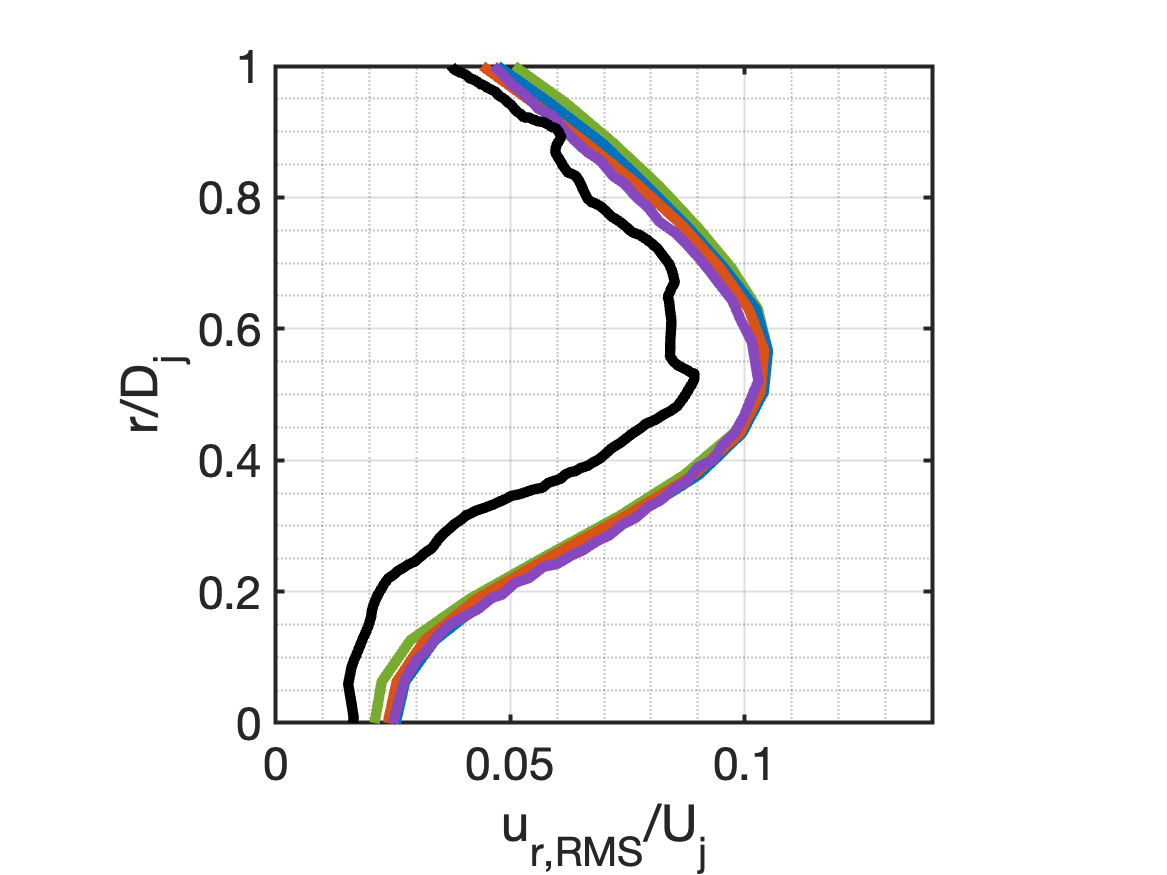}
	\label{res:s1s3_radialj}	
	}
\subfloat[$x/D_j=10.0$]{
	\includegraphics[trim = 20mm 0mm 32mm 8mm, clip, height=0.2\linewidth]{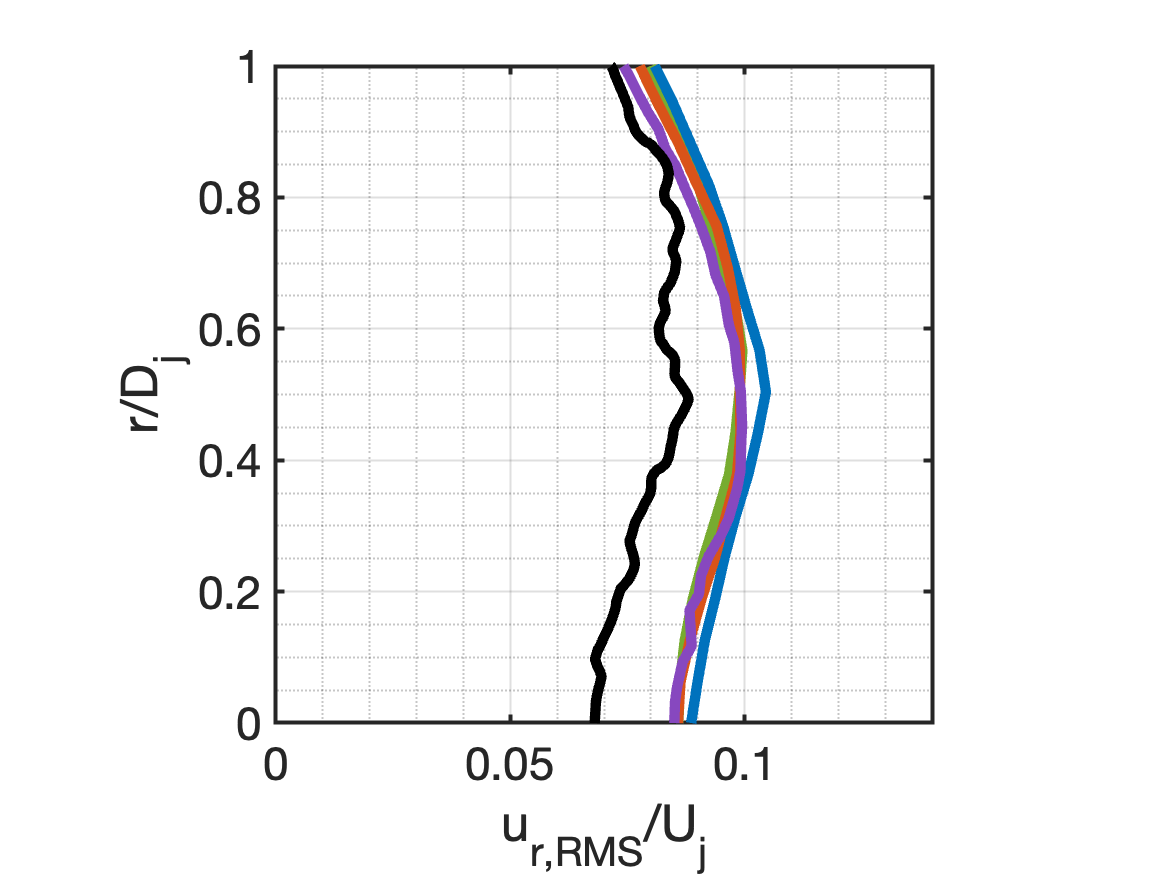}
	\label{res:s1s3_radialk}
	}
\subfloat[$x/D_j=15.0$]{
	\includegraphics[trim = 20mm 0mm 32mm 8mm, clip, height=0.2\linewidth]{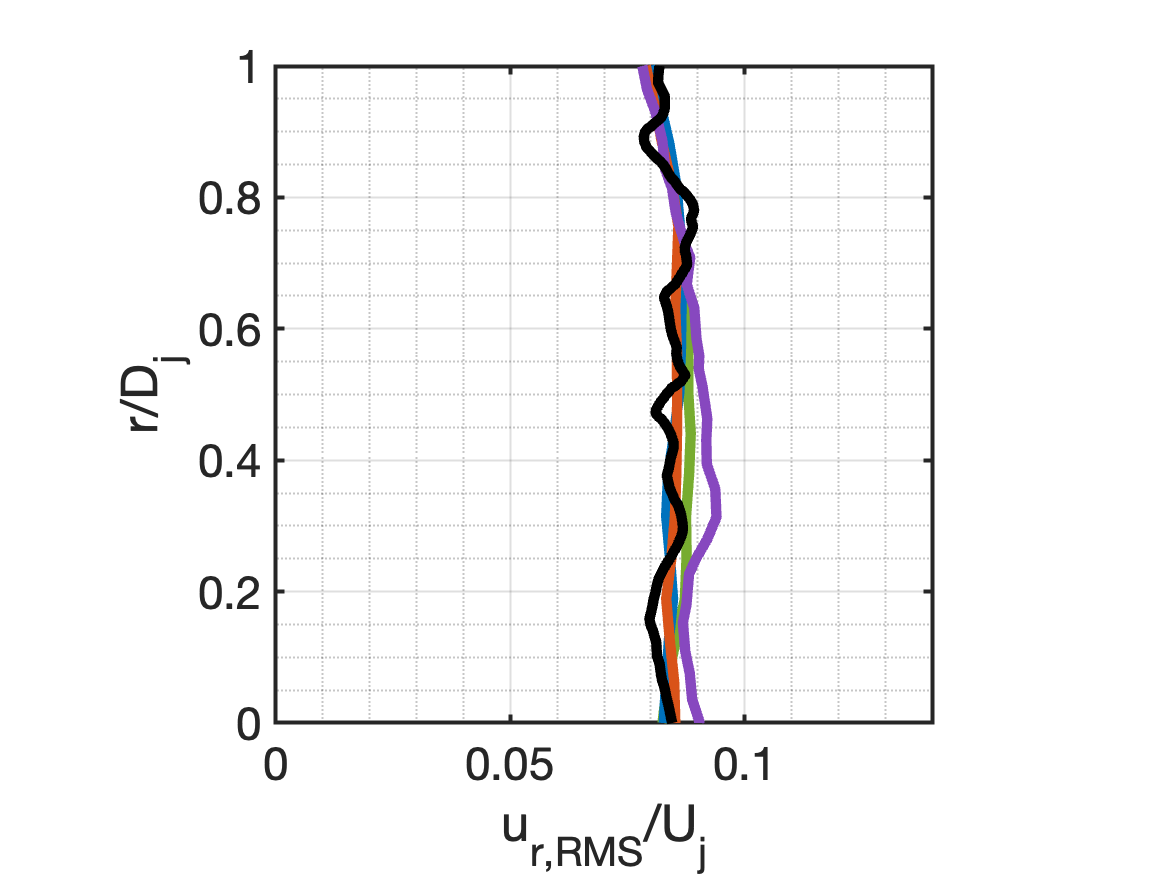}
	\label{res:s1s3_radiall}	
	}
\\
\subfloat[$x/D_j=2.5$]{
	\includegraphics[trim = 20mm 0mm 32mm 8mm, clip, height=0.2\linewidth]{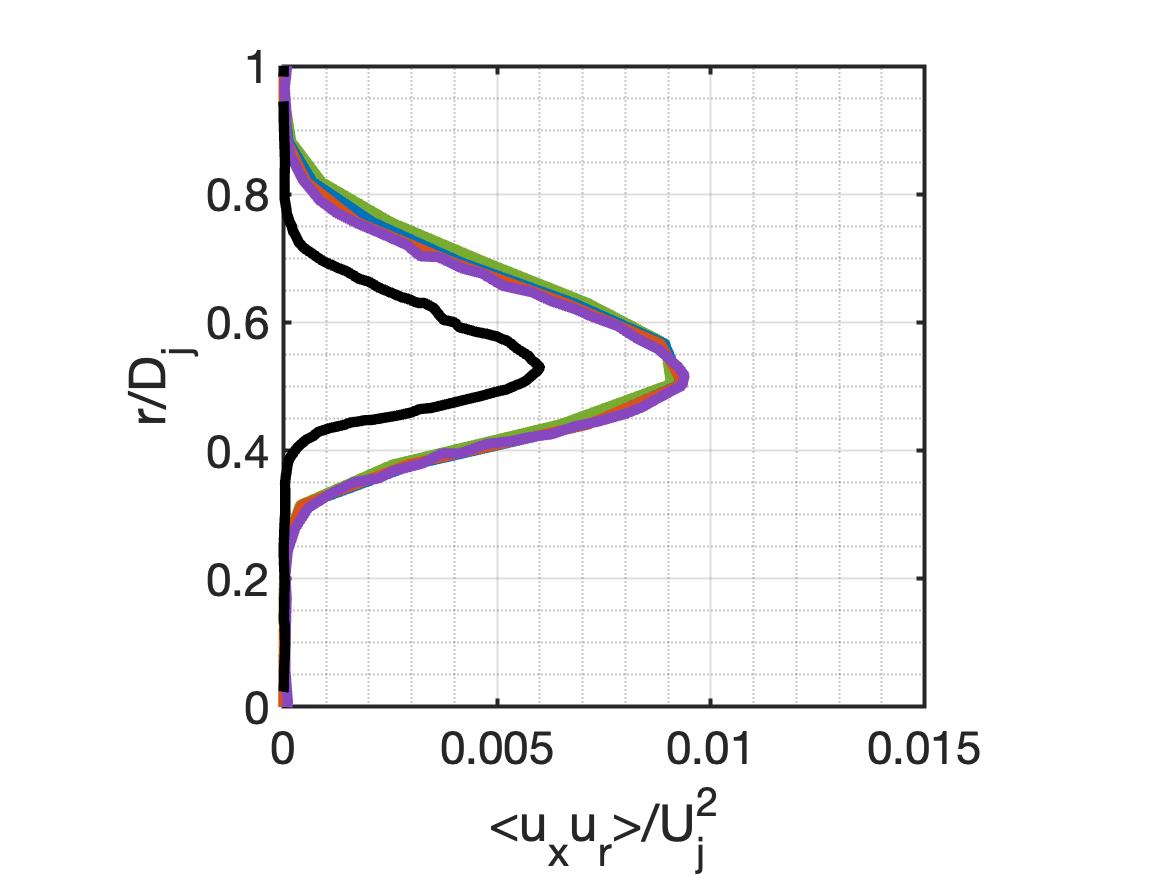}
	\label{res:s1s3_radialm}	
	}
\subfloat[$x/D_j=5.0$]{
	\includegraphics[trim = 20mm 0mm 32mm 8mm, clip, height=0.2\linewidth]{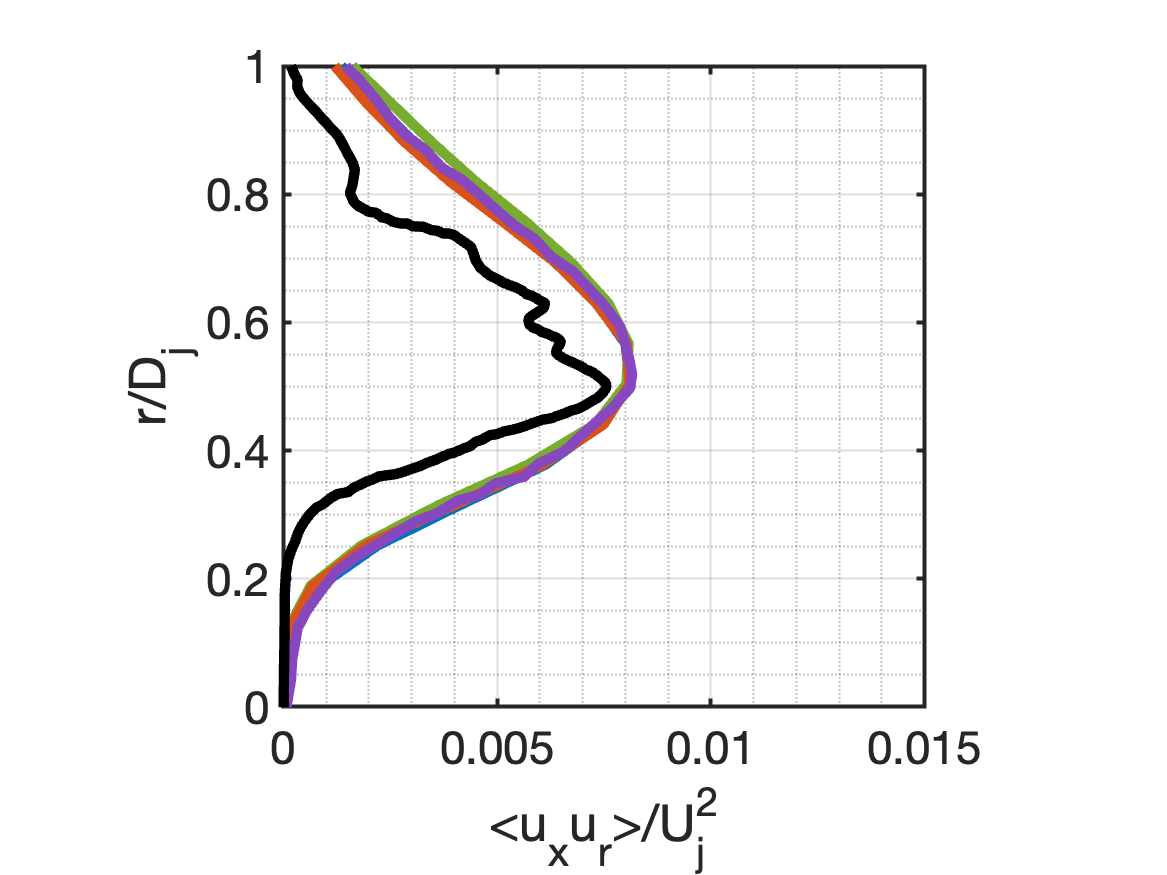}
	\label{res:s1s3_radialn}	
	}
\subfloat[$x/D_j=10.0$]{
	\includegraphics[trim = 20mm 0mm 32mm 8mm, clip, height=0.2\linewidth]{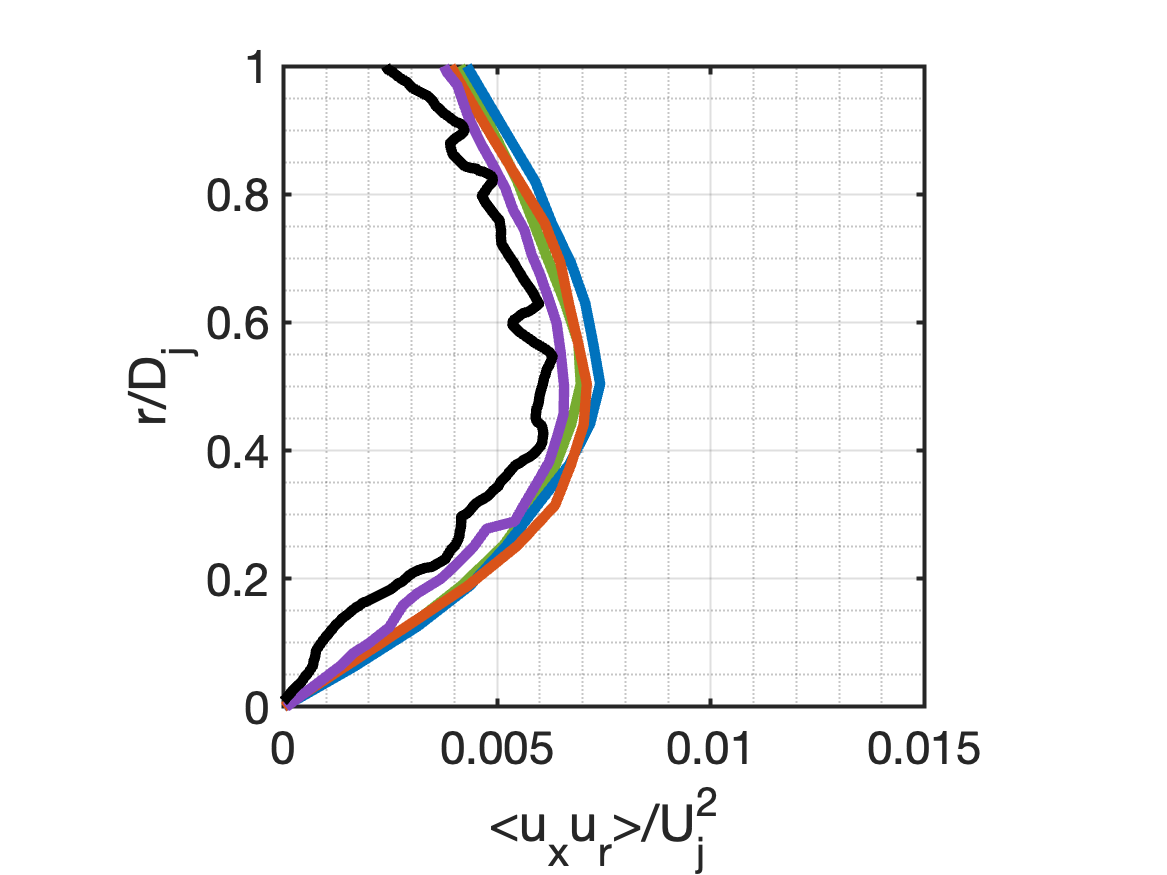}
	\label{res:s1s3_radialo}
	}
\subfloat[$x/D_j=15.0$]{
	\includegraphics[trim = 20mm 0mm 32mm 8mm, clip, height=0.2\linewidth]{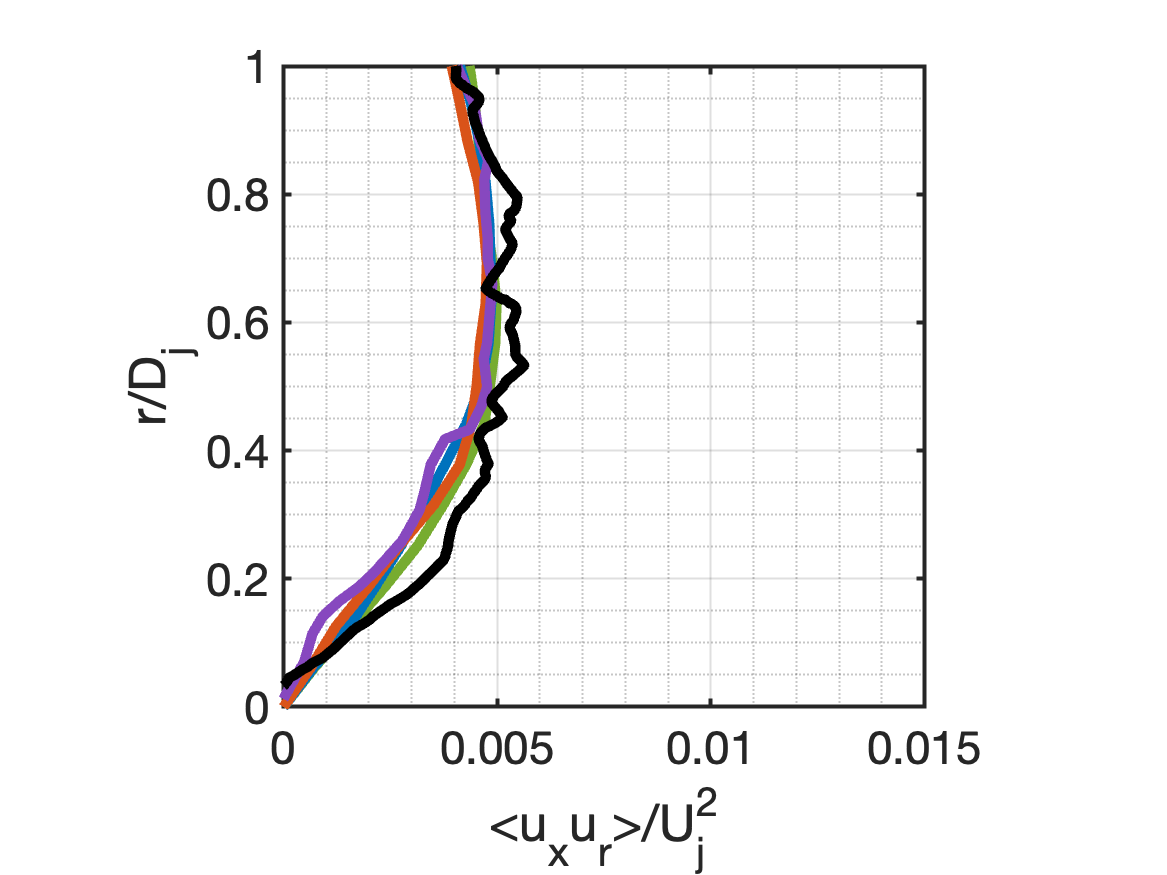}
	\label{res:s1s3_radialp}	
	}
\\
\subfloat{
	\includegraphics[trim = 9mm 8.5mm 6mm 122.5mm, clip, height=0.055\linewidth]{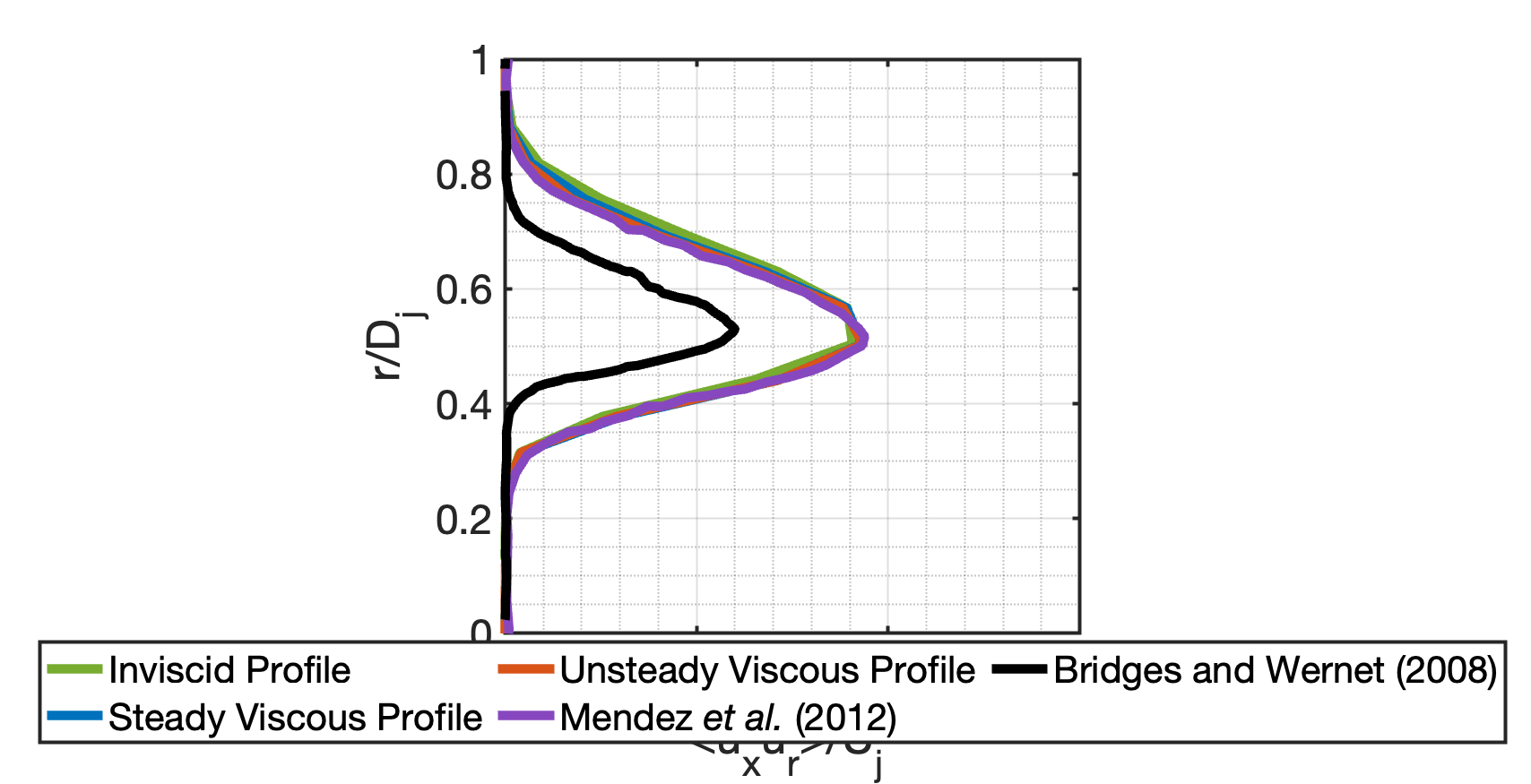}
	}
\caption{Radial profiles of the mean longitudinal velocity component, RMS values of the longitudinal velocity component fluctuations, RMS values of the radial velocity component fluctuations, and mean shear stress tensor component are presented in four longitudinal stations: $x/D_j=2.5$, $x/D_j=5.0$, $x/D_j=10.0$, and $x/D_j=15.0$.}
\label{res:s1s3_radial}
\end{figure*}

This section continues the study evaluating the radial profiles of the mean longitudinal velocity component, the RMS values of the longitudinal and radial velocity fluctuations, and the shear stress tensor component at four streamwise locations: $x/D_j=2.5$, $x/D_j=5.0$, $x/D_j=10.0$, and $x/D_j=15.0$. As in the previous discussion, the profiles obtained in the present work using different inlet boundary conditions are compared with both numerical results from \citet{Mendezetal2012}, calculated at a Reynolds number of $1.5 \times 10^{5}$,  and experimental data from \citet{BridgesWernet2008}, as shown in Fig.\ \ref{res:s1s3_radial}.

The profiles of the mean longitudinal velocity component are shown in Figs.\ \ref{res:s1s3_radiala} to \ref{res:s1s3_radiald}. The results from the jet simulations are in good agreement with each other. At the first two stations, $x/D_j=2.5$ and $x/D_j=5.0$, the profiles show good agreement in terms of peak values, although slight differences in shape are observed. At the last two stations, $x/D_j=10.0$ and $x/D_j=15.0$, the peak values predicted by the numerical simulations are underpredicted compared to the experimental reference. The overall shape of the profiles is well captured.

The radial profiles of the fluctuating quantities $u_{x,\mathrm{RMS}}/U_j$, $u_{r,\mathrm{RMS}}/U_j$, and $\langle u_x u_r \rangle$, shown in the second, third, and fourth rows of Fig.\ \ref{res:s1s3_radial}, exhibit consistent trends and good agreement among all four simulations. At the first two stations, Figs.\ \ref{res:s1s3_radiale} and \ref{res:s1s3_radialf}, the simulations accurately capture the profile shapes of the experimental data for all three fluctuating quantities. Although they slightly overpredict the peak values around $r/D_j = 0.5$. At the third station, the numerical $u_{x,\mathrm{RMS}}/U_j$ profiles agree well with the experimental data for $r/D_j \geq 0.4$, but show discrepancies for $r/D_j \leq 0.4$. At the last station, Fig.\ \ref{res:s1s3_radialh}, all simulations underpredict the peak of $u_{x,\mathrm{RMS}}/U_j$ compared to the experimental data in the range $0.2 < r/D_j < 0.9$. The numerical $u_{r,\mathrm{RMS}}/U_j$ profiles accurately capture the profile shapes of the experimental data with a slight overprediction of the peak values around $r/D_j = 0.5$ at the first two stations, Figs.\ \ref{res:s1s3_radiali} and \ref{res:s1s3_radialj}. The $u_{r,\mathrm{RMS}}/U_j$ profiles are overestimated numerically at $x/D_j = 10.0$, Fig.\ \ref{res:s1s3_radialk}, but match well with the experimental reference at $x/D_j = 15.0$, Fig.\ \ref{res:s1s3_radiall}. A similar behavior is observed in the shear stress tensor component, $\langle u_x u_r \rangle/U_j^2$, where all simulations slightly overpredict the experimental profiles for $x/D_j = 2.5$, $x/D_j = 5.0$, and $x/D_j = 10.0$, Figs.\ \ref{res:s1s3_radialm} to \ref{res:s1s3_radialo}, but show good agreement with the experimental data at $x/D_j = 15.0$, Fig.\ \ref{res:s1s3_radialp}.

\subsubsection{Power Spectral Density Analyses}

\begin{figure*}[htb]
\centering
\subfloat[$x/D_j=2.5$]{
	\includegraphics[width=0.48\linewidth]{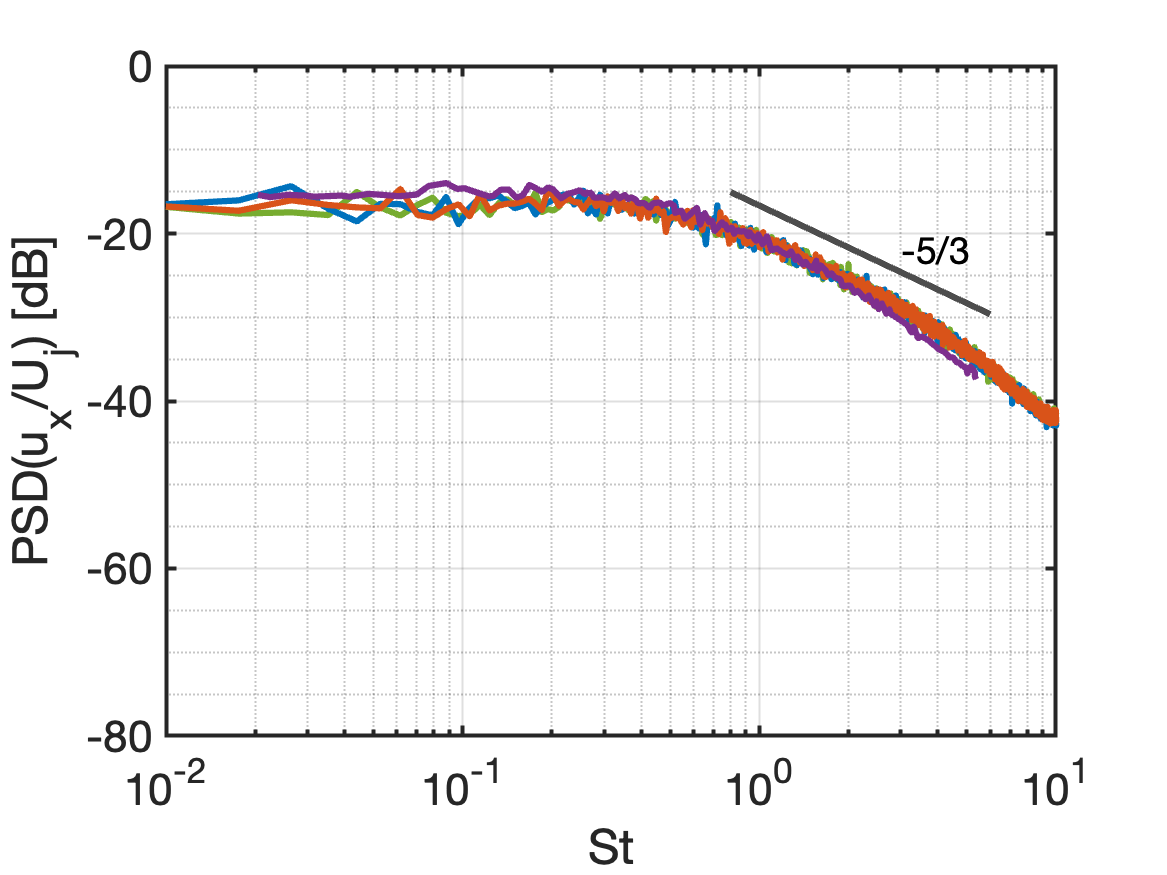}
	\label{res:lpl_uxd1_s1s3}
	}
\subfloat[$x/D_j=5.0$]{
	\includegraphics[width=0.48\linewidth]{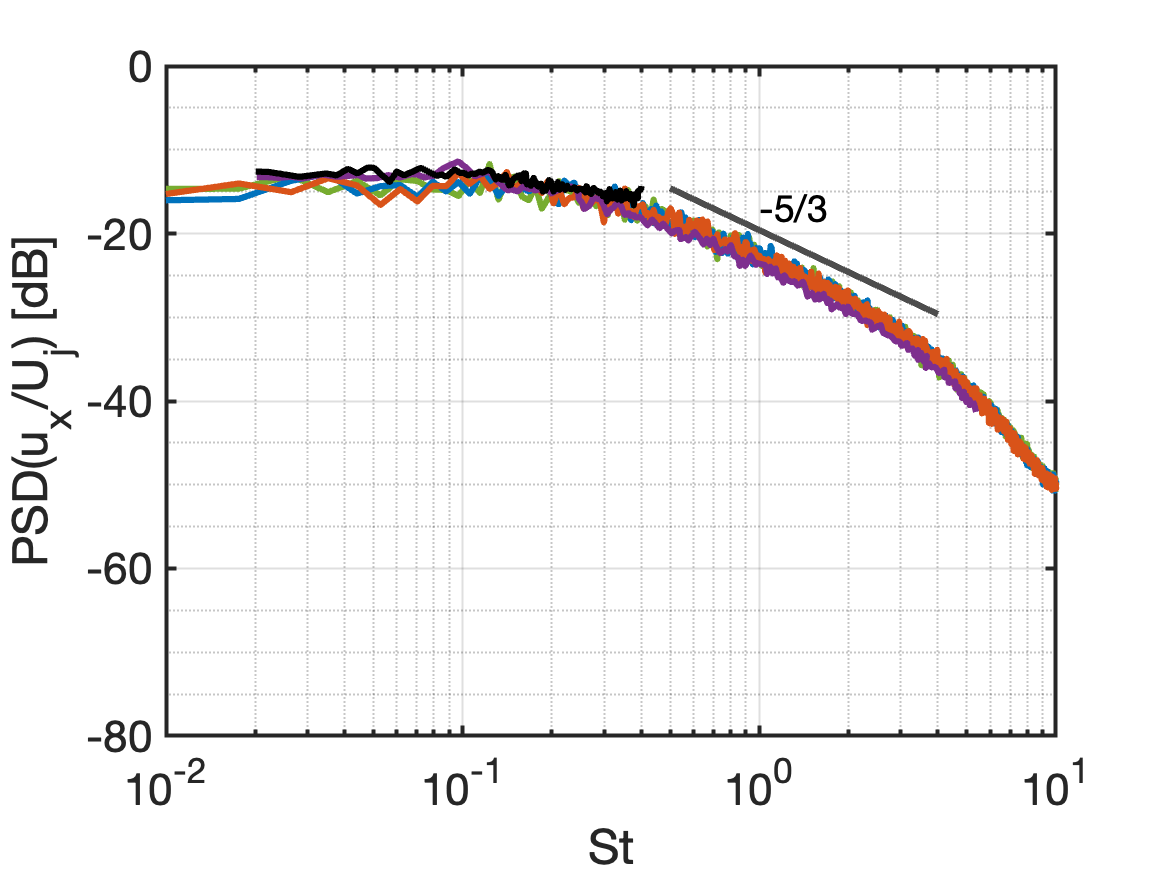}
	\label{res:lpl_uxd2_s1s3}
	}
\\
\subfloat[$x/D_j=10.0$]{
	\includegraphics[width=0.48\linewidth]{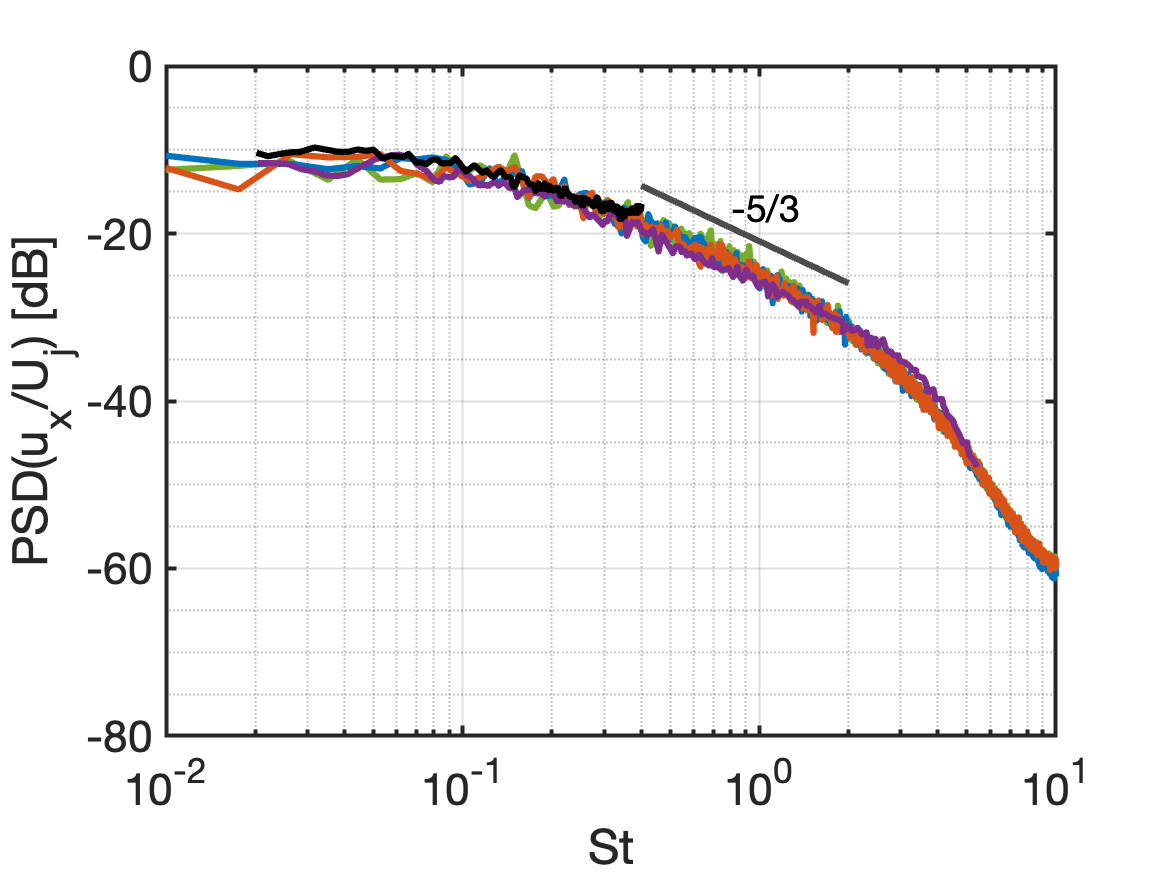}
	\label{res:lpl_uxd3_s1s3}
	}
\subfloat[$x/D_j=15.0$]{
	\includegraphics[width=0.48\linewidth]{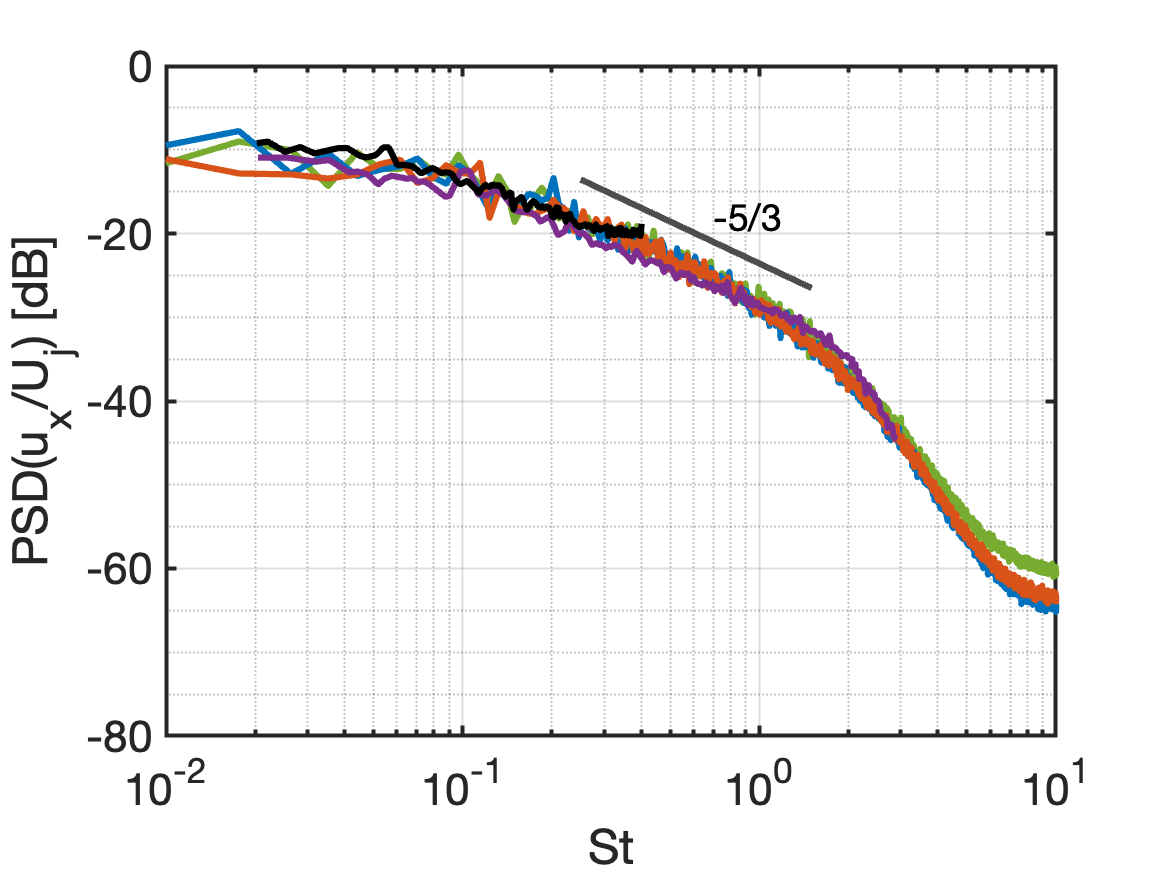}
	\label{res:lpl_uxd4_s1s3}
	}
\\
\subfloat{
	\includegraphics[trim = 9mm 8.5mm 6mm 122.5mm, clip, height=0.055\linewidth]{legend_s1tos3.png}
	}
\caption{Power spectral densities of the longitudinal velocity component fluctuation at four streamwise planes. PSD values obtained along the lipline of the jet flow.}
\label{res:lpl_u_s1s3}
\end{figure*}

\begin{figure*}[htb]
\centering
\subfloat[$x/D_j=2.5$]{
	\includegraphics[width=0.47\linewidth]{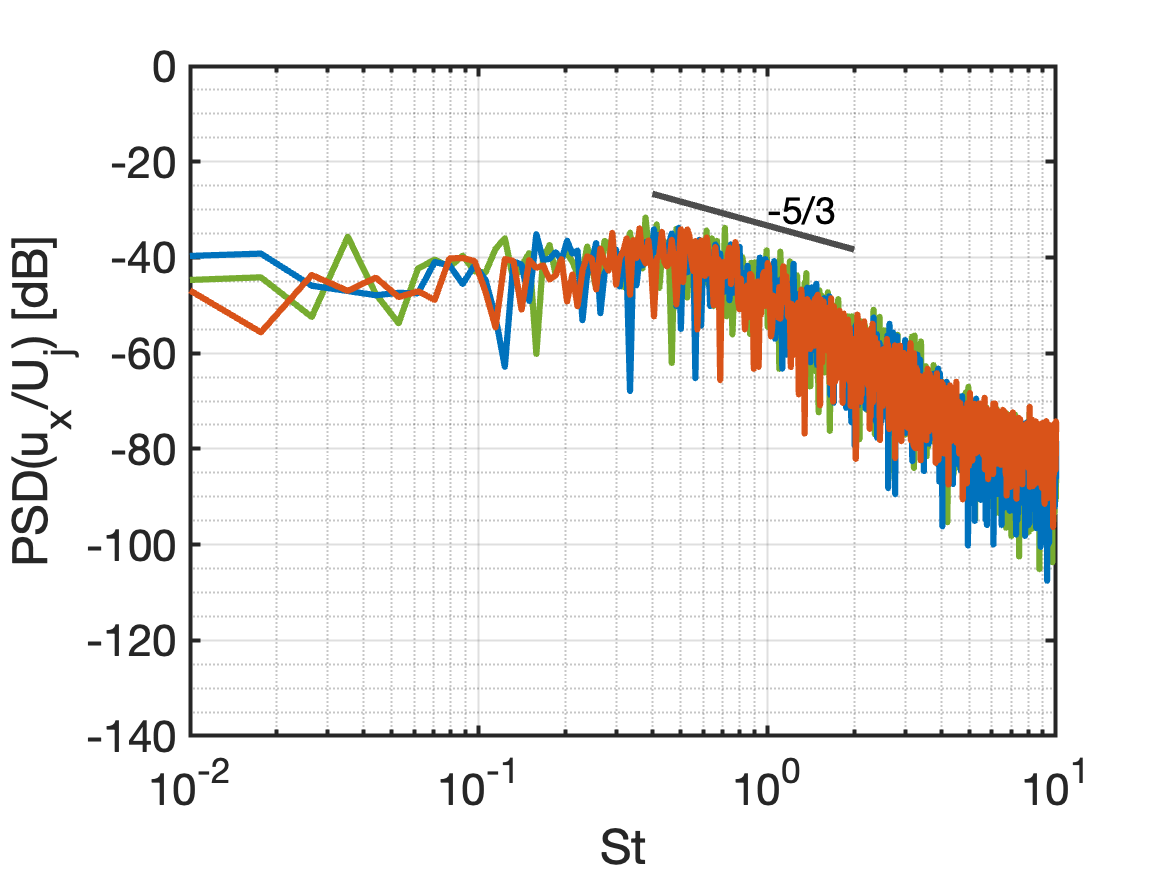}
	\label{res:ctl_uxd1_s1s3}
	}
\subfloat[$x/D_j=5.0$]{
	\includegraphics[width=0.47\linewidth]{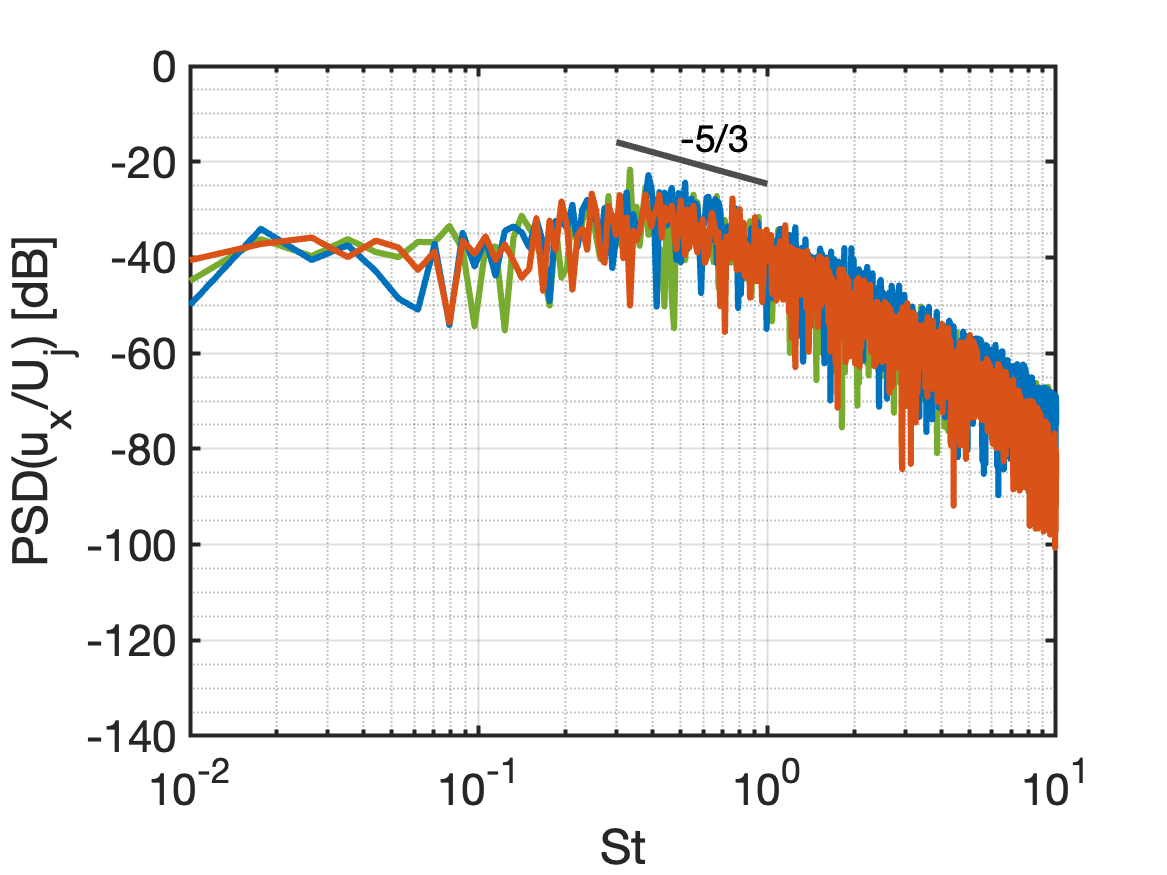}
	\label{res:ctl_uxd2_s1s3}
	}
\\
\subfloat[$x/D_j=10.0$]{
	\includegraphics[width=0.47\linewidth]{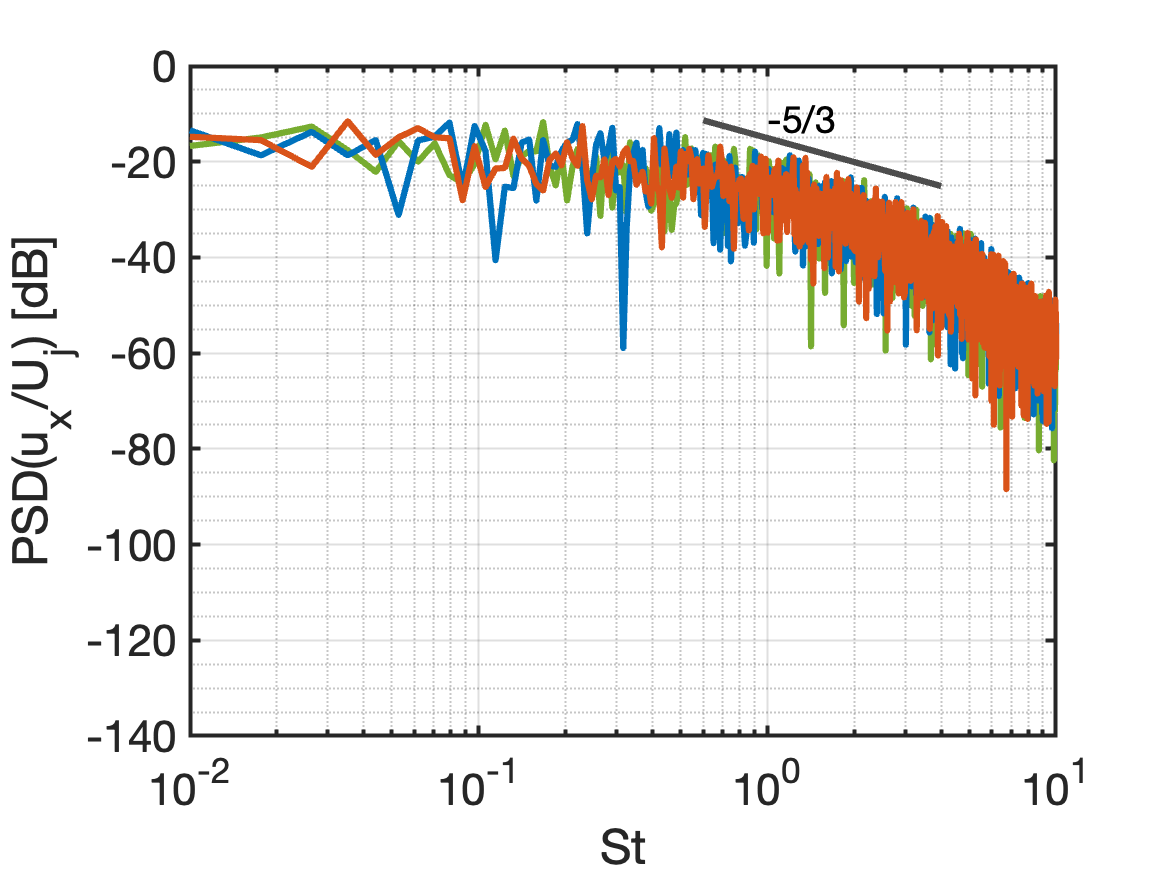}
	\label{res:ctl_uxd3_s1s3}
	}
\subfloat[$x/D_j=15.0$]{
	\includegraphics[width=0.47\linewidth]{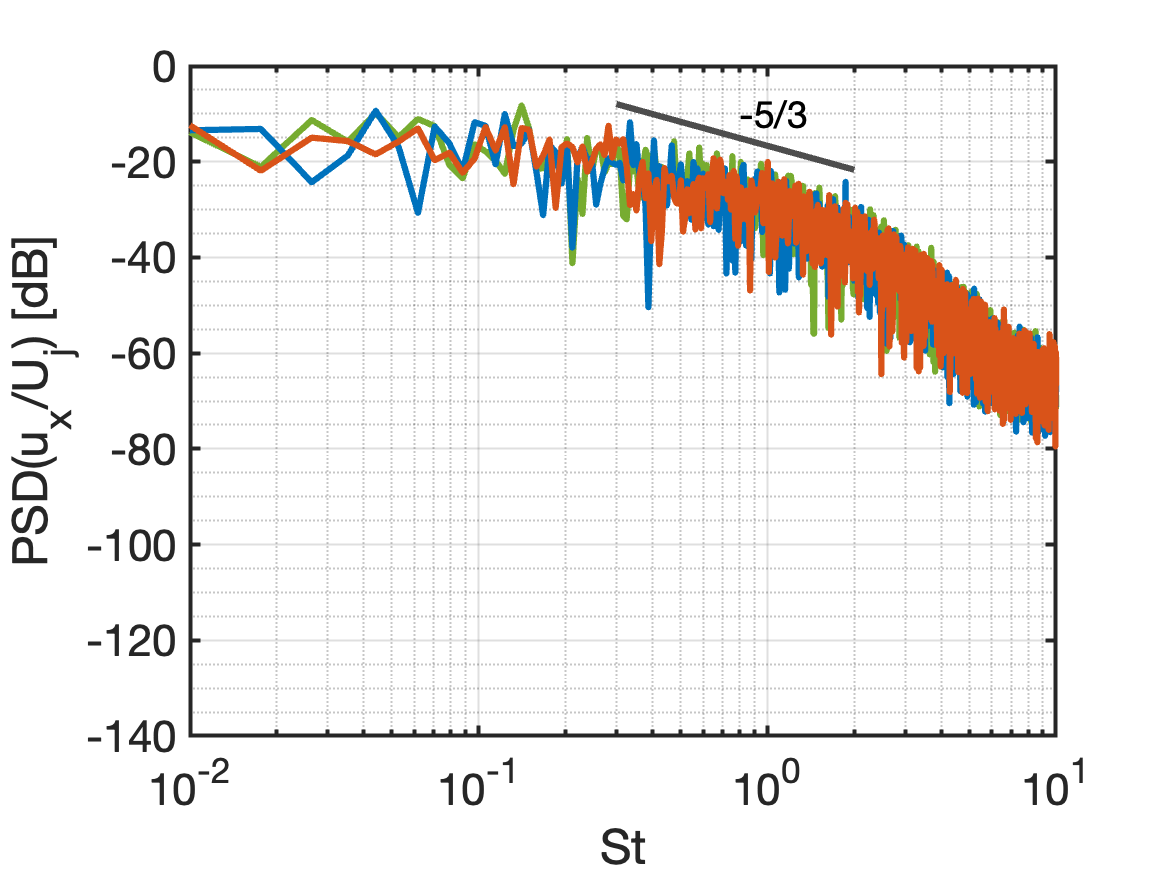}
	\label{res:ctl_uxd4_s1s3}
	}
\\
\subfloat{
	\includegraphics[trim = 19mm 8.5mm 17mm 122.5mm, clip, height=0.055\linewidth]{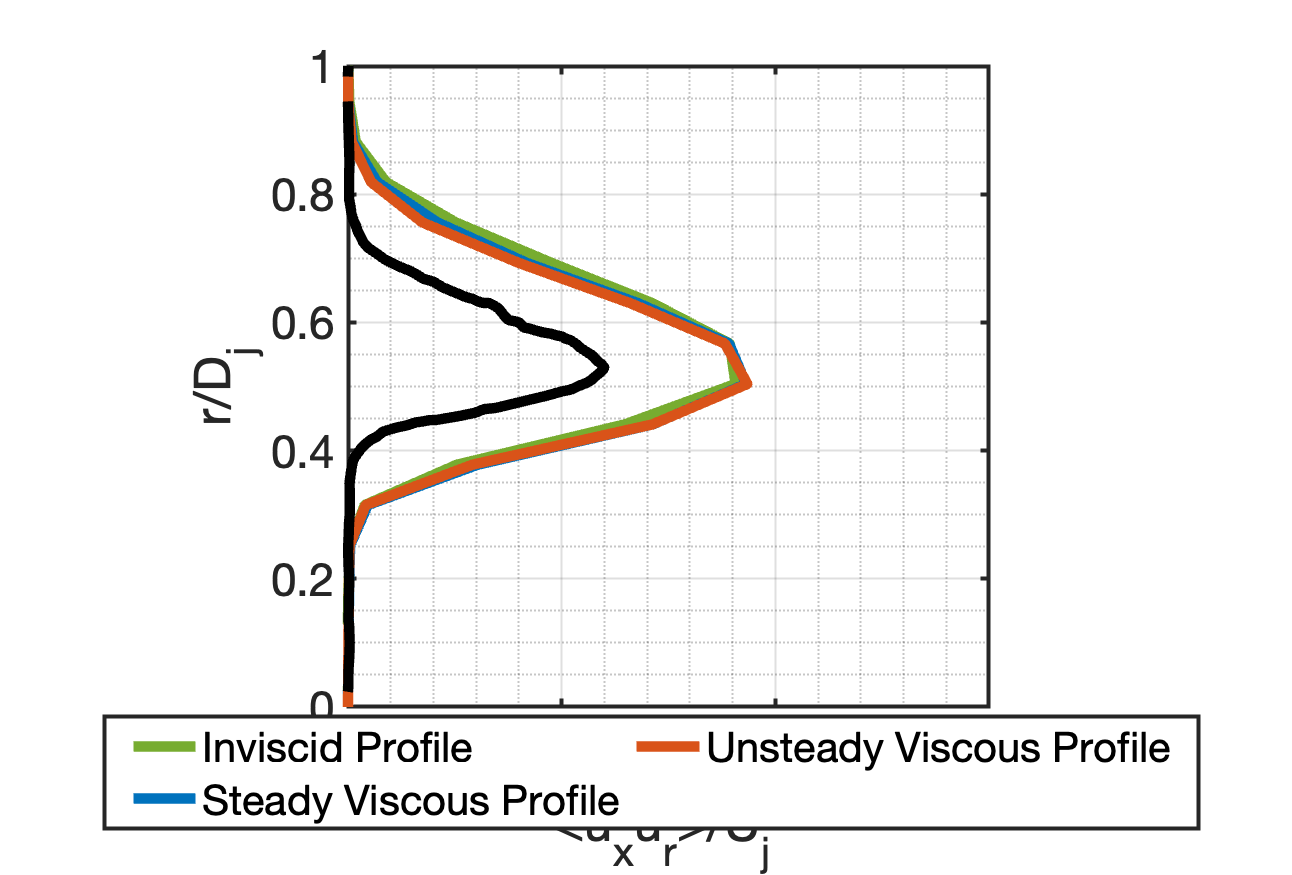}
	}
\caption{Power spectral densities of the longitudinal velocity component fluctuation at four streamwise planes. PSD values obtained along the centerline of the jet flow.}
\label{res:ctl_u_s1s3}
\end{figure*}

Power spectral density (PSD) is applied here to understand how turbulent kinetic energy is distributed across frequencies. The longitudinal velocity component is used to evaluate the three different inlet boundary conditions against numerical and experimental data reported in the work of \citet{Mendezetal2012}. The data are collected at the same four axial stations where the radial profiles were analyzed, {\it i.e.}, $x/D_j=2.5$, $x/D_j=5.0$, $x/D_j=10.0$, and $x/D_j=15.0$, both along the lipline, Fig.\ \ref{res:lpl_u_s1s3}, and the centerline, Fig.\ \ref{res:ctl_u_s1s3}, of the jet flow. 
The numerical values from the three numerical simulations investigated in the present work along the jet lipline consist of an average over $120$ evenly distributed azimuthal positions. The centerline profile consists of only one set of data. Since the time sample is long enough to compare with experimental and other numerical data, no additional treatment is applied to the velocity signals.

It is important to note that the experimental reference of \citet{BridgesWernet2008} does 
not include PSD data for the isothermal jet configuration, although data are available for 
a heated jet case. Unfortunately, the format in which these data were reported in the 
original publication does not allow for the direct extraction of the information required 
for comparison. As reported by \citet{Mendezetal2012}, the authors of that study obtained 
access to the original experimental dataset of \citet{BridgesWernet2008} and were, therefore, 
able to compare their numerical results with the experimental data. Consequently, the 
experimental data presented in Fig.\ \ref{res:lpl_u_s1s3} of the present paper were taken 
from \citet{Mendezetal2012}. However, since these data were originally generated in the 
experiments by \citet{BridgesWernet2008}, they are referred to as the Bridges and Wernet 
experimental data in Fig.\ \ref{res:lpl_u_s1s3}.
Therefore, the present analysis relies on data from the perfectly expanded heated jet 
configuration reported by \citet{Mendezetal2012}, corresponding to a temperature ratio of 
approximately $1.76$. 
As previously pointed out, the computations of \citet{Mendezetal2012} for the heated jet configuration considered a Reynolds number of $7.6 \times 10^{4}$, which is significantly smaller than the Reynolds number for the flow in the experiments and for the simulations performed in the present work. 
Moreover, the references cited here only provide PSD data along the jet lipline. Hence, in summary, we are comparing here PSDs for an isothermal jet at a high Reynolds number, calculated in the present work, with experimental data for a heated jet, also obtained at a high Reynolds number, and computational data for a heated jet calculated at a much lower Reynolds number.

According to \citet{Pope2000}, three frequency fluctuation ranges can be defined and related to the turbulent structures in the flow: energy-containing, inertial, and dissipation ranges. The energy-containing range corresponds to the largest eddies, associated with low-frequency values. This is the region where most of the turbulent kinetic energy is concentrated, and its characteristics depend on the specific flow configuration. The inertial range is identified by the $-5/3$ power law behavior of the velocity spectrum. It is in this frequency range that energy from large eddies is transported to small eddies. The dissipation range, linked to the smallest turbulent structures, appears at higher Strouhal numbers, where the velocity spectrum departs from the $-5/3$ slope. It is in this region that most of the turbulent energy dissipation is observed. Due to the spatial development of the jet flow, the Strouhal number at which these transitions occur may vary across different regions of the flow field.

The PSD analysis of longitudinal velocity fluctuations at the jet lipline, Fig.\ \ref{res:lpl_u_s1s3}, shows that the three inflow profiles yield similar trends, with results in good agreement with the numerical reference. The Strouhal number range accessible in physical experiments is limited by the measurement apparatus, reaching a maximum of $St \approx 0.4$. As a result, the experimental data barely extend into the inertial range, though agreement with the numerical results is still evident. At the jet lipline, the inertial range shifts toward lower Strouhal numbers with increasing distance from the nozzle exit, accompanied by higher power density levels in the energy-containing range. The reference numerical data of \citet{Mendezetal2012} display abrupt changes in spectral slope near $St \approx 4$ at $x/D_j = 10.0$ and $St \approx 1$ at $x/D_j = 15.0$, whereas the present simulations exhibit smooth spectral behavior across frequencies.

The PSD values at the jet centerline depicted in Fig.\ \ref{res:ctl_u_s1s3} show consistent behavior across the three inflow conditions. The results indicate that the inflow condition has only a minor influence on the spectral distribution of velocity fluctuations. Data from successive stations reveal an increase in power density within the energy-containing range as the jet develops downstream. The inertial range appears near $St \approx 1.0$ at all four investigated stations.

The power spectral density analyses do not show any relevant difference in the spectral behavior of the velocity signals with the three imposed inflow conditions. At the four investigated stations along the lipline and the centerline, the power levels in the energy-containing region are similar to those from the physical experiments and the numerical reference, and the frequency signals of the velocity fluctuations are all very similar. The comparison of the numerical data from the three numerical simulations with the numerical reference and the experimental data endorses the quality of the large-eddy simulations performed in the present work.

\section{Concluding Remarks}
\label{sec:conclusion}

Large-eddy simulations of a perfectly expanded free jet flow were performed to investigate the influence of different inflow boundary conditions on the flow field and time-averaged properties. The simulations employed a nodal discontinuous Galerkin spatial discretization, and the solver was validated against both numerical and experimental data from the literature. All numerical data generated in this study are made available to the scientific community.

Validation against the solver of \citet{Mendezetal2012}, as well as the experimental dataset of \citet{BridgesWernet2008}, showed good agreement in the velocity profiles. In particular, the results obtained with the nodal discontinuous Galerkin scheme were consistent with those of \citet{Mendezetal2012}, and both datasets reproduced the experimental trends, especially in the region farther downstream of the inlet, {\em i.e.} $x/D_j > 5.0$.

In line with the objectives stated in the introduction, the present study 
was designed to assess, in a controlled and systematic manner, the sensitivity of a 
perfectly expanded supersonic jet to progressively enriched inflow boundary conditions, 
ranging from a simplified inviscid profile to steady and unsteady viscous 
representations. By adopting this gradient of inflow complexity, the results demonstrate 
that incorporating realistic mean viscous effects at the nozzle exit has a measurable 
impact on near-field jet development, whereas the additional introduction of inflow 
unsteadiness through the present tripping approach produces only marginal changes in 
time-averaged and second-order statistics.

Three inflow boundary conditions were examined: an inviscid profile with constant properties, a steady viscous profile obtained from a RANS nozzle simulation, and an unsteady viscous profile generated by applying a tripping method \cite{BogeyMarsdenBailly2011} to the viscous boundary layer. The centerline velocity distributions indicated that the inviscid inflow produced a slightly longer potential core compared with the steady and unsteady viscous cases. Both viscous profiles introduced a nearfield velocity deficit at the lipline and reduced the peak RMS values of longitudinal velocity fluctuations by approximately $10\%$, shifting the location of maximum fluctuations closer to the nozzle exit.
Radial profiles at four downstream stations revealed only minor differences among the three inflow conditions, which did not alter the overall trends.

The RMS profiles of radial velocity fluctuations and the shear stress component show similar behavior, with both, steady and unsteady, viscous inflows consistently reducing the peak values. However, these values remain higher than the experimental measurements in the near-nozzle region. The unsteady viscous inflow produces results comparable to the steady viscous case, indicating that the tripping method implemented here is not as effective as that used in the original work of \citet{BogeyMarsdenBailly2011}. 
Future studies should investigate alternative tripping strategies or parameter variations to better assess their influence on jet development. 
In particular, recycling-rescaling inflow techniques \cite{Morganetal11,Xiao2017} could be used to generate fully developed turbulent boundary layers at the nozzle exit, while preserving the correct mean profiles and integral turbulence quantities. On the other hand, synthetic turbulence methods \cite{Patruno18,Patterson_Balin_Jansen_2021,Fukami_etal_19,DiMare_06} offer fine control over fluctuation levels and spectra at the inflow, enabling systematic studies of how nozzle-exit turbulence structure affects jet development and near-nozzle RMS levels.

Power spectral density analyses of the streamwise velocity fluctuations confirmed that the imposed inflow conditions had little effect on spectral distributions. Agreement with numerical and experimental PSD values was observed within the accessible Strouhal range, although the experiments were limited to lower frequencies. These findings indicate that enriching inflow conditions with additional physical content improves agreement with nozzle-included simulations, particularly near the jet inlet.

Finally, the present work provides a comprehensive high-fidelity database of supersonic jet simulations, including both the present inflow-condition study and the resolution study of \citet{Abreuetal2024}. The database offers a large time series of high-frequency data in regions of interest, significantly larger than existing experimental or numerical datasets. Files are supplied in structured \textit{vtk} \cite{schroeder98} format, ensuring compatibility with open-source post-processing tools. Beyond serving as a benchmark for future numerical validation, this database supports emerging data-driven approaches. In particular, it enables the training of machine learning models for turbulence modeling, which have shown potential to reduce computational costs while preserving predictive accuracy \cite{DEZORDOBANLIAT24,Amarloo23,Brunton20,Ling16,Fukami19,Oulghelou2025}.

\section*{Acknowledgements}
The authors acknowledge the financial support provided by Fundação de Amparo à Pesquisa do Estado de São Paulo, FAPESP, under the Research Grants No.\ 2013/07375-0 and 2021/11258-5. The authors also acknowledge the support provided by Conselho Nacional de Desenvolvimento Científico e Tecnológico, CNPq, under the Research Grant No.\ 315411/2023-6. The availability of the computational resources from the Center for Mathematical Sciences Applied to Industry, CeMEAI, funded by FAPESP under the Research Grant No.\ 2013/07375-0, and from the SDumont supercomputer from the National Laboratory for Scientific Computing, LNCC/MCTI, under the HFWBTF project, are also gratefully acknowledged. The work was also granted access to the HPC resources of IDRIS under the allocations A0132A12067, A0152A12067, A0172A12067, and A0192A12067 made by GENCI. The first author acknowledges authorization by his employer, Embraer S.A., which has allowed his participation in the present research effort.

\section*{Data Availability}

The data that support the findings of this study are openly available in Zenodo at 
\url{https://doi.org/10.5281/zenodo.13909193},
\url{https://doi.org/10.5281/zenodo.13908216}, and
\url{https://doi.org/10.5281/zenodo.13908446}, with respective reference numbers {\em 13909193}, {\em 13908216}, and {\em 13908446}.

\bibliographystyle{aipnum4-1-custom}
\bibliography{pof}

\end{document}